\documentclass[preprint2]{proto}
\usepackage{times}
\usepackage{xcolor}
\usepackage{graphicx}
\usepackage{wasysym}
\usepackage{float}
\usepackage{dblfloatfix}

\graphicspath{{./}{Figures/}}

\newcommand\psj{Planetary Science Journal}
\newcommand\maps{Meteoritics and Planetary Science}

\usepackage{url}

\usepackage{enumitem}

\newcommand{\Rosetta}{{\it Rosetta}}

\newcommand{\jwst}{{\it JWST}}
\newcommand{\spitzer}{{\it Spitzer}}
\newcommand{\Spitzer}{{\it Spitzer}}
\newcommand{\SpitzerIRS}{{\it Spitzer}~IRS}
\newcommand{\sofia}{{\it SOFIA}}

\newcommand{\isosws}{{\it ISO~SWS}}
\newcommand{\DI}{{\it Deep Impact}}
\newcommand{\Stardust}{{\it Stardust}}
\newcommand{\Herschel}{{\it Herschel}}
\newcommand{\ALMA}{{\it ALMA}}

\newcommand{\tempelone}{9P/Tempel~1} 
\newcommand{\nineP}{9P/Tempel~1} 
\newcommand{\CG }{67P/C-G} 
\newcommand{\Chury}{67P/Churyumov-Gerasimenko} 
\newcommand{\WMone}{C/2000~WM1~(LINEAR)}
\newcommand{\Halley}{1P/Halley} 
\newcommand{\HB}{Hale-Bopp} 
\newcommand{\HaleBopp}{C/1995~O1 (Hale-Bopp)} 
\newcommand{\Hartley}{103P/Hartley~2} 
\newcommand{\PWildtwo}{81P/Wild~2} 
\newcommand{\Wild}{Wild~2} 
\newcommand{\Holmes}{17P/Holmes} 

\newcommand{\nm}{$nm$}
\newcommand{\um }{$\mu m$}
\newcommand{\wvnum}{$cm^{-1}$}

\newcommand{\rh }{$r_h$}
\newcommand{\km }{$km$}
\newcommand{\au }{$au$}

\newcommand{\qabs}{Q$_{\lambda, abs}$}
\newcommand{\qsca}{Q$_{\lambda, sca}$}
\newcommand{\qabsir}{Q$_{abs, IR}$}
\newcommand{\qabsvis}{Q$_{abs, UVIS}$}

\newcommand{\Ga}{$G(a)$} 

\newcommand{\HtwoO}{H$_2$O}
\newcommand{\COtwo}{CO$_2$}

\newcommand{\ltsimeq}{\raisebox{-0.6ex}{$\,\stackrel
        {\raisebox{-.2ex}{$\textstyle <$}}{\sim}\,$}}
\newcommand{\gtsimeq}{\raisebox{-0.6ex}{$\,\stackrel
        {\raisebox{-.2ex}{$\textstyle >$}}{\sim}\,$}}

\begin{document}

\title{\textbf{\LARGE CHEMICAL AND PHYSICAL PROPERTIES OF COMETARY DUST}} 

\author {\textbf{\large C. Engrand}}
\affil{\small\em IJCLab, Université Paris-Saclay/CNRS-IN2P3, UMR9012, 91405 Orsay Campus, France}

\author {\textbf{\large J. Lasue}}
\affil{\small\em IRAP, Université de Toulouse, CNRS, CNES, 31400 Toulouse, France}

\author {\textbf{\large D. H. Wooden}}
\affil{\small\em NASA Ames Research Center, MS 245-3, Moffett Field, CA
94035-0001, USA}

\author {\textbf{\large M. E. Zolensky}}
\affil{\small\em NASA Johnson Space Center, ARES, X12 2010 NASA Parkway,
Houston, TX 77058-3607, USA}

\begin{abstract}

\begin{list}{ } {\rightmargin 0.3in}
\baselineskip = 11pt
\parindent=1pc
{\small 
Cometary dust particles are the best preserved remnants of the matter present at the onset of the formation of the Solar System. Space missions, telescopic observations and laboratory analyses advanced the knowledge on the properties of cometary dust. The only samples with an ascertained cometary origin were returned by the {\it Stardust} space mission from comet 81P/Wild 2. The ``chondritic porous" (here called ``chondritic anhydrous") interplanetary dust particles (CA-IDPs) and micrometeorites (CP-MMs), and the ultracarbonaceous Antarctic MMs (UCAMMs) also show strong evidence for a cometary origin. Astronomical observations show that cometary infrared (IR) spectra can generally be modeled using five families of particles : amorphous minerals of olivine and pyroxene compositions, crystalline olivines and pyroxenes, and amorphous carbon. From the analyses of space missions {\it Giotto}, {\it Vega-1\&2}, \Stardust, {\it Rosetta} and of CA-IDPs, CP-MMs and UCAMMs, the elemental composition of cometary dust is generally consistent with the chondritic composition (as defined by the composition of the CI, or Ivuna-type, carbonaceous chondrites), with the notable exception of elevated contents in carbon (and possibly nitrogen) compared to CI. As seen in CA-IDPs CP-MMs and UCAMMs, the organic matter of cometary dust is mixed with minor amounts of crystalline (10\% to 
\gtsimeq 25\% of the minerals) and amorphous mineral phases. The most abundant crystalline minerals are ferromagnesian silicates (olivine and low-Ca pyroxenes), but high-Ca pyroxenes, refractory minerals and low-Ni Fe sulfides are also present. The crystalline olivine and low-Ca pyroxene compositions can vary from their Mg-rich end-member (forsterite and enstatite) to relatively Fe-rich compositions. {\it Stardust} samples from comet 81P/Wild 2 contain in particular olivines and pyroxenes with minor element abundances linking them to unequilibrated ordinary chondrites. Refractory minerals as well as secondary minerals like low-Fe, Mn-enriched (LIME) olivines, unusual Fe sulfides or mineral aggregates of specific compositions like Kosmochloric high-Ca pyroxene and FeO-rich olivine (KOOL grains) are also found in CA-IDPs, CP-MMs and Stardust samples. The presence of carbonates in cometary dust is still debated (both in astronomical observations and in samples analyzed {\it in situ} or in the laboratory), as well as the presence of hydrated silicates, as proposed after the {\it Deep-Impact} mission. A phyllosilicate-like phase was however observed in a UCAMM, and a CP-MMs shows a mixing texture with an hydrated part. The abundance of pyroxene to olivine (in numbers) in CA-IDPs, CP-MMs and UCAMMs is larger than in primitive meteorites (e.g.\ the Px/Ol ratio is usually larger than 1). GEMS phases (glass with embedded metals and sulfides) are abundant in cometary dust, although not systematically found. Some of the organic matter present in cometary dust particle resembles the insoluble organic matter (IOM) present in primitive meteorites, but amorphous carbon and exotic (e.g.\ N-rich in UCAMMs) organic phases are also present. The hydrogen isotopic composition in the organic matter of cometary dust particles analyzed in the laboratory is usually rich in deuterium, tracing a formation at very low temperatures, either in the protosolar cloud or in the outer regions of the protoplanetary disk. The presolar dust concentration in cometary dust can reach about 1\% (in CA-IDPs collected during  dust streams of comets 26P/Grigg-Skjellerup and 21P/Giacobini-Zinner), which is the most elevated value observed in extraterrestrial samples. 
The size distribution of cometary dust in comet trails is well represented by a power-law distribution (differential size distribution) with a mean power index N typically ranging from -3 to -4. Polarimetric and light scattering studies of cometary dust suggest mixtures of porous agglomerates of sub-micrometer minerals with organic matter, which is compatible with the {\it in situ} analyses of 67P/Churyumov-Gerasimenko dust particles by MIDAS (Rosetta) and with the studies of {\it Stardust} samples, CA-IDPs, CP-MMs and UCAMMs. Cometary dust particles have low tensile strength, and low density, as deduced from  observations and analyses of during the {\it Rosetta} mission.
The presence of high temperature minerals in cometary dust highlights the need for a large scale transport mechanism in the early protoplanetary disk. Many gaps in the understanding of the formation of cometary dust remain, such as the incorporation of minerals from evolved asteroids, the possibility of aqueous alteration on the comet, the formation of organic matter as well as the mixing process(es) with minerals and ices, the apparent small presolar heritage... 
\\~\\~\\~}
\end{list}
\end{abstract}


\label{sec:two}

\section{COMETARY DUST : FROM SPACE MISSIONS TO GROUND-BASED OBSERVATIONS AND TO THE LABORATORY }

\label{sec:two.missions}

\begin{figure*}[t!]
\centering
\includegraphics[width=0.9\textwidth]{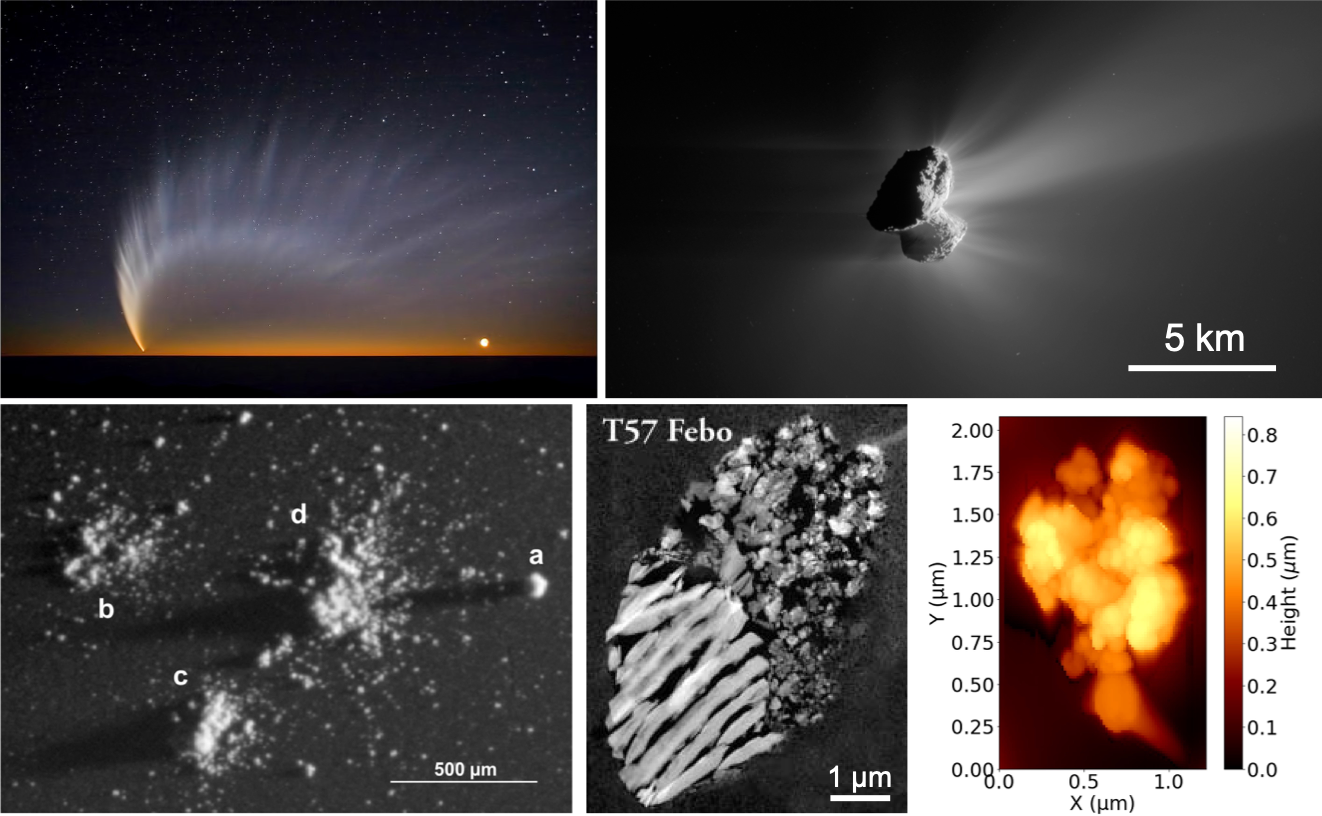}
\caption{\it \small Cometary dust seen at increasing size resolution (from top left to bottom right):  Comet C/2006 P1 McNaught observed from Paranal (commons license); comet 67P/Churyumov-Gerasimenko (67P/C-G) by Rosetta (commons license), cometary dust particles of 67P/C-G collected by COSIMA \citep{Langevin2016}, section of a Stardust particle from 81P/Wild 2 track 57 \citep{Matrajt2008}, 67P/C-G dust particles collected by MIDAS on Rosetta \citep{mannel2019}.}
\label{fig:comet}
\end{figure*}

Space missions are very powerful in advancing the understanding of comets. However, the cost and timeline of such missions have only permitted the characterization of a limited number of comets so far. Spacecraft flybys of comet \Halley{} by {\it Giotto} and {\it Vega-1\&2} with, respectively, the PIA and PUMA 1\&2 instruments, comet \PWildtwo{} (\Stardust), comet \nineP{} (\DI), comet 103P/Hartley 2 (\DI-Extended) and the long duration rendezvous with comet \Chury{} {\it (Rosetta)} provided key advancements in our understanding about the composition and structure of the cometary dust particles, complementing less costly telescopic observations and laboratory work on cometary materials captured in the stratosphere and from polar regions and those returned by the \Stardust \ mission to comet \PWildtwo. 

Ground-based and space-borne observations of comets encompass a wide range of spatial scales, spectral ranges and spectral resolutions. A small perihelion event like with comet C/2006~P1 (McNaught) (Fig.~\ref{fig:comet}) can produce a spectacular release of small to large particles into the coma and dramatic comet tail structures that can be modeled to assess coma particle sizes and modulations in dust production rates.

As the decades advanced from the late 1970s, the instruments and telescopes provided increases in sensitivity and wavelength coverage that allowed multi-epoch studies in the '10~\micron \ window' and limited studies near 20~\micron \ from ground-based telescopes as well as uninterrupted wavelength coverage from $\sim$5--40\micron \ from \isosws , \sofia , \spitzer , and with the majority being single-epoch observations.  The James Webb Space Telescope (\jwst)\ will provide a greater span of wavelength coverage, which facilitates the simultaneous use of scattered light and thermal emission studies to characterize the dust composition and particle properties in cometary comae. Thus, the decline in the availability in mid- and far-IR instrumentation for photometric and spectroscopic studies of comets on ground-based telescopes is complimented by higher sensitivity and broader wavelength coverage airborne and space-based telescopes. Multi-epoch studies have thus far revealed that the dust composition of the coma of an individual can vary significantly with heliocentric distance, potentially due to changes in seasonal illumination that changes the 'active' areas and/or changes in jet activity. Also, the coma's dust composition can appear to change composition, i.e., increase the relative abundance of crystalline silicates because a decrease in heliocentric distance (\rh ) causes increases in solar insolation whereby dust components that are less absorbing of sunlight (e.g., more transparent Mg-rich crystalline silicates that are less optically active) may warm sufficiently to gain spectral contrast with respect to more active species (e.g., amorphous carbon and Mg:Fe amorphous silicates).

\begin{figure*}[t!]
\centering
\includegraphics[width=0.9\textwidth]{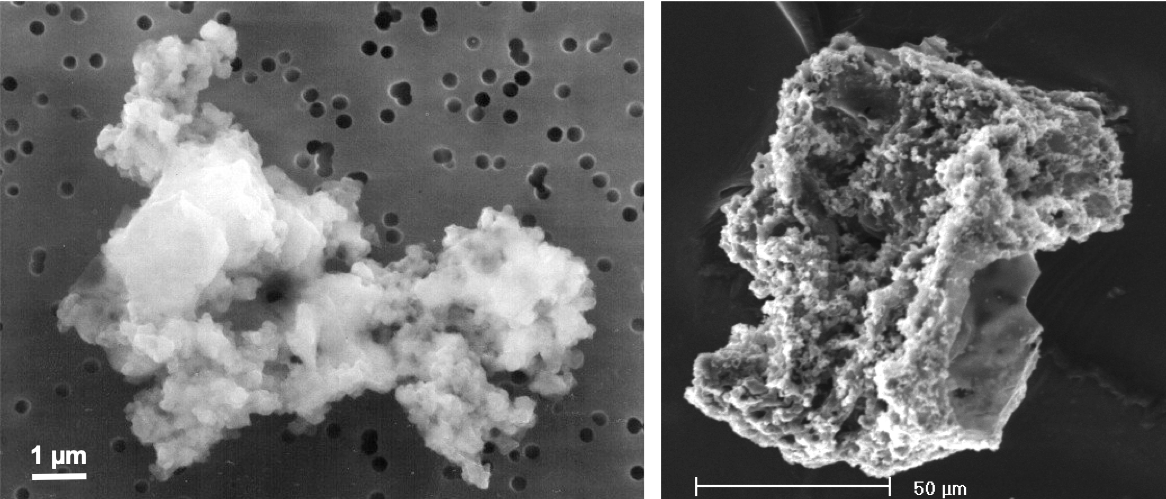}
\caption{\it \small 
Secondary electron images of dust collected on Earth with a probable cometary origin: a) chondritic anhydrous interplanetary dust particle (CA-IDP) collected in the stratosphere by NASA; b) ultracarbonaceous Antarctic Micrometeorites (UCAMM) from the Concordia collection \citep{Duprat2007a,Duprat2010}.
}
\label{fig:IDP_UCAMM}
\end{figure*}

NASA has been collecting Interplanetary Dust Particles (IDPs) in the stratosphere 
\citep{Warren1994}, following the pioneering work of Don Brownlee \citep{Brownlee1977}.  
Several lines of evidence point to a cometary origin for chondritic anhydrous (CA) IDPs \citep{Bradley1986, Bradley1994, Bradley1999} (Fig.~\ref{fig:IDP_UCAMM}), including higher atmospheric entry velocities (as determined by noble gas measurements, \citep{Nier1993}), high particle porosity, anhydrous nature, high bulk carbon content associated with a high abundance of pyroxenes minerals \citep{Thomas1993}, short solar exposure histories, high presolar grain concentrations \citep{Rietmeijer1998, Palma2005, Nguyen2007, Busemann2009,Brownlee:1995ab, Bradley2014a}. We chose here to name these IDPs ``chondritic anhydrous" whereas they are usually quoted as ``chondritic porous" (or ``chondritic porous anhydrous") in the literature. This choice was made following the observation that many hydrous IDPs are equally porous \citep{Zolensky1992} so the term `chondritic porous' is somewhat misleading. Moreover, IDPs having a fluffy-like texture when observed as whole particles do not always contain significant porosity when examined in their interior (e.g.\ by sectioning with ultramicrotomy). In the search for cometary IDPs, timed collections in the stratosphere were performed with the aim of collecting IDPs during dust streams of comets 26P/Grigg-Skjellerup and 21P/Giacobini-Zinner, but the expected fraction of collected particles arising from those particular sources was only a few  \%  \citep{Busemann2009, Bastien2013}.  Because of limitations of the collection technique, no particular IDP can be unambiguously identified with a specific small body, so the question of their origin(s) is not yet completely resolved. After 50 years of investigation, links between IDPs and comets remain hazy, and some IDPs might originate from icy asteroids \citep[e.g.][]{Vernazza2015a}, which could in turn sample the "asteroid-comet" continuum proposed by \cite{Gounelle2011}. Nevertheless, a tentative consensus has been formed that the chondritic anhydrous IDPs (CA-IDPs) probably are mainly of cometary origin.  Are any of the hydrous chondritic IDPs from comets? The discovery of Ca-Al-rich refractory inclusions (CAIs) among \PWildtwo \ particles could  suggest a cometary origin for some refractory IDPs \citep{Zolensky1987, McKeegan1987}, which have long been ignored. These IDPs are finer grained than typical meteoritic CAI.

Larger interplanetary dust particles called micrometeorites (MMs) have also been recovered from polar ice and snow.  They were originally found by \citet{Maurette1986,Maurette1987} in Greenland ice, and later collected at lower temperatures from Antarctic ice, and then snow \citep[e.g.][]{Maurette1991a,Duprat2007a,Dobrica2009,Noguchi2015}.  These larger particles are generally more strongly heated during atmospheric entry than IDPs and may be altered in the terrestrial environment, especially by leaching when collected from ice where they can reside for several tens of thousands of years before collection. The samples collected from snow however do not show evidence for extensive aqueous alteration \citep{Duprat2007a}. Extraterrestrial dust particles in the size range of MMs ($\sim$ 200 \micron) constitute the dominant input of extraterrestrial matter on Earth \citep{Love1993a,Rojas2021}, and they could have played a role in the formation of the terrestrial hydrosphere and the origin of life on Earth \citep[e.g.][]{Maurette2006a}. Numerical modelling suggests that $\sim$80\% of micrometeorites could originate from comets \citep{Carrillo-Sanchez2016}, but they probably derive from both asteroids and comets. Some MMs have been found to have identical fine-grained components to chondritic-porous IDPs (here called CA-IDPs) and, thus, these ``CP-MMs" sample sources that are most likely cometary \citep{Noguchi2015,Noguchi2017}. Ultracarbonaceous Antarctic Micrometeorites (UCAMMs – Fig. 2) also constitute a new family of micrometeorites that was recently discovered in the Concordia and Dome Fuji collections \citep{Nakamura2005b,Duprat2010,Yabuta2017}. They are dominated by organic matter with a variable (but minor) stony component, show large anomalies of their hydrogen isotopic composition, and most probably originate from comets. They contain an unusual N-rich organic matter that could have formed by Galactic cosmic ray irradiation of N-rich ices in the outer regions of the protoplanetary disk \citep{Dartois2013a,Dartois2018,Auge2016}.

This chapter will describe the chemical and physical properties of cometary dust particles, based on the information recovered from space missions, ground based telescopic observations and analyses in the laboratory of CA-IDPs, CP-MMs and UCAMMs. Since this chapter gathers information from very different collecting and/or analytical methods, the following possible biases should be kept in mind: i) the {\it in situ} mass spectrometers at comet 1P/Halley mostly detected very small particles (nanometers) - possibly sub-constituents of larger particles. The {\it Rosetta}/COSIMA mass spectrometer performed analyses of dust from 67P/Churyumov-Gerasimenko at the $\sim$25 µm scale, thus averaging the composition of sub-components; ii) samples returned from the Stardust mission were more or less altered by the high-speed collection in the aerogel, depending on the mechanical strength of the initial particles; iii) the identification of cometary dust compositions from astronomical IR spectroscopy requires a careful control on observational and analytical parameters, as presented in section 2.5; iv) information deduced from CA-IDPs, CP-MMs and UCAMMs can suffer from small number statistics, and from the poor knowledge of their initial parent bodies.
\section{\textbf{CHEMICAL PROPERTIES}}
\label{sec:chem}

\subsection{Elementary composition}
\label{sec:three.elemcomp}

The elementary composition of cometary dust particles can be determined from ground-based, space missions and dust samples of very probably cometary origin that are collected on Earth : the chondritic-anhydrous IDPs (CA-IDPs) \citep{Bradley2014a}, the CP-MMs \citep{Noguchi2015,Noguchi2017}, and ultracarbonaceous Antarctic micrometeorites (UCAMMs) \citep{Duprat2010}.

\begin{figure*}[t!]
\centering
\includegraphics[width=0.9\textwidth]{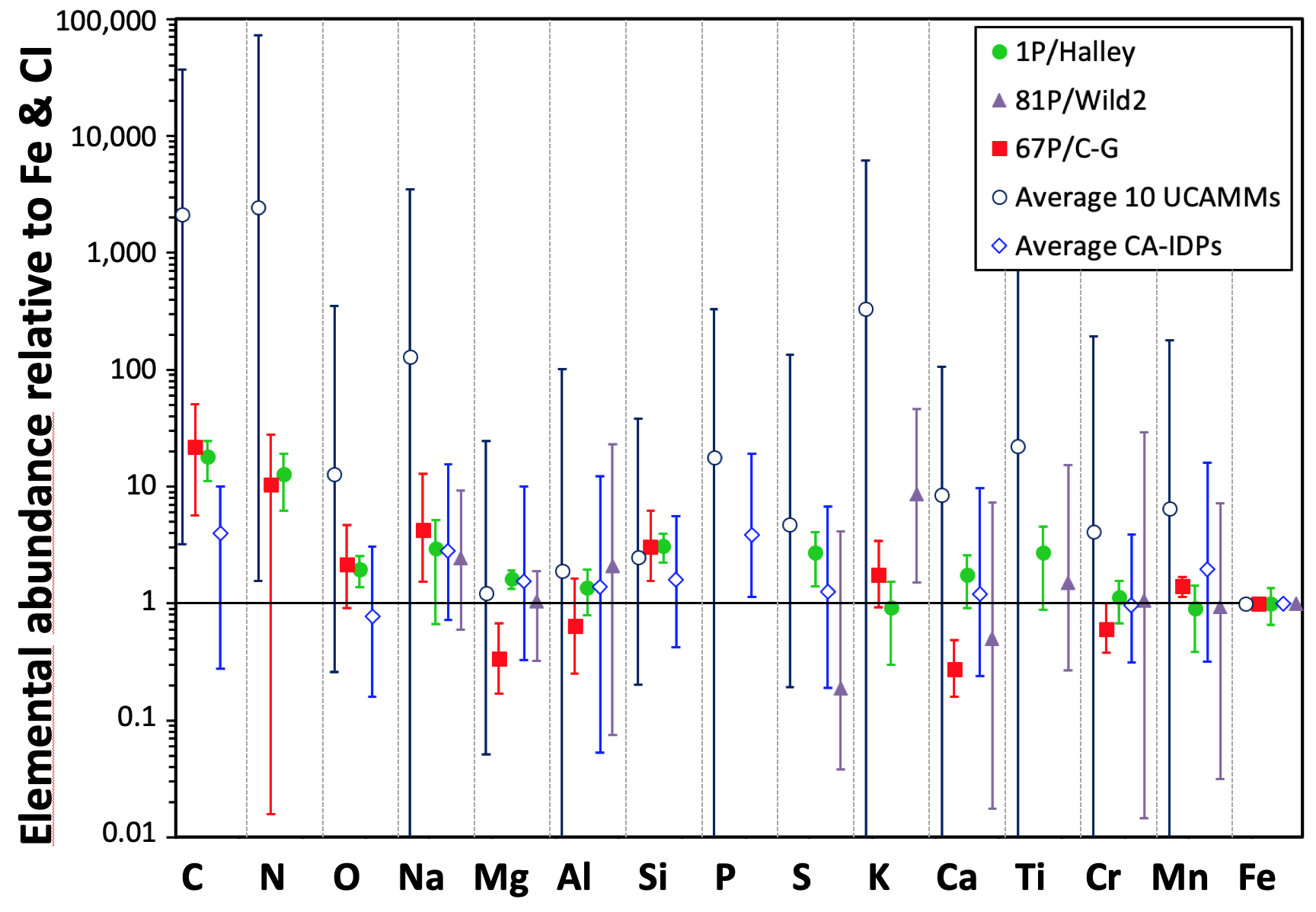}
\caption{\it \small 
Average elemental ratios relative to Fe and to CI \citep{Lodders2010} for dust particles from comets 1P/Halley \citep{Jessberger1988}, \PWildtwo{} in aerogel \citep{Flynn2006, Ishii2008, Lanzirotti2008, Leroux2008, Stephan2008, Stephan2008a}, \Chury{} (\CG){} \citep{Bardyn2017}, for 10 ultracarbonaceous Antarctic micrometeorites (UCAMMs) \citep[and unpublished data]{Dartois2018} and for 115 chondritic anhydrous IDPs (CA-IDPs) \citep{Thomas1993, Keller2004,Schramm1989} (not all elements were measured for all CA-IDPs, see text). Error bars represent the variation range of the elemental compositions among cometary dust particles for the CA-IDPs and UCAMMs samples and for comets 1P/Halley and 81P/Wild2. In the case of comet 67P/C-G, the error bars represent uncertainties on the values, which are higher or equal to the variation of composition between particles.}

\label{fig:comp}
\end{figure*}

Insight into the elemental composition of cometary dust particles was gathered so far for 4 comets visited by space missions (1P/Halley, 9P/Tempel 1, 103/Hartley 2, 67P/Churyumov-Gerasimenko), and from a comet sample return (\Stardust \ mission to 81P/Wild2).
For comet 1P/Halley, mass spectrometry from {\it Giotto}, {\it Vega-1} and {\it Vega-2}, of cometary dust impacting at high speeds ($\sim$70-75 km.$s^{-1}$) determined two populations : the ``rocky” (elements of Mg, Fe, O, S of approximately solar composition) and the ``CHON” particles with elemental abundances enhanced over the CI chondritic composition (as defined by the Ivuna-type carbonaceous chondrites) and the Sun \citep{Jessberger1986, Jessberger1999}. All particles were in fact, on some level, mixtures of ``rocky" and ``CHON" materials with about one-quarter being predominantly ``rocky", one-quarter predominantly ``CHON", and half being mixed with a span of 0.1--10 times CHON/rock elemental ratios \citep{Fomenkova1992a, Lawler1992}.  The very smallest particles had the greatest C abundances \citep{Lawler1992}. The bulk composition of Halley dust particles was chondritic within a factor of two, with the exception of carbon, nitrogen and hydrogen, which were enriched with regard to CI by a factor of 11, 8 and 4, respectively (Fig.~\ref{fig:comp}).

The \Stardust \ {\it Mission} flyby showed that particle streams in the coma resulted from the disintegration (called ``autobrecciation") of larger particles released from the nucleus at slower speeds \citep{Clark2004}. \Stardust \ returned samples revealed the presence of a greater fraction of 'hot inner disk materials' than assessed from Halley, including high temperature CAIs, micro-chondrules (100 \micron - size) spanning Mg- to Fe-rich olivine (crystals), plagioclase, nepheline and graphitic carbon, (see sections \ref{Organics} and \ref{sec:two.mineral} for more details) \citep{Zolensky2006, Nakamura2008, De-Gregorio2017}.

The impact on comet \nineP, which was created by the \DI \ {\it Mission}, released particles into the coma, with coma-gas-accelerated speeds of 200 m/s, that spectrally appeared to be more similar to the submicron-sized silicate-rich and crystal-rich comet Hale-Bopp. In the hours after impact and from ground-based studies of the inner coma, dust compositions varied between highly silicate- and forsterite-rich to poor relative to (what is fitted as) dark carbonaceous species \citep{Harker2005, Harker2007, Sugita2005}. Visible polarization studies revealed the ejection of surface dark carbonaceous particles \citep{Furusho2007}. Spectral studies at lower spatial resolution also revealed smaller and more crystal-rich materials in the coma after impact \citep{Lisse2005a}. The fortuitous explosive release of matter into the coma of \Holmes \ similarly revealed smaller particles and more crystal-rich compositions that were hitherto thought to be associated with Oort cloud comets like Hale-Bopp \citep{Reach2010}. 

The flyby over comet \Hartley{} showed that the two sources of volatile gas 'activity drivers', \HtwoO \ and \COtwo , produced a different size and composition in the coma whereby, compared to the \HtwoO \ rich mid-region, the \COtwo -rich end was ejecting ice chunks and organic gases and probably solid state organics \citep{AHearn2015, Feaga2021}. 

\begin{figure}[t!]
\centering
\includegraphics[width=0.5\textwidth]{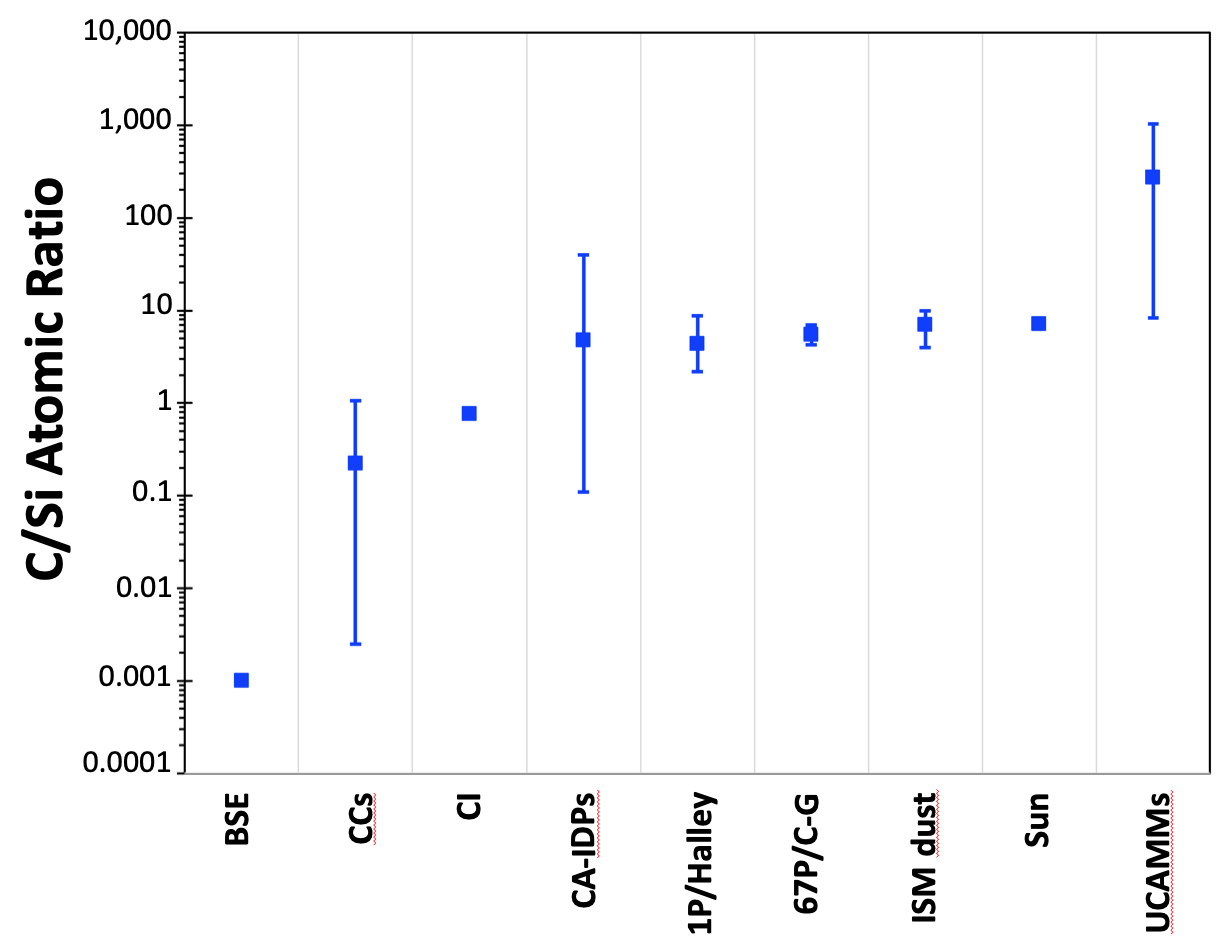}
\caption{\it \small 
Atomic C/Si ratios for bulk silicate Earth (BSE) \citep{Bergin2015}, carbonaceous chondrites (CCs) \citep{Jarosewich1990}, the CI value \citep{Lodders2010}, the average of 24 CA-IDPs \citep{Matrajt2005a,Thomas1993}, comet 1P/Halley \citep{Jessberger1988}, comet 67P/Churyumov-Gerasimenko (67P/C-G) \citep{Bardyn2017}, ISM dust \citep[e.g.][]{Dartois2018}, the Sun \citep{Lodders2010}  and the average of 10 UCAMMs \citep[and unpublished data]{Dartois2018}.
}
\label{fig:c_Si}
\end{figure}

The comet \Chury{} (hereafter \mbox{67P/C-G}) was studied in detail during the rendezvous with the {\it Rosetta} mission, which orbited the comet from August 2014 to September 2016. The knowledge base about particle structures and compositions was considerably expanded by the {\it Rosetta} investigations. Cometary dust in \mbox{67P/C-G} coma consists mainly of mm-size and slow-moving (few~km.$s^{-1}$ \citet{Rotundi2015}) hierarchical aggregates \citep{mannel2019} that collapsed to various degrees upon collection \citep{Langevin2016, Lasue2019c} (see also section \ref{Morpho}). The dust mass analyser COSIMA \citep{Kissel2007} collected more than 35,000 particles \citep{Merouane2016}, and about 250 of them were analyzed. The ROSINA gas mass spectrometer occasionally captured rock-forming elements like Na, Si, K, Ca, and Fe in relative abundances suggesting a composition enriched in Si compared to CI \citep{Wurz2015,Rubin-M2022}. These atomic species in the gas phase could originate from solar wind sputtering of dust particles, or directly from nanoparticles present in the coma. Fe and Ni atoms could be associated with organics (see section \ref{Organics}). The composition of \CG{} dust particles as measured by COSIMA shows some variability \citep{Bardyn2017,Sansberro2022} and the average composition of $\sim$ 30 of these particles was quantified and showed that dust was composed of stony material mixed with high molecular weight solid state organics (45\% by mass) \citep{Bardyn2017, Fray2016}. The bulk composition of \CG{} dust particles is rather chondritic, except for higher content of C and possibly N (Fig.~\ref{fig:comp}). A carbon to silicon atomic ratio $\rm C/Si = 5.5^{+1.4}_{-1.2}$ was measured in \CG{} dust particles \citep{Bardyn2017} (Fig.~\ref{fig:c_Si}). This value is about one order of magnitude larger than the CI value $(0.76 \pm 0.10)$, and close to the protosolar value $(7.19 \pm 0.83$, \citet{Lodders2010}). The H/C ratio is $1.04 \pm 0.16$ \citep{Isnard2019}, which is higher than in IOM extracted from the most primitive meteorites, but lower than the value measured in \CG{} high molecular weight organic molecules that have an average H/C ratio of 1.56 $\pm$ 0.04  \citep{Haenni2022}, which is compatible with that of the soluble organic matter in the Murchison carbonaceous chondrite \citep{Schmitt-Kopplin2010}.  The average nitrogen to carbon ratio of \CG{} dust particles is $\rm N/C = 0.035 \pm 0.011$ \citep{Fray2017}. This value is in turn compatible with the chondritic value $\rm (N/C~= \sim 0.04)$ \citep{Alexander2017}, but about one order of magnitude lower than the protosolar value $\rm (N/C = 0.3 \pm 0.1$, \citet{Lodders2010}). The discovery of ammonium salts (in particular ammonium hydrosulphide and fluoride) in \CG{} \citep{Altwegg2020,Poch2020,Altwegg2022} could account for this missing nitrogen reservoir, as these salts would have sublimated before analysis in COSIMA. Phosphorus and fluorine were detected by COSIMA in the dust particles \citep{Gardner:2020wy}.

The composition of cometary dust in comet \PWildtwo{} was measured in samples returned by the \Stardust \ mission \citep{Brownlee2006,Horz2006}. Because of the high speed collection of the samples, the light elements could not be quantified, and volatile elements like S probably were redistributed around the tracks \citep{Ishii2008}. Other elements show a chondritic composition within a factor of two \citep{Flynn2006, Ishii2008, Lanzirotti2008, Leroux2008, Stephan2008, Stephan2008a}.
The bulk composition of cometary dust as discussed here is displayed in Fig.~\ref{fig:comp}, with the abundances normalized to Fe and to CI. This kind of representation allows comparison with reference values, but should be taken with a hint of caution, as apparent enrichment/depletions could depend on the normalizing element (Fe was chosen here). Na and Si seem systematically enriched in cometary dust. The Na enrichment was indeed used as a tracer in COSIMA elementary maps to pinpoint the location of the dust particles, and was observed during the entry of comet C/2013 A1 (Siding Spring) in the Martian atmosphere \citep{Benna:2015aa}.

The composition of CA-IDPs was measured for 24 IDPs \citep{Thomas1993,Keller2004} for all elements displayed in Fig.~\ref{fig:comp}, except for N, K, and Ti, and for 91 IDPs for major elements (Mg, Al, Si, S, Ca, Fe) \citep{Schramm1989}. These IDPs show a fairly chondritic composition, with an enrichment in C of about 5 times the CI value, which seems correlated with a mineralogy dominated by pyroxenes \citep{Thomas1993} (see also section \ref{CA-IDPs_mineralo}).
The composition of 10 UCAMMs was measured by electron microprobe and show large ranges of variations, as seen in Fig.~\ref{fig:comp}. Within this large variation range, the compositions are compatible with CI, except for C and N which are markedly enriched in UCAMMs with regard to the CI composition. 

Figure~\ref{fig:c_Si} displays the C/Si atomic ratio in different kinds of Solar System material, ordered by increasing values. We can note that data available for cometary dust and CA-IDPs are compatible with that of the Sun and of ISM dust. Objects formed in the inner Solar System show lower values than the Sun as noted by \citet{Bergin2015}. The C/Si atomic ratio of UCAMMs is very high, even higher than that of ISM dust, suggesting a local accumulation process of organics with regard to minerals in the formation regions of UCAMMs \citep[e.g.][]{Dartois2018}.

\subsection{Organics} \label{Organics}
\label{sec:two.organics}

Cometary dust particles are rich in organic matter. These organics are present both as volatile compounds mixed with the host ice phase of the dust particles, and as solid organic matter in the dust particles themselves. The organics that are present in the dust particles remain solid when the comet approaches the Sun, and are thus quoted as “refractory” organic matter. They can be studied in samples in the laboratory, by astronomical observations (depending on the observations conditions and size of the organics) or by {\it in situ} analyses during space missions. The volatile organic compounds that have been identified so far in cometary dust particles are associated with ice that sublimates when the comet approaches the Sun, and can thus be present in the cometary coma. The {\it Rosetta} mission around comet 67P/C-G allowed a detailed characterization of the volatiles around the comet, as a function of heliocentric distance. These volatiles species consist mainly in CH(N)O-bearing molecules, with a great variety of CH-, CHN-, CHS-, {CHO\textsubscript{2}-} and CHNO-bearing species, both saturated and unsaturated \citep{Altwegg2017,Haenni2022}, as well as the CN radical and related molecules \citep{Haenni2020,Haenni2021}. Ammonium salts (in particular ammonium hydrosulphide and fluoride) were also found in \CG{} coma and at the nucleus' surface \citep{Altwegg2020,Altwegg2022,Poch2020}. Glycine was identified in the gas phase, with a distribution compatible with the sublimation of ices associated with the cometary dust particles, rather than a direct sublimation of ices from the nucleus \citep{Altwegg2016, Hadraoui2019, Hadraoui2021}. Glycine had also been identified at the surface of Al foil exposed to the coma of 81P/Wild 2 \citep{Elsila2009}.

The ``refractory” organic matter was identified as being present in comet 1P/Halley, but its nature could not be studied by the {\it Giotto} and {\it Vega} mass spectrometers. Spectral evidence for aromatic organic molecules, i.e., PAHs were suggested from UV spectral analyses of comet 1P/Halley \citep{Moreels1994, Clairemidi2008}. The very smallest particles had the greatest C abundances \citep{Lawler1992}. CHON particles had a range of H, O, and N ratios to C \citep{Fomenkova1992a, Fomenkova1994a}.

\begin{figure*}[t!]
\centering
\includegraphics[width=0.9\textwidth]{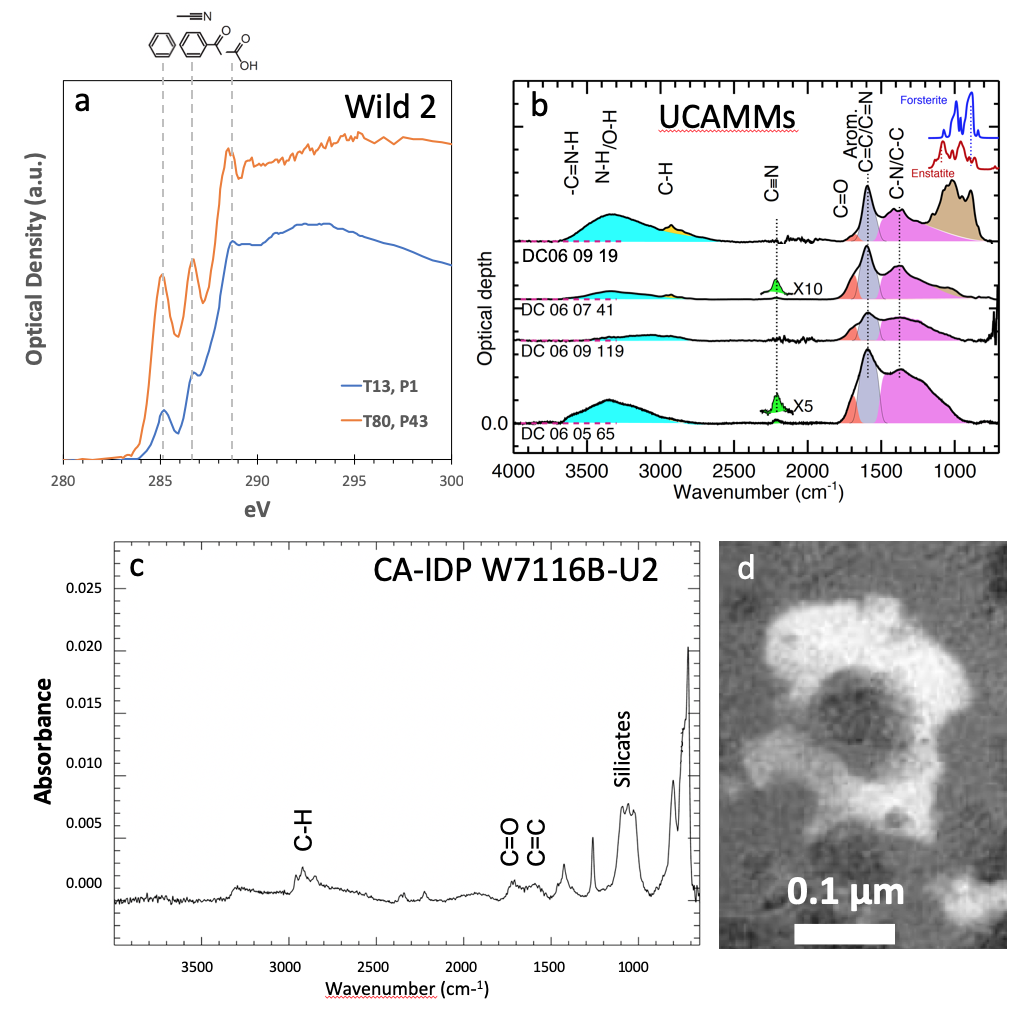}
\caption{\it \small 
(a) Spectral absorption signatures of organic matter in comet 81P/Wild 2 samples at the carbon K-edge in STXM-XANES (adapted from \citet{De-Gregorio2011}); (b) µFTIR spectra of 4 representative UCAMMs \citep{Dartois2018}; (c) µFTIR spectrum of organic matter extracted from a chondritic anhydrous IDP \citep{Matrajt2005a}. (d) Energy filtered image at the carbon edge of a nanoglobule in  comet 81P/Wild 2 samples obtained by transmission electron microscopy \citep{Matrajt2008}.
}
\label{fig:organics}
\end{figure*}

The high speed collection of 81P/Wild 2 cometary dust did not allow a good preservation of the organics \citep{Brownlee2014, Keller2006, Sandford2010}. However, some organic matter, including clumps that were `behind’ terminal particles, and therefore somewhat protected from the heat generated by impact during collection, revealed a suite of complex organic bonds including mainly alkenes, aromatic C=C and carboxyl C=O as well as a variety of textures for the organic matter including organic nanoglobules (Fig.~\ref{fig:organics}) \citep{Matrajt2012, De-Gregorio2011, De-Gregorio2017}. The spectral signature of preserved organic matter in \Stardust \ samples show similarities with that of insoluble organic matter extracted from meteorites \citep{De-Gregorio2011}, although a reduced form of carbon was also observed in one \Stardust \ sample \citep{De-Gregorio2017}. The concentration of carbon could not be quantified in 81P/Wild 2 samples due to the collection method in aerogel. The low concentration observed in the samples is interpreted as a consequence of the harsh collection of the samples. It could also represent a collection bias of dust from a portion of the coma which was poor in carbon and  not representative of the whole comet \citep{Westphal2017}.

The solid organic matter identified in \CG{} dust particles also shows similarities with insoluble organic matter extracted from meteorites \citep [e.g.] [] {Alexander2017}, although with a higher atomic H/C ratio ($1.04 \pm 0.16$) \citep{Fray2016, Fray2017, Isnard2019}. The higher H/C ratio found in \CG{} could suggest less processing and a more primitive origin of the organics present in cometary particles than in IOM. This comparison is however made between the bulk organic matter for \CG{} and the IOM obtained from meteorites by acid treatments, after removal of soluble organics. Soluble organics represent only a small fraction of the carbon in meteorites ($\sim$ 20\%), but they could account for some of the difference observed between the H/C ratios of organic matter in \CG{} and in IOM. The O/C atomic ratio in \CG{} dust is likely higher than in meteoritic IOM \citep{Bardyn2017}. From the COSIMA analyses, the abundance of organic matter in \CG{} dust particles was estimated at $\sim$45wt\%, or $\sim$70 vol\% \citep{Bardyn2017,Levasseur-Regourd2018b}. The lower abundance of 52~$\pm$~8~vol\% of organic matter deduced by \cite{Fulle2016c} from density measurement is due to the assumption by these authors of an IDP-like composition that underestimated the carbonaceous content of \CG{}. This solid state organic matter is reminiscent of ”CHON” in Halley but the techniques available for analyses better reveal the complexity and details of this organic matter. The ROSINA gas mass spectrometer revealed a huge host of complex molecules, including many S species, through the fortuitous impact with a dust particle that occurred during a close flyby of the nucleus \citep{Altwegg2017}. Glycine, methylamine and ethylamine, as well as phosphorus, were detected by ROSINA in the gas phase in \CG{} \citep{Altwegg2016}. The observations of glycine in the coma can be explained by the presence of this amino acid in sublimating water ice in the dust particles \citep{Hadraoui2019, Hadraoui2021}.

Reflectance spectra of the surface of \CG{} also suggest a high abundance of organic matter in the surface material, with a darkening material that could be submicrometer-sized Fe sulfides \citep{Quirico2016, Rousseau2018, Capaccioni2015}.

Recent observations showed that Fe and Ni atoms are ubiquitous in cometary atmospheres, even at large heliocentric distances \citep{Bromley2021,Hutsemekers2021,Manfroid2021,Guzik2021}. The observed Fe/Ni ratio is about an order of magnitude higher than the solar value. These elements could not result for the sublimation of minerals like silicates or Fe sulfides, due to the low equilibrium temperature of the cometary atmospheres. The correlations observed between the productions rate of Fe and Ni and of carbon oxides by \cite{Manfroid2021} suggest that these Fe and Ni emissions could be produced by the sublimation of Fe and Ni carbonyls (Fe(CO)$_5$ and Ni(CO$_4$), which are possible constituents of cometary or interstellar matter.

CA-IDPs, CP-MMs and UCAMMs are enriched in organic matter compared to primitive meteorites. Organic matter in CA-IDPs was studied by Fourier transform infrared microscopy (µFTIR) \citep{Flynn2003, Keller2004, Matrajt2005a, Munoz-Caro2006, Merouane2014}, Raman microscopy \citep{Wopenka1988, Quirico2003, Quirico2005a, Bonal2006b, Munoz-Caro2006, Sandford2006, Rotundi2007, Rotundi2008, Busemann2009} and transmission electron microscopy with electron energy loss spectroscopy (TEM/EELS) \citep{Flynn2003, Keller2004}. The organic content of CP-MMs was studied by TEM/EELS, Raman microscopy, and X-Ray absorption microspectroscopy (STXM-XANES) \citep{Noguchi2015, Noguchi2017,Noguchi2022,Yabuta2015}. The organic matter of UCAMMs was studied by µFTIR, Raman microscopy, IR nanospectroscopy (AFM-IR) and STXM-XANES \citep{Dartois2018, Mathurin2019,Guerin2020}. Carbonaceous materials in CA-IDPs range from hydrocarbon nanoglobules \citep{Wirick2009} to completely graphitized carbon \citep{De-Gregorio2017}, which have been used as thermometers \citep{Matrajt2013}. 100 \nm-thick coatings of organics observed on some anhydrous crystals and GEMS (glass with embedded metals and sulfides) are proposed to have facilitated grain aggregation \citep{Flynn2003}. Figure~\ref{fig:organics} displays the µFTIR signature of organic matter in a CA-IDP after acid treatment (Fig.~\ref{fig:organics}c) and in UCAMMs without any chemical treatment (Fig.~\ref{fig:organics}b). The signature of the organics in CA-IDPs shows a large abundance of polyaromatic organic matter, with aromatic carbon, ketone (C=O), carboxylic groups (COOH) and aliphatic C-H contributions. 
 The organics in CP-MMs show the signature of aromatic/olefinic carbon, aromatic ketone, and carboxylic carbon. Carbonaceous nanoglobules are usually present \citep{Noguchi2015,Noguchi2017,Noguchi2022,Yabuta2015}.
 The signature of organic matter in UCAMMs is unusual, with large amounts of N-bearing species (including nitrile), and low signature of C=O and aliphatic C-H \citep{Dartois2013a, Dartois2018}. STXM-XANES analyses of UCAMMs show that the organics in UCAMMs consist in fact of three distinct organic phases, with different spectroscopic signatures and different amount of nitrogen \citep{Engrand2015a, Charon2017aa, Guerin2020}. The first organic phase of UCAMMs is smooth and N-rich, with N/C atomic ratios up to 0.2. This phase has no equivalent in meteorites, and could result from the irradiation by Galactic cosmic rays of methane and nitrogen-rich ices at the surface of small bodies in the outer regions of the protoplanetary disk \citep{Dartois2013a, Dartois2018, Auge2016, Auge2019, Rojas2020}. The other two organic phases identified in UCAMMs bear similarities with that of chondritic IOM. A carbon-rich clast identified as a cometary xenolith in the LaPaz Icefield 02342 meteorite also shows spectroscopic similarities with meteoritic IOM \citep{Nittler2019}. The peak intensity ratios of CH$_{2}$/CH$_{3}$  measured by µFTIR in aliphatic C-H in CA-IDPs \citep{Flynn2003, Keller2004, Matrajt2005a, Munoz-Caro2006, Merouane2014} and in UCAMMs \citep{Dartois2013a,Dartois2018} are higher than the value measured in dust from the diffuse interstellar medium, which is around 1 \citep{Sandford1991,Pendleton1994,Dartois2007a,Godard2012}.

The Raman signature of OM in CA-IDPs, CP-MMs and UCAMMs confirm the polyaromatic nature of the solid organics in these particles, and the low thermal metamorphic grade of their organic matter \citep{Wopenka1988, Quirico2003, Quirico2005a, Bonal2006b, Munoz-Caro2006, Busemann2007, Busemann2009, Rotundi2007, Rotundi2008, Brunetto2011, Noguchi2015, Dobrica2011, Dartois2013a, Dartois2018, Starkey2013}. The potential effect of atmospheric entry heating on the degree of disorder of the organic matter cannot be ruled out, but specific experiments would be needed to study these effects in detail.

Organic nanoglobules seem to be ubiquitous in samples of cometary origin. They are found in 81P/Wild 2 samples \citep{Matrajt2008, De-Gregorio2010}, CA-IDPs \citep{Matrajt2012}, CP-MMs \citep{Noguchi2015}, and UCAMMs \citep{Charon2017aa} (and also in chondritic AMMs \citep{Maurette1995a}, for which models predict a cometary origin for 80\% of them \citep{Carrillo-Sanchez2016}).

\subsection{Isotopes}
\label{sec:three.isotopes}

The {\it Giotto} and {\it Vega} missions during a flyby around comet 1P/Halley led to the rough measurement of carbon isotopes in cometary dust, but most of the data on the isotopic compositions of cometary dust particles were gathered from laboratory analyses of returned cometary samples (\Stardust \ mission – 81P/Wild 2 comet), from {\it in situ} analyses ({\it Rosetta} mission on \Chury {} – \CG) or from the analysis of CA-IDPs, CP-MMs and UCAMMs. 

\subsubsection{Hydrogen, carbon, nitrogen and sulfur isotopes}

\begin{figure}[t!]
\centering
\includegraphics[width=0.5\textwidth]{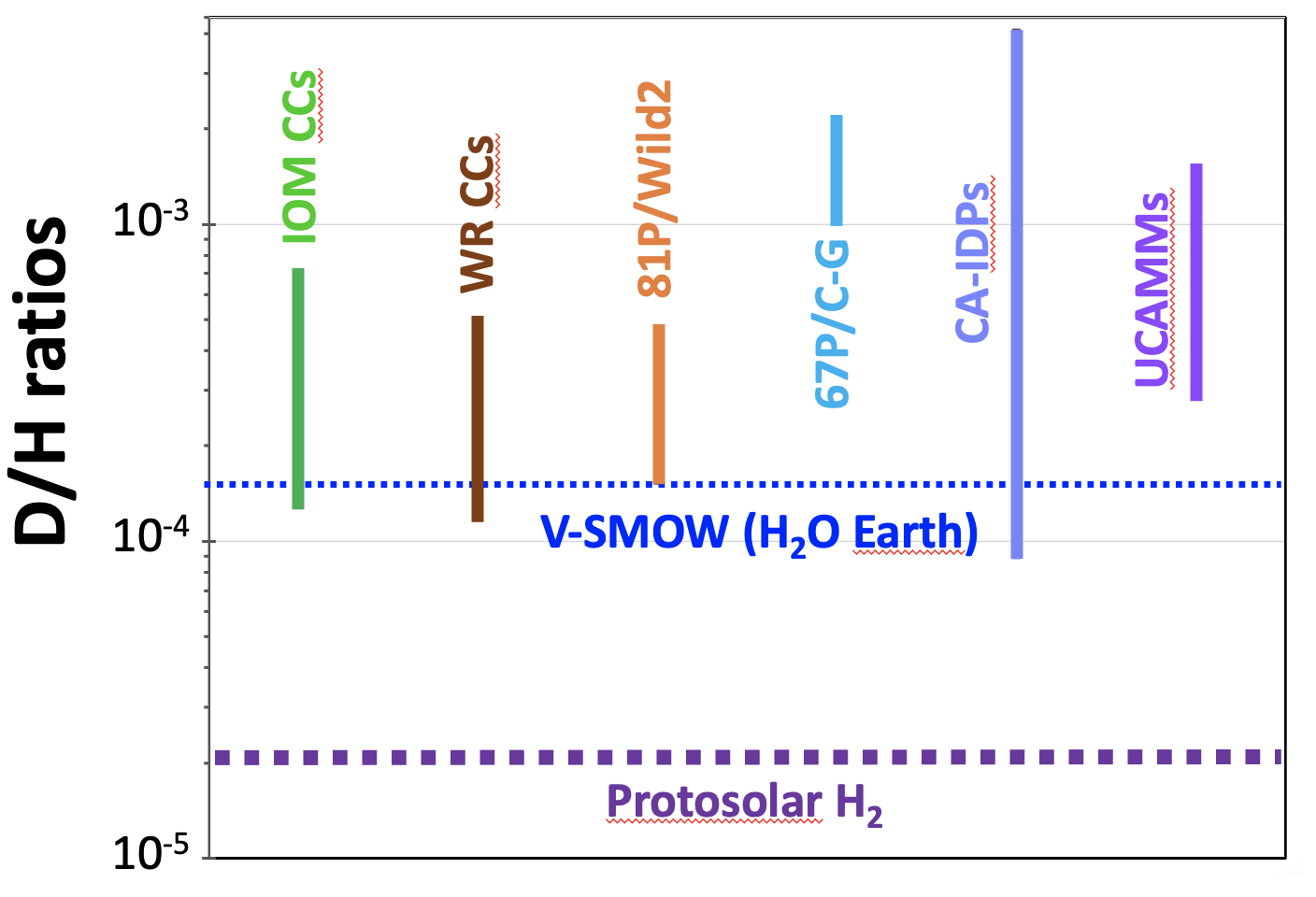}
\caption{\it \small 
D/H ratio measured in solid phase in cometary dust particles measured in the Stardust samples (\PWildtwo), in refractory organics by the Rosetta/COSIMA instrument (\Chury – \CG), in chondritic anhydrous IDPs (CA-IDPs) and UCAMMs. The range of composition of D/H ratios measured in insoluble organic matter extracted from carbonaceous chondrites (IOM CCs) and in whole-rock carbonaceous chondrites (WR CCs), as well as the terrestrial value (V-SMOW, in water) and protosolar value (in H2) are also shown for reference. Data from \citet{Alexander2007, Alexander2012a, McKeegan2006, Paquette2021, Zinner1983, McKeegan1985, Messenger2000, Aleon2001a, Busemann2009, Duprat2010, Rojas2022a, Geiss1998}.
}
\label{fig:DH}
\end{figure}

The hydrogen isotopic composition of comets has been measured in the gas phase of comets for an increasingly large number of species \citep [e.g.] [and references therein] {Bockelee-Morvan2015,Altwegg2015, Mueller2022a}, but the {\it in situ} measurement of D/H ratios in cometary dust was only made possible by the {\it Stardust and Rosetta} missions. The D/H ratios of cometary dust particles is displayed in Fig.~\ref{fig:DH}. The D/H ratio of samples from comet \PWildtwo{} (\Stardust \ mission) were measured by secondary ion mass spectrometry (SIMS) \citep{McKeegan2006,De-Gregorio2010,De-Gregorio2011,Matrajt2008,Stadermann2008}. The bulk D/H isotopic ratios of \Stardust \ samples vary from the terrestrial D/H value (V-SMOW) of 1.5576 × 10$^{-4}$ to sub-µm sized hotspots that can reach values up to $\delta$D $\sim$ 2000 permil, which corresponds to three times V-SMOW. The hydrogen isotopic composition was also measured in dust particles from comet \CG{} by COSIMA. The average value measured in the organic matter of 25 cometary particles from \CG{} is D/H = (1.57 ± 0.54) × 10$^{-3}$ \citep{Paquette2021}, which is about one order of magnitude larger than the terrestrial value, and marginaly compatible with values measured in organic molecules in cometary comae \citep{Meier1998b,Mueller2022a}.
The hydrogen isotopic composition of CA-IDPs varies between D/H $\sim$10$^{-4}$ and D/H $\sim$ 4 × 10$^{-3}$  \citep{Aleon2001a,Busemann2009,McKeegan1985,Zinner1983}. The bulk D/H value measured for eight UCAMMs vary between $\sim$3 × 10$^{-4}$ to 1.5 × 10$^{-3}$, with µm-sized regions that reach up to 30 times the terrestrial value \citep{Duprat2010,Rojas2022a}.

The {\it Giotto} and {\it Vega} missions to comet Halley allowed the discovery of isotopically light carbon in the dust particles ($^{12}$C/$^{13}$C $\sim$ 5000), providing a possible link to presolar graphite \citep{Amari:1993aa} or SiC grains, for which a few such light values have been measured \citep{Hoppe:2000aa, Lin:2002aa, Nittler:2003ab}. The C isotopic composition of comet Wild2 dust particles in \Stardust \ samples varies between $\delta$$^{13}$C $\sim$ -20 and $\sim$ -50 permil \citep{McKeegan2006}, which are values compatible with that observed in carbonaceous chondrites \citep{Alexander2007} and CA-IDPs \citep{Messenger2003a}. This value is slightly higher than the solar value determined by the Genesis mission at d $\delta$$^{13}$C = -105 $\pm$ 20 permil \citep{Hashizume2004}. The bulk carbon isotopic composition was measured in three UCAMMs, and vary from $\sim$25 permil to $\sim$ -85 permil. A noticeably low isotopic composition at $\sim$ -120 permil is found as a “cold spot” in one UCAMM \citep{Rojas2022a}.

\begin{figure*}[t!]
\centering
\includegraphics[width=0.9\textwidth]{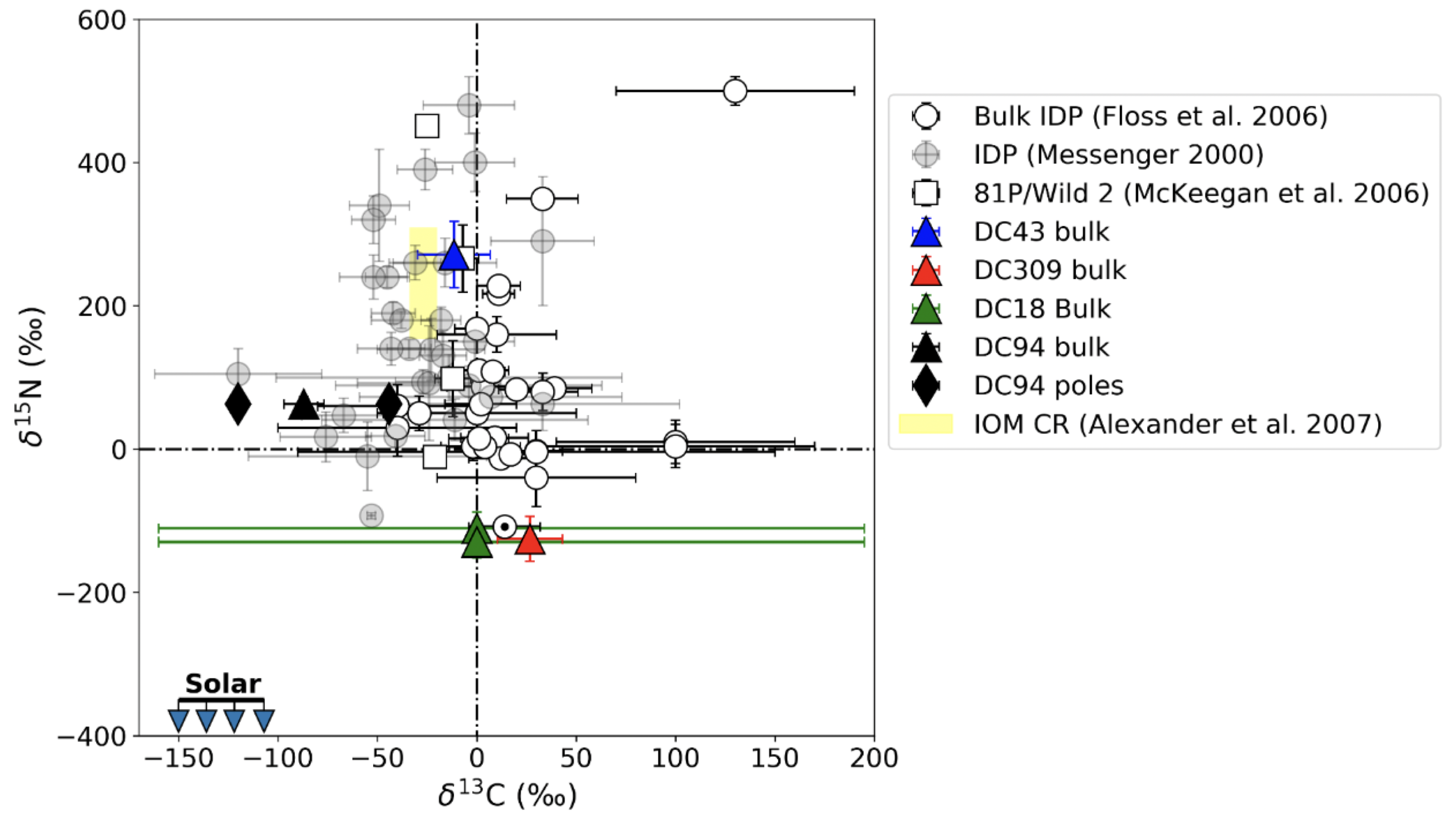}
\caption{\it \small 
Nitrogen and carbon isotopic compositions of IDPs, 91P/Wild2 samples, UCAMMs (DC43, DC309, DC18, DC94) and IOM extracted from CR chondrites (from \citet{Rojas2022a})
}
\label{fig:Niso_Ciso}
\end{figure*}

The nitrogen isotopic composition of \Stardust \ samples shows moderately elevated values, with hotspots of sub-micrometric sizes reaching values up to $\sim$ 500 permil \citep{McKeegan2006}, which are compatible with values measured in CA-IDPs \citep{Messenger2003a, Aleon2003a, Busemann2009}. The bulk nitrogen isotopic composition in five UCAMMs vary from $\sim$ -130 permil to $\sim$ 270 permil \citep{Rojas2022a} (see Fig.~\ref{fig:Niso_Ciso}). In most cases, there is no correlation between the nitrogen and hydrogen isotopic compositions of cometary dust.

The sulfur isotopic composition measured in comet 81P/Wild 2 samples is compatible with the solar value, showing an extraterrestrial origin of the impact residue and of the sulfide measured  \citep{Heck2012,Ogliore2012a}. The sulfur isotopic composition of a cosmic symplectite was analysed in \Stardust \ samples, which showed enrichment in $^{33}$S \citep{Nguyen2017a}. This composition could result from photochemical irradiation of solar nebular gas. The sulfur isotopic composition ($^{34}$S and $^{32}$S) of \CG{} dust was measured during the {\it Rosetta} mission by COSIMA, and is compatible with the reference value \citep{Paquette2017a}. Due to mass interferences, the $^{33}$S isotope could not be quantified, so a potential $^{33}$S excess could not be ruled-out for \CG{} dust particles.

\subsubsection{O, Si isotopes}
The oxygen isotopic composition in extraterrestrial matter is used as a taxonomic tool, as most meteorite classes own a given (range of) oxygen isotopic composition(s). 
The oxygen isotopic composition of \Stardust \ samples plot on a slope 1 line and show values which are compatible with carbonaceous chondrite signatures, including $^{16}$O enrichments for refractory minerals identified in the Wild2 samples \citep{Nakamura2008, Nakashima2012a, Joswiak2014, Ogliore2015, Defouilloy2017, Zhang2021}. The oxygen isotopic composition of 81P/Wild 2 samples shows a correlation between $\Delta$$^{17}$O and the Mg content of the analyzed minerals, as in CR chondrites \citep[e.g.][and references therein]{Nakashima2012a, Zhang2021}.

The $^{18}$O/$^{16}$O isotopic ratio could be measured by COSIMA in \CG{} dust particles, at $^{18}$O/$^{16}$O =  2.0 × 10$^{-3}$ $\pm$ 1.2 × 10$^{-4}$  \citep{Paquette2018a}. Given the large error bar associated to this value (due to limitations of measuring isotopic compositions with a ToF-SIMS method), this value is compatible with the terrestrial V-SMOW value, and covers the whole range of values found in meteorites. For reference, the oxygen isotopic composition of H$_2$O and CO$_2$ in \CG~coma are 2.25 × 10$^{-3} \pm$ 1.77 × 10$^{-4}$ and 2.02 × 10$^{-3} \pm$ 3.28 × 10$^{-5}$, respectively \citep{Schroeder2019, Hassig2017}. Other gaseous oxygen-bearing molecules in \CG{} can contain heavy oxygen isotopes compared to V-SMOW, especially for S-rich molecules \citep{Altwegg2020a}.

The oxygen isotopic composition of CA-IDPs is compatible with that of 81P/Wild2 samples \citep{McKeegan1987, Aleon2009a, Nakashima2012, Zhang2021}. 

The silicon isotopic composition of dust at comet \CG{} could be measured by the ROSINA instrument \citep{Rubin-M2017} and showed a depletion of heavy silicon isotopes $^{29}$Si and $^{30}$Si compared to the solar value. Such depletions in heavy isotopes are rare, and only found in rare presolar grains identified in meteorites \citep{Hynes2009}.

\subsubsection{Mg isotopes – $^{26}$Al}
The magnesium isotopic composition of 81P/Wild2 samples was measured to search for the past presence of $^{26}$Al at the time of mineral formation in comet Wild2. No resolvable $^{26}$Mg excess resulting from the decay of $^{26}$Al was found in Wild2 samples, suggesting either : i) a late formation (a few Myr after CAI formation) of minerals in Wild 2, ii) a protoplanetary disk heterogeneous in $\rm ^{26}Al$, or iii) the formation of Wild 2 minerals before injection of $\rm ^{26}Al$ in the protoplanetary disk \citep{Matzel2010, Nakashima2015}.  

The magnesium isotopic composition measured in olivines from \Stardust \ samples shows small variations in $\delta$$^{26}$Mg and $\delta$$^{25}$Mg values that are compatible with small mass-dependent fractionation from a chondritic reservoir with respect to the Mg isotopes \citep{Fukuda2021}. 

\subsubsection{Presolar grains}
Presolar grains are found in minute amounts in interplanetary material. They are grains that were present in the molecular cloud that led to the formation of the solar system and survived in the extraterrestrial samples that can be analyzed in the laboratory. At this time, we can only identify the isotopically anomalous stardust grains that were synthesized in previous generations of stars and got incorporated in the protosolar molecular cloud \citep{Hynes2009, Stephan2020}. Stricto sensu, grains that formed in the protosolar molecular cloud before the birth of the Sun are also “presolar”, but cannot be identified by isotopic methods, as they carry the solar signature of the initial cloud.

In meteorite samples, the most abundant identified presolar grains are silicates, whereas SiC and graphite were historically the first ones to be identified in the acid residue extracted from meteorites.

As they formed far from the Sun at cold temperatures, comets are expected to have preserved a large abundance of presolar grains. After correction for possible partial destruction during the harsh collection conditions of Wild 2 samples, isotopically anomalous grains remain rare among analyzed Wild 2 materials, occurring at initial abundances of $\sim$700 ppm \citep{Nguyen2020b, Stadermann2008, Floss2013}. 
Presolar grains are also found in CA-IDPs, CP-MMs and UCAMMs in abundances that can reach up to about 1\% \citep{Busemann2009, Floss2012, Floss2016}.

\subsection{Mineralogy}
\label{sec:two.mineral}

\subsubsection{Comets 1P/Halley and \CG{} : hints at their mineralogy}
In comet 1P/Halley, the ``rocky" particles had a wide range of Mg/Fe but with a narrower range of Mg/Si with similarities to Mg-rich silicates (40\%--60\% by number of particles), specifically Mg-rich pyroxenes, Fe(Ni) sulfides, with little Fe metal, and $<$1\% Fe-oxides \citep{Schulze1997}. CAI-like materials were not found in Halley. Few particles could be directly traced to pure mineral grains although the 11.2~\micron \ spectral feature of forsterite was first identified in Halley \citep{Bregman1987, Campins1989}. 

There was no instrument on {\it Rosetta} that allowed unambiguous identification of minerals in dust from \mbox{67P/C-G}. The very low reflectance of the nucleus surface suggest the present of opaque minerals, that could be Fe sulfides \citep{Quirico2016,Rousseau2018,Capaccioni2015}. The bulk composition and density of \CG{} dust particles are also compatible with the presence of silicates, Fe sulfides and carbon \citep{Bardyn2017, Fulle2016b}.

\subsubsection{\PWildtwo \ Mineralogy}
To date the only samples that are unambiguously derived from a comet are the \PWildtwo \ coma dust grains collected by the \Stardust \ spacecraft, returned to Earth in 2006. Well-preserved coma grains from comet \PWildtwo \ are dominated by the coarsest components. Fine-grained materials, representing perhaps 90\% of the impacting cometary coma grains, were severely altered or vaporized during high speed (6.1~$\rm km~sec^{-1}$) capture in the aerogel capture media \citep{Brownlee2006}. 
Fine-grained material was only preserved in a minority of cases, but probably sufficiently well to permit elucidation of its general nature \citep{Ishii2008}. It is also possible that the apparent lack of fine-grained amorphous solids could be an artifact of this collection bias. GEMS, frequently abundant in chondritic anhydrous IDPs \citep{Bradley1994}, have not been reliably identified among \Stardust \ materials (although there are unverified reports by \citet{Gainsforth2016}, possibly also due to destruction during collection and compositional and structural similarities to melted silica aerogel.

As expected, in the \Stardust \ samples the coarse-grained mineral phases are dominated by olivines, pyroxenes and sulfides. The compositions are distinct from meteorites \citep{Joswiak2012, Frank2014a, Joswiak2017}. Olivines exhibit practically the entire range from forsterite to fayalite, with no significant compositional peak. Some terminal olivines are thought to be ``micro chondrules" by their similarity to type II (Fe$>$10\%) chondrules in primitive chondrites such as CRs and CMs \citep{Frank2014a, Wooden2017}.  The minor element compositions of \Stardust \ olivines link only a subset to LIME olivine (condensates) \citep{Joswiak2017}. One \Stardust \ micro-chondrule `Iris' reveals its rapid cooling at high oxygen fugacity and high Na enrichment in the gas phase \citep{Gainsforth2010}. The lack of a pronounced compositional peak at forsterite was very unexpected, as this is a hallmark of chondritic IDPs \citep{Rietmeijer1998} and the least equilibrated carbonaceous chondrites \citep{Frank2014a}. In terms of the olivine and pyroxene compositions, the closest meteoritic analogues are the unequilibrated ordinary chondrites 
\citep{Frank2014a}. 
The high abundance of relatively coarse-grained  ($>$30~\micron \ size) \citep{Frank2014a, Wozniakiewicz2015}, well-crystalline ferromagnesian silicates  was also unexpected, given laboratory analogue experiments and interstellar dust spectroscopic studies which indicated that silicates should be mostly amorphous \citep[e.g.][]{Kemper2005}. 

Refractory minerals found in meteoritic calcium aluminium rich inclusions (CAIs) were identified in \Wild \ samples, which contain olivines, pyroxenes, sulfides, and refractory oxides. Mineral assemblages, mineral chemistries and measured bulk particle compositions reveal that these grains are similar to refractory materials in chondrites, with mineral chemistries most similar to CAI from CR2 and CH2 chondrites \citep{Zolensky2006, Simon2008, Chi2009} and Al-rich chondrules \citep{Bridges2012,Joswiak2014}.

Chondrule fragments were also found in Wild 2. They are similar to chondrules found in carbonaceous chondrites, but with interesting differences. Few type I chondrules (FeO- and volatile-poor) have been found from \Wild , though these are the most abundant type in meteorites.  To date mainly FeO-, MnO-, volatile-rich type II chondrules have been identified from \Wild \ \citep{Nakamura2008,Matzel2010,Joswiak2012, Ogliore2012, Frank2014a, Gainsforth2015}. One Al-rich, $^{16}$O rich chondrule fragment has also been identified from Wild 2, as found in carbonaceous chondrites \citep{Bridges2012}.

\begin{figure*}[t!]
\centering
\includegraphics[width=0.9\textwidth]{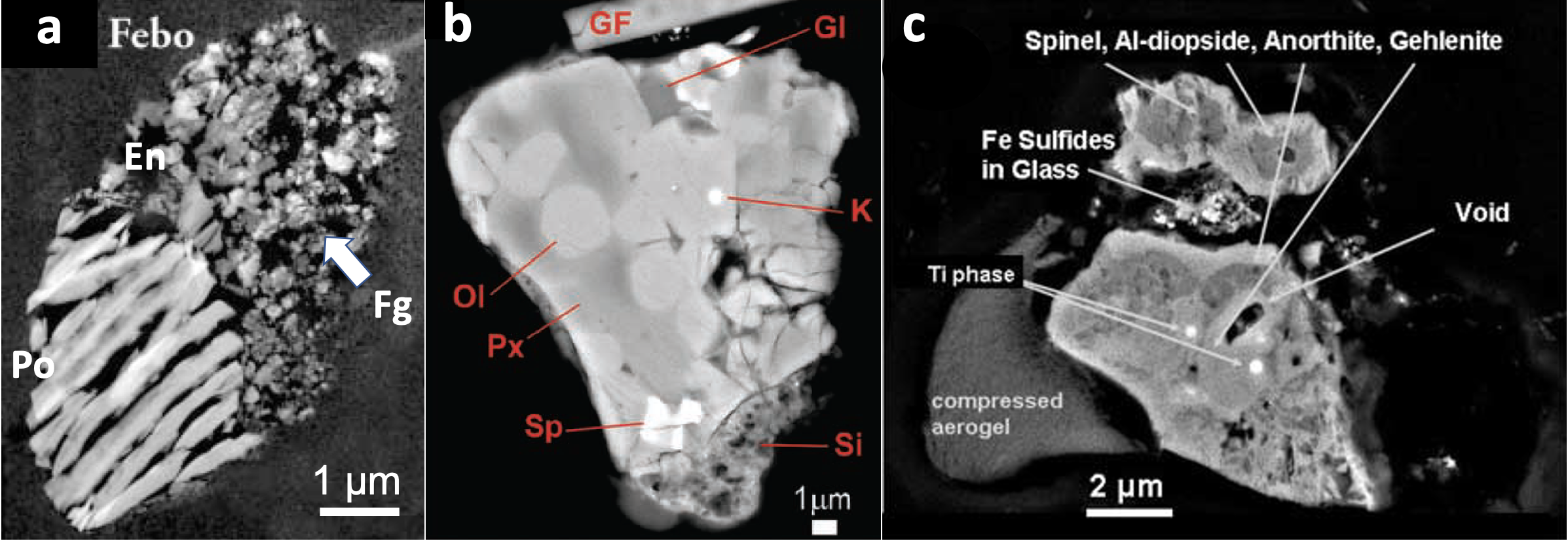}
\caption{\it \small Mineral diversity observed in 81P/Wild 2 samples brought back by the Stardust spatial mission : a) Scanning transmission electron microscopy dark field image of Track 57 terminal grain. Large pyrrhotite (Po) and enstatite (En) crystals are annotated, as well as fine-grained material (Fg). b) Backscattered electron micrograph of a chondrule-like fragment in particle Torajiro, containing olivine (Ol), low-Ca pyroxene (Px), Cr-spinel (Sp), glass (Gl), kamacite (K), silica aerogel from the collector (Si). A glass fiber (GF) holds the sample \citep{Nakamura2008}. c) Backscattered electron image of a CAI-like particle  from track 25 \citep{Zolensky2006}.}
\label{fig:mineralo_W2}
\end{figure*}

Minor element compositions of Wild 2 olivine and pyroxenes, particularly Cr and Mn, suggest that Wild 2 experienced mild secondary thermal metamorphism \citep{Frank2014a}, to approximately 400°C.  Some Wild 2 olivines and pyroxenes show compositional similarities to those in ordinary chondrites (L/LL), CH, Rumuruti and aubrite meteorites \citep [] [and submitted]{Frank2019}. In addition to diverse nebular components associated with multiple chondrite types, Wild 2 apparently incorporated materials that were liberated from evolved, internally heated asteroids, another hint for the presence of an early large scale transport mechanism acting from the inner to the outer regions of the protoplanetary disk.

A mineral assemblage found in Wild 2 samples consists of FeO-rich olivines and Na- and Cr-rich clinopyroxenes (typically augites), sometimes with poorly crystallized albite or albitic glass with spinel \citep{Joswiak2009}. These assemblages have been named “KOOL” (Kosmochloric high-Ca pyroxene and FeO-rich olivine) grains and are observed in more than half of all \Stardust \ tracks. KOOL grains are also observed in CA-IDPs and CP-MMs. The textures and mineral assemblages of KOOL grains are suggestive of formation at relatively high temperatures by igneous or metamorphic processes (its unclear which) and may have formed under relatively high $f_{O_2}$ conditions.  KOOL grains have not been observed in chondrites, however the oxygen isotopic composition of a single Wild 2 KOOL grain is similar to some type II (FeO-rich) chondrule olivines from OC, R, and CR chondrites \citep{Kita2011a, Isa2011}. One type II microchondrule in \Wild \ shows kosmochloric enhancement possibly reinforcing the link between KOOL grains and chondrule forming processes \citep{Gainsforth2015}. KOOL grains may represent an important precursor material for FeO-rich chondrules.  

While no large carbonate grains have been identified among \Wild \ samples, submicron carbonate grains have been reported \citep{Flynn2009}, including Mg-Fe-carbonates associated with amorphous silica and iron sulfides \citep{Mikouchi2007}. The observation is interesting because carbonates are typically products of aqueous processes. While Ca carbonate could plausibly be a manufacture contaminant in aerogel, Mg carbonates are unlikely \citep{Mikouchi2007}. However, in principle carbonates also can be formed without the presence of liquid water, in gas-phase reactions in the nebula \citep{Toppani2005, Wooden2002, Wooden2017}, so the presence of carbonates is not an unambiguous signature of cometary aqueous alteration.  

Sulfides are abundant in \Wild , at all sizes \citep{Zolensky2006}.  
These are predominantly pyrrhotite (Fe${_{\rm (1-x)}}$S, x = 0 to 0.2), but unusual sulfides are abundant.  Some pyrrhotites dominate terminal particles often within assemblages with igneous textures \citep{Joswiak2012, Gainsforth2013, Gainsforth2014}.
As is generally the case, pyrrhotite often occurs in association with pentlandite (FeNi)$_9$S$_8$, and Fe-Ni metal \citep{Joswiak2012}.
ZnS (probably sphalerite) is unusually abundant in \Wild \ as compared to chondrites. A single report of cubanite (CuFe$_2$S$_3$) has been interpreted as evidence for aqueous processing 
\citep{Berger2011}, however this mineral can form in non-aqueous environments. 

The iron oxide magnetite (Fe$_3$O$_4$), including a Cr-rich variety, has been identified in a few \Wild \ grains \citep{Bridges2015}. 
Although magnetite in carbonaceous chondrite meteorites is often ascribed to a secondary origin by aqueous alteration \citep{Kerridge1979} more detailed observation of \Wild \ magnetite is necessary in order to reliably assess its origin.

It is clear that carbonates, sulfides and oxides trace a diverse range of formation and processing environments and possibly provide direct evidence for aqueous alteration within \Wild , although the rarity of these particular phases and the lack of any report of phyllosilicates \citep{Brownlee2017} limits the overall extent of aqueous alteration.

\subsubsection{Chondritic Anhydrous IDP Mineralogy} \label{CA-IDPs_mineralo}

\begin{figure*}[t!]
\centering
\includegraphics[width=0.9\textwidth]{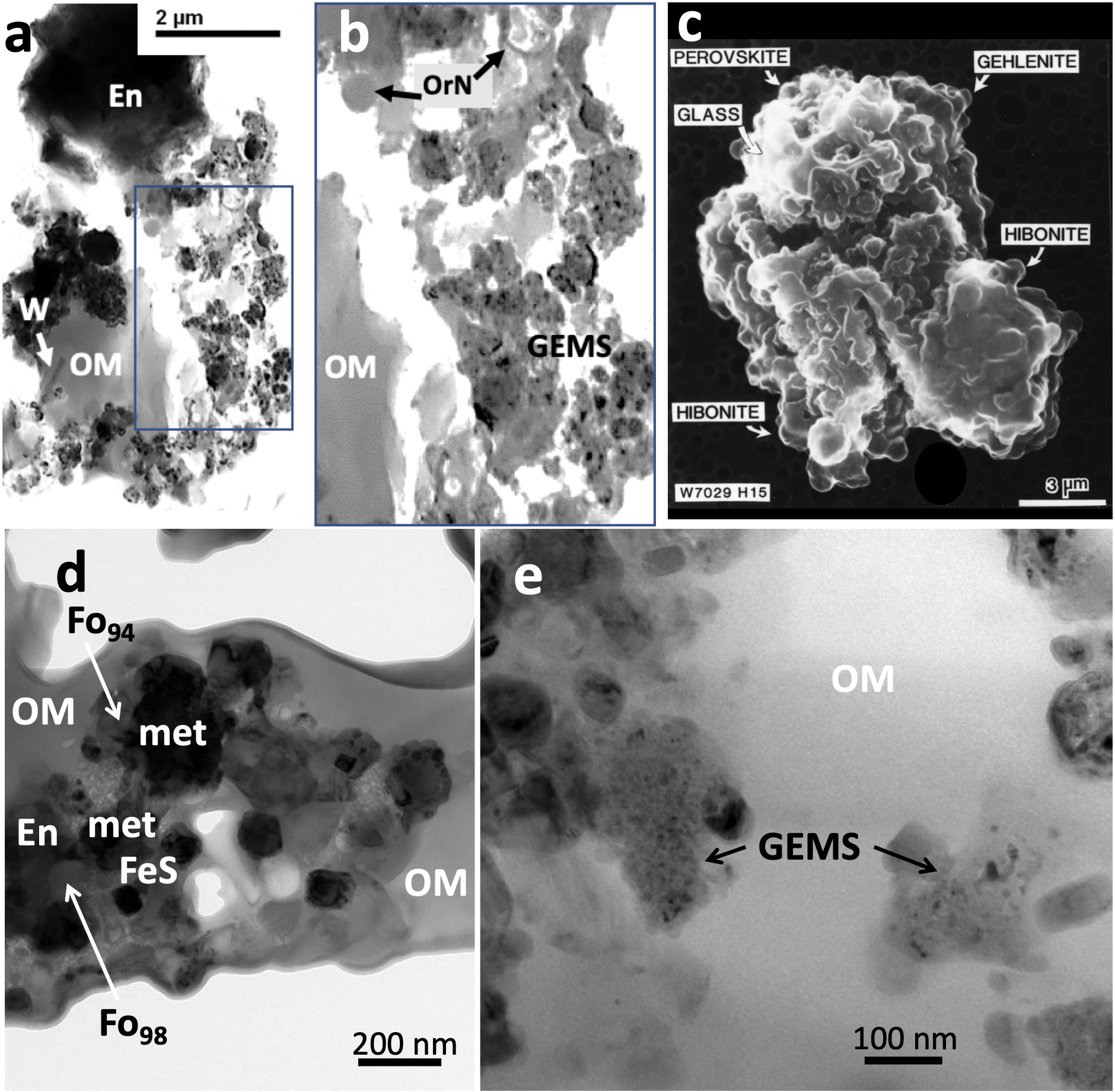}
\caption{\it \small Bright field tranmission electron microscopy (TEM) images (a,b,d,e) ad secondary electron image (c) illustrating the mineral diversity observed in CA-IDPs (a,b,c) and UCAMMs (d,e). (a,b) Anhydrous chondritic IDP U2153 Cluster particle 1; (b) area outlined at in (a).  Enstatite (En), an enstatite whisker (W), organic material (OM), GEMS, and two organic nanoglobules (OrN) are indicated (TEM image courtesy of K. Nakamura-Messenger);(c) Refractory IDP W7029 H15, containing perovskite, hibonite, gehlenite and a glass (After \citep{Zolensky1987}. (d,e) UCAMM DC06-05-94. Mg-rich olivines (Fo94 and Fo98), enstatite (En), Fe-Ni metal (met), Fe sulfides (FeS), organic material (OM)  are indicated (TEM images courtesy of H. Leroux).}
\label{fig:mineralo_IDP_UCAMM}
\end{figure*}

Individual IDPs are under 100~\um \ 
in diameter, and consist of tens to hundreds of thousands of grains, with greatly varying mineralogy and composition, i.e., non equilibrium phase assemblages. The mineralogy of the anhydrous chondritic IDPs evidences a wide range of protoplanetary disk locations and processes. The most abundant crystalline phases are ferromagnesian silicates, mainly olivine and low-Ca pyroxene with lesser amounts of high-Ca pyroxene, plagioclase, and  Fe-Ni-Zn-sulfides \citep{Rietmeijer1998}.
The (crystalline) olivine and low-Ca pyroxene compositions range from almost pure forsterite and enstatite to relatively high Fe-compositions. 
While olivine and low-Ca pyroxene in some IDPs is predominantly Mg-rich, a census of anhydrous and hydrous IDPs shows a slight preponderance of Fe-contents  $\sim$60\% 
\citep{Zolensky2008}. This is unlike the flat distribution of Fe-contents for terminal olivines ('micro-chondrules') reported for \Wild \ \citep{Frank2014a}. However additional olivine and pyroxene compositional data for IDPs is required to verify these apparent trends.

Low Fe-, Mn-enriched (LIME) olivines are proposed to be high-temperature nebular condensates \citep{Klock1989}.
Enstatite `whiskers' in chondritic IDPs, elongated along the [100] crystallographic axis, are consistent with rapid growth from a vapor phase \citep{Bradley1994b}.
Most anhydrous chondritic IDPs also contain nanoscale beads of glass with embedded metal and sulfides, called “GEMS” \citep{Bradley1994a}.  

The origins of GEMS is debated, with proposed formation mechanisms including 
irradiation of crystalline grains (olivine, pyroxene, etc.),  formation in the ISM \citep{Bradley2013}, or in the protosolar molecular cloud or outer solar nebula (protoplanetary) disk such that GEMS experienced inheritance of `solar composition' \citep{Keller2011}, or the formation by cold processes  precursor to the aggregation of IDPs \citep{Ishii2018}. 
In meteorites, GEMS-like phases have been reported in few meteorites, for example the Paris CM chondrite \citep{Leroux2015}, but this identification is disputed \citep{Villalon2016}. Only in the Ningqiang C3 chondrite is a radiation damage origin demonstrated \citep{Zolensky2003a}. GEMS  
are therefore believed to be more typical of comets than asteroids.
It is therefore very unfortunate that GEMS apparently cannot be reliably recognized in \PWildtwo \ 
samples because of their similarity to melted silica aerogel \citep{Ishii2011}. 

The unanticipated (to say the least) discovery of numerous CAIs among the recovered \PWildtwo \ grains has refocused attention to refractory IDPs \citep{Zolensky1987, McKeegan1987}, which had been ignored as they were erroneously assumed to have purely asteroid origins. These refractory IDPs differ from meteoritic analogues principally in being much finer grained, although they still await detailed characterization. 

Through seeking IDPs for comparison to \Stardust \ terminal olivine grains or 'micro-chondrules', the Giant IDPs became a focus of state-of-the-art studies  because they were found to possess a wide range of Mg:Fe- as well as minor element Mn-, Cr-, and Ca- compositions, potentially similar to the \Stardust \ olivines \citep{Brownlee2017}, although additional IDP olivine analyses are still required to demonstrate similarity. As laboratory techniques advanced and focused on Giant IDPs, the studies of anhydrous chondritic IDPs with only high-Mg content olivines were set aside.  In order to compare the formation conditions of the olivine and pyroxene in these anhydrous chondritic IDPs, similar studies to the Giant IDPs, at high spatial resolution and sensitivity, e.g.\ of elemental compositions of individual grains, are necessary.

\subsubsection{CP-MMs mineralogy}
Chondritic porous micrometeorites (CP-MMs) were first reported by \cite{Noguchi2015}. Their mineralogy is similar to that of CP-IDPS (here quoted as CA-IDPs), and is dominated by GEMS, low-Ca pyroxene (including enstatite whisker/platelet), olivine, and pyrrhotite. These minerals have angular to subrounded shapes and range from 200~nm to 1 µm in size. The olivines and pyroxenes are Fe poor, with compositions ranging from (Mg/Mg+Fe) = 0.7 to 1. Olivines and low-Ca pyroxenes in CP-MMs have compositional peaks at Mg end members (Forsterite and Enstatite). CP-MMs contain low-iron manganese-enriched (LIME) and low-iron chromium-enriched (LICE) ferro-magnesian silicates. The most Mn-enriched LIME mineral was a low-Ca pyroxene, with 2.9  wt\% MnO, and the most Cr-enriched LICE mineral was also a low-Ca pyroxene with 1.7 wt\% Cr2O3. Kosmochlor pyroxenes (Ca- and $\rm NaCrSi_2O_6$-rich are occasionally found in CP-MMs. Roedderite with co-existing low-Ca pyroxene and amorphous silicate was also found \citep{Noguchi2015, Noguchi2017}.  \cite{Noguchi2022} reports a unique particle showing both an (anhydrous) CA-IDP like composition with an hydrated part, suggesting that a partial hydration of this particle.

\subsubsection{UCAMM Mineralogy}
The mineral components of UCAMMs consist of isolated minerals or small mineral assemblages embedded in the organic matter \citep{Dobrica2012,Charon2017aa,Guerin2020,Yabuta2017}. Both crystalline and amorphous phases are present. Crystalline minerals consist of low-Ca Mg-rich pyroxenes (with stoichiometry ranging between En$_{60}$ and En$_{97}$) and Mg-rich olivines (stoichiometry comprised between Fo$_{75}$ and Fo$_{99}$) with rare Ca-rich pyroxenes and Fe(Ni) sulfides. Several hypocrystalline-like (“chondrule-like”) mineral assemblages were identified in several UCAMMs. Low Ni-Fe metal and Fe sulfides are present in mineral assemblages. In several cases, Fe metal inclusions show a rim of Fe sulfide, suggesting the occurrence of an incomplete sulfidization process. Pentlandite is occasionnaly observed. Secondary minerals also include Mn-, Zn-rich sulfide, perryite, as well as small iron oxides, carbonates and phyllosilicate-like phases \citep{Dobrica2012,Guerin2020}. Small Na-rich inclusions with stoichiometry close to Na$_2$S have been observed in the organic matter \citep{Guerin2020}. 
Glassy phases have been found in several UCAMMs \citep{Charon2017aa,Guerin2020,Yabuta2017} that resemble GEMS found in primitive IDPs \citep{Keller2011,Bradley2014a}. The GEMS-like phases in UCAMMs however tend to lack the metal inclusions of GEMS in IDPs, Fe being mostly in the form of Fe sulfides nano-inclusions. As with the Wild 2 grains, the close association of high-temperature crystalline phases with low-temperature carbonaceous matter in UCAMMs supports the hypothesis of a large-scale radial mixing in the early solar nebula \citep{Brownlee2006}. \\

UCAMMs have an olivine to low-Ca pyroxene ratio of approximately 0.5, similar to that of chondritic anhydrous IDPs \citep{Zolensky1994} and CP-MMs \citep{Noguchi2022}. This ratio is compatible with that of P- and D-type asteroids and comets \citep{Vernazza2015a}.
Although amorphous minerals like GEMS are ubiquitous in CA-IDPs, CP-MMs and UCAMMs, their crystalline mineral abundance is higher than the the upper limit of a few percent of crystallinity observed in the interstellar medium \citep{Kemper2004, Kemper2005}. \cite{Bradley2014a} reports that GEMS represent up to 70 vol\% of CA-IDPs. The value is not quoted for CP-MMs, but the authors draw a similarity with CA-IDPs. \cite{Dobrica2012} report that crystalline materials represent at least 25\% of the mineral phases analyzed in UCAMMs.

It is worth noting that in CA-IDPs, CP-MMs and UCAMMs, olivines and low-Ca pyroxenes have compositional peaks at Mg end-members. This is not observed for Wild 2 samples. Either Wild 2 ferromagnesian silicates are not typical of comets, or CA-IDPs, CP-MMs and UCAMMs derive from parent bodies different from Wild 2-type comets.

\subsection{Cometary dust compositions from astronomical IR spectroscopy}
\label{sec:three.IRspectra}
\subsubsection{IR spectroscopy}

IR spectroscopic spectral energy distributions (SEDs, $\lambda F_\lambda$ {\it versus} $\lambda$) of dust thermal emission from cometary comae ($\sim $3--40~\micron), when fitted with thermal emission models, provides constraints for the composition of the dust particles, their structure (porosity or crystal shape) and their size distribution. The `grain size distribution' (GSD), or equivalently the dust differential size distribution (DSD), has parameters of power law slope $N$, and either smallest radii limit $a_0$ or small particle radii $a_p$ where the DSD peaks before rolling off at yet smaller particle radii (often called the Hanner GSD, \citet{Hanner1983}). 
The particle porosity is often parameterized by a particle fractal dimension $D$ (see Table~\ref{tab:dusteq}). The particle porosity and DSD slope ($D$ and $N$) are co-dependent parameters \citep{Wooden2002}. With broad wavelength coverage, the mid-IR (MIR,~$\lambda\lambda $5--13~\micron) and far-IR (FIR,~$\lambda\lambda$14--40~\micron ) resonances can be sought and fitted by thermal models to  constrain the dust mineralogies.

From IR SEDs, the dust compositions are better determined for the more numerous smallest particles (in the DSD) and/or hottest particles because smaller particles produce stronger contrast spectral features (referring to \qabs) and hotter particles produce more relative flux. 
The more distinct spectral features arise from submicron to micron-sized solid particles ($P=0\%$) or from up to $\sim$20~\micron \ moderately porous particles ($P=85\%$), which has some implications for comparisons between IR SEDs and cometary samples. For example, the 20~\micron \ and larger \Stardust \ terminal olivine particles would not produce distinct spectral features. Larger extremely porous particles, which are as hot as their small monomers \citep{Xing1997}, if present in comae even if in smaller mass fractions can produce resonances \citep{Kolokolova2007} and  contribute to the SEDs \citep{Bockelee-Morvan2017a} (section~\ref{sec:two.dust.67Poutbursts}).  Particles in the DSD larger than $\sim$20~\micron \ are present in cometary comae because observed FIR SEDs decline more slowly than a single color temperature assessed for the MIR. As a compliment to this, comae do not solely possess large extremely porous particles because the presence of solely porous aggregate particles as warm as their submicron-sized monomers could not explain cometary IR SEDs. 
Thermal model parameters of composition, which is more sensitive to the properties of the smaller particles, and DSD parameters ($P$, $N$, and $a_0$ or $a_p$) are constrained by fitting the entire IR SED with a DSD that extends to particle radii out to which the IR SED is sensitive, e.g., mm-size particles for $\lambda\lambda$7--40~\micron . Above mm-size particles, longer wavelengths such as by \Herschel \ \citep{Kiss2015} and \ALMA \ are required, but the mineralogy is constrained by resonances at MIR--FIR wavelengths discussed here. 

Identification of dust compositions may be made by comparison of observed spectral features to laboratory absorption spectra but quantifying the relative abundances requires computing the thermal emission models to predict and compare an emission spectrum of an ensemble of particles to the observed IR SED using standard minimization techniques ($\chi^2$-minimization). A key aspect of determining the dust composition, when feasible, is fitting a broad wavelength SED so that multiple spectral resonances (spectral features) can be fitted from a material's vibrational stretch and vibrational bending modes that span, respectively, MIR to FIR wavelengths.

\subsubsection{{Five primary dust compositions}}
\label{sec:two.dust_five_comp}

The compositions of cometary dust as determined by IR spectral analyses has five primary components that suffice to allow the IR SED to be well-fitted by thermal models for most comets: two Mg:Fe amorphous silicates that produce the broad 10~\micron \ and 20~\micron \ features, two Mg-rich crystalline silicates that produce multiple narrow spectral peaks in the range of $\sim$ 8--33~\micron , and a highly absorbing and spectrally featureless and warmer dust component that is ubiquitously observed to dominate the dust's thermal emission 
in the NIR ($\sim$3--7.5~\micron). There is a consensus amongst modelers that this NIR dust emission, sometimes called the NIR dust `continuum' or `pseudo-continuum)', is produced by a highly absorbing, carbonaceous dust component. This component is well fitted by the optical properties of amorphous carbon  \citep{Woodward2021, Bockelee-Morvan2017a, Harker2022}. Aliphatic carbonaceous dust materials are relatively transparent compared to amorphous carbon, and graphitic carbonaceous matter is rare in cometary samples. The potential connections between cometary dust and amorphous carbon, versus other species, e.g.\ organics dominated by aromatic bonds, by hydrogenated amorphous carbon (HACs), or by graphite, are discussed in \citet{Wooden2017, Woodward2021}.

For dielectric materials like silicates, there are spectral features from which we can deduce the mineralogy in combination with modeled radiative equilibrium temperatures and modeled spectral emission features. 
These four primary siliceous compositions are: 
Mg:Fe~$\simeq$~50:50 amorphous pyroxene-like (Mg$_{0.5}$,Fe$_{0.5}$)SiO$_3$ and 
Mg:Fe~$\simeq$~50:50 amorphous olivine-like  (Mg$_{0.5}$,Fe$_{0.5}$)$_2$SiO$_4$, hereafter referred to as Mg:Fe amorphous pyroxene and Mg:Fe amorphous olivine \footnote{Mg:Fe~$\simeq~$50:50 ratios for each of Mg:Fe amorphous olivine and Mg:Fe amorphous pyroxene were determined from radiative  equilibrium temperature calculations and SED fitting of comet Hale-Bopp \citep{Harker2002,Harker2004}, using optical constants from \cite{Dorschner1995}, and so Mg:Fe~$\simeq$~50:50 is used by dust modelers \citep[e.g.,][]{Ootsubo2020, Bockelee-Morvan2017a}.}; 
Mg-olivine (forsterite) (Mg$_x$,Fe$_{(1-x)}$)$_2$SiO$_4$ for 1.0$\leq$x$\leq$0.8, which produces sharp resonances (spectral `peaks') at or near 11.1--11.2, 19.5, 23.5, 27.5 and 33.5~\micron \ and weaker peaks at 10.5 and 16.5~\micron \ \citep{Crovisier1997, Hanner2010, Wooden2017, Koike2003, Koike2010}; 
Mg-orthopyroxene (enstatite) (Mg$_y$,Fe$_{(1-y)}$)SiO$_3$ for 1.0$\leq$y$\leq$0.9 that produces resonances at or near 9.3 and 10.0~\micron \ and with a set of FIR resonances near 20~\micron , which is also where Mg:Fe amorphous pyroxene has a broad feature. Depending on the laboratory data and optical constants, the FIR resonances for Mg-pyroxene may be at these sets of wavelengths: (18.5, 19.2, 20.3~\micron) for ellipsoidal shapes (in our plots) using optical constants from \citet{jaeger1998}, which gives peaks measured at (18.2, 20.6, 21.6~\micron); also, laboratory spectra of ortho-enstatite give peaks at (9.3, 10.7, 19.5, 20.7~\micron) from \citet{Chihara2001}. Cometary features from Mg-pyroxene are not yet detected at high spectral contrast in the FIR. For a spectrum with strong Mg-pyroxene one may look at FIR  \SpitzerIRS \ spectrum of Herbig Ae/Be star HD179218 \citep[cf.][Fig.~13]{Juhasz2010}. Mg-olivine is a well established cometary dust component by either a peak at 11.1-11.2~\micron \ or by a shoulder on the broad `10~\micron ' feature, which makes the feature look `flat-topped'. Also, the FIR peaks for Mg-olivine are well separated from the center of the broad Mg:Fe amorphous olivine `20~\micron ' feature, whose Mg:Fe composition is established by the radiative equilibrium temperatures \citep{Harker2002,Harker2004}. When detected, the mass fraction of Mg-olivine present in the coma is \gtsimeq 20\%. 

Crystalline silicate feature wavelengths and feature relative intensities depend on composition and crystal shape. Mg-rich (100\% -- 90\% Mg) crystalline Mg-olivine (forsterite) is tri-refringent (3 optical axes or 3 sets of indices of refraction) and has spectral peaks that can shift to somewhat shorter or longer wavelengths depending on crystal shape \citep{Koike2010, Lindsay2013}. The observed spectral features of forsterite are better fitted by rectangular prisms slightly elongated or flattened along the b-axis \citep{Lindsay2013} or by ellipsoidal shapes flattened also along the b-axis \citep{Harker2002,Harker2004, Harker2007}; note that \citet{Fabian2000} shows ellipsoidal shapes and quotes elongation along the b-axis but actually, as computed, these particles are b-axis flattened when the ($L_b$) parameter used in computing the ellipsoidal particles, as given as by \citep{Bohren1983}, is larger. The optical constants used for forsterite (in figures shown here) are from \citet{Steyer1974}. See \citet{Juhasz2009} for a comparison of other optical constants for Mg-olivine. 
Optical constants of Mg-olivine that are derived from measurements of polished single-crystal samples \citep{Steyer1974, Suto2006} are preferred for modeling crystal shapes \citep{Lindsay2013} rather than optical constants derived from ground samples that inherently have shape-dependencies \citep{Fabian2001, Koike2010, Imai2009}. 
Specific examples of crystal shape are revealed in the absorption spectra of forsterite powders that are prepared by hand-grinding or ball-grinding, where ball-grinding imparts greater sphericity to the particles \citep{Koike2010, Lindsay2013, Imai2009, Tamanai2006, Tamanai2009}. A long standing discrepancy in lab data was resolved when it was realized that the preparation of the sample as well as the medium in which the ground mineral sample is embedded (KBr {\it vs.} PE) affects the wavelength positions and relative depths of absorption bands and the degree of sensitivity varies for different spectral features as well as with embedding medium \citep{Tamanai2009}; particles lofted in air also affect the absorption spectra because particles electrostatically cling and agglomerate \citep{Tamanai2006}.

With increasing Fe-content, olivine peaks shift to longer wavelengths \citep{Koike2003}. The observed range of wavelengths for the cometary 11.1-11.2~\micron \ feature restricts Mg-contents from 100--80\% for Mg-olivine (see Fig.~\ref{fig:sil_resid_wave_vs_Mgcontent}).

Silicates, being dielectrics, emit strongly through their MIR vibrational stretching and FIR vibrational bending modes and possess little absorptivity at wavelengths shorter than $\sim$7.5~\micron \ so, we repeat, the thermal emission at shorter wavelengths is likely from a carbonaceous dust component. Iron sulfide (FeS) mineral is dark but not as absorbing as amorphous carbon. FeS has yet to be modeled as a major dust component contributing in the NIR for reasons that include: the optical constants are lacking for the full range of wavelengths and/or are contested and are specifically lacking at 3--10~\micron , FeS is not yet firmly detected spectroscopically (section~\ref{sec:two.dust_not_detected_yet}), and FeS does not yield the correct scattered light color even if held to 6\% volume of the particle \citep{Bockelee-Morvan2017a} (see section~\ref{sec:two.dust.67Poutbursts}).

Each of the five dust compositions that dominate SEDs have analogous materials in cometary samples and in anhydrous chondritic IDPs (CA-IDPs). Spectral features are measured in bulk and thin sections of CA-IDPs \citep{Bradley1992, Wooden2000,Matrajt2005}, in CP-MMs \citep{Noguchi2015} and in UCAMMs \citep{Dartois2018}. The amorphous silicates are thought to be akin to the GEMS \citep{Bradley1992, Bradley1999, Bradley2013,Noguchi2015,Dobrica2012}.  The Mg-olivine and Mg-pyroxene are spectrally akin to forsterite Fo$_{100}$--Fo$_{80}$ and ortho-enstatite En$_{100}$--En$_{90}$. The amorphous carbon is akin to some phases of disordered carbon and the occasionally quoted amorphous carbon in CA-IDPs \citep{Munoz-Caro2006, Wirick2009, Woodward2021}. A fraction of the smallest particles observed in the {\it in situ} measurements of \Halley \ were composed solely of carbon \citep{Lawler1992, Fomenkova1994a}. Alternatively, we note that possibly aromatic-bonded carbon ($\pi$-bonds or C$=$C bonds) could produce a significant absorptivity because of its higher UV-VIS cross sections.
Organic matter in CA-IDPs and CP-MMs is generally similar to IOM from primitive meteorites, however with intrinsic characteristics like a higher aliphatic/aromatic ratio in CA-IDPs compared to meteoritic IOM \citep{Flynn2003, Keller2004, Matrajt2005, Munoz-Caro2006, Wirick2009, Matrajt2012, Noguchi2015}. Organic matter in UCAMMs is dominated by polyaromatic matter like in IOM, however it can show higher abundances of N than in meteoritic IOM \citep{Dartois2018} (see section \ref{sec:two.organics}).

\begin{figure}[t!]
\centering
\includegraphics[width=0.41\textwidth]{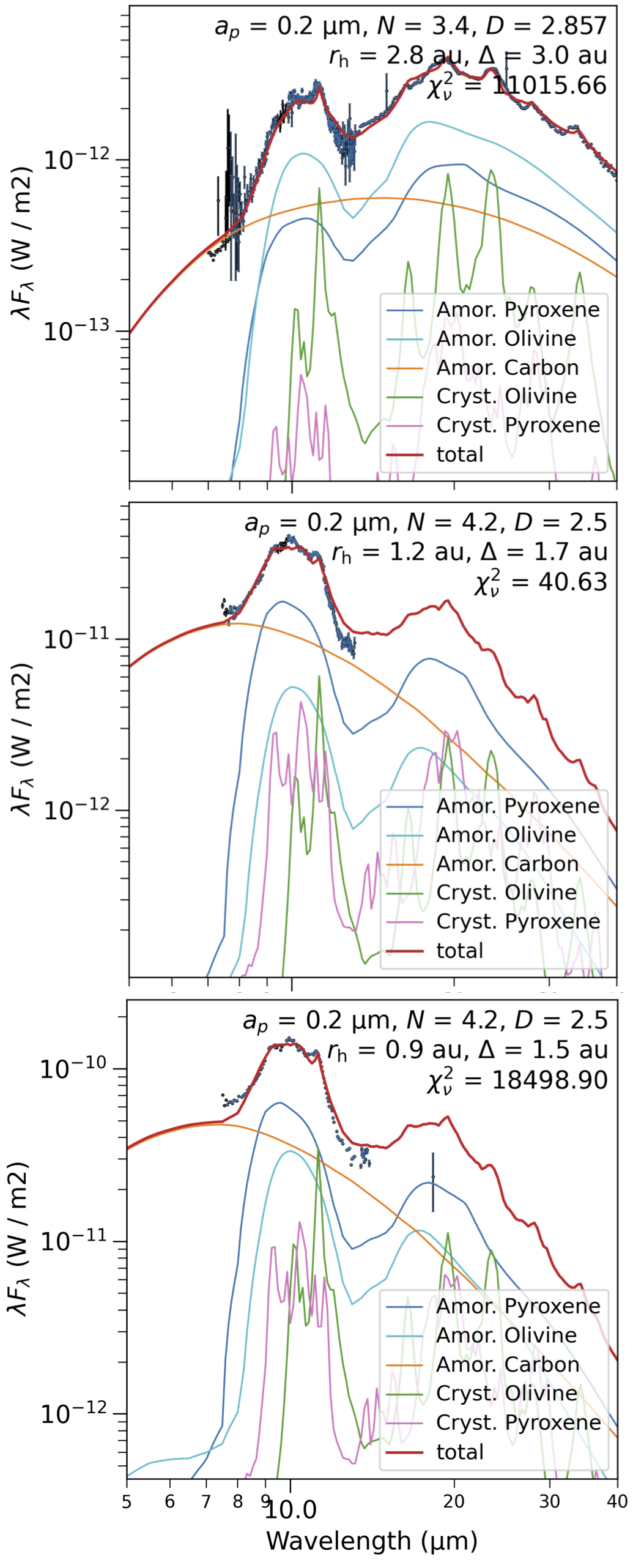}
\caption{Comet C/1995 O1 (Hale-Bopp) at three epochs with thermal models \citep{Woodward2021} fitted with 5 compositions ({\it Top to Bottom}): 1996-10-11~UT pre-perihelion \rh = 2.75~\au \ from \emph{IRTF}+HIFOGS and \emph{ISO}+SWS \citep{Crovisier1997}, \emph{IRTF}+HIFOGS on 1997-02-14~UT and 1997-04-11~UT at \rh=1.2~\au \ and \rh=0.93~\au \ 
\citep{Wooden1999, Harker2002,Harker2004}.  Mg-olivine (crystalline) has peaks at 10.0, 11.1-11.2, 16.5, 19.5, 23.5, 27.5, and 33.5~\micron . Mg-pyroxene (crystalline) has a triple of peaks at 9.3, 10.5, and 11.0-11.3~\micron \ and have modeled far-IR peaks near 19.6~\micron , 18.7~\micron \ and $\sim$ 17.9~\micron \ and in $\lambda$-vicinity of the FIR Mg:Fe amorphous pyroxene broad feature.}
\label{fig:halebopp}
\end{figure}

Figure~\ref{fig:halebopp} shows comet \HaleBopp , which is the best example of an IR SED of a coma with a plethora of submicron silicate crystals and a higher silicate-to-amorphous carbon ratio (high contrast silicate features relative to the `featureless' emission from a distribution of porous amorphous carbon particles and distribution of larger porous amorphous silicates). 
Clear and distinctive spectral peaks from (crystalline) Mg-olivine and (crystalline) Mg-pyroxene provide a benchmark for modeling crystalline silicates (shapes and temperatures). Specifically, the `hot crystal' model \citep{Harker2002,Harker2004, Harker2007, Woodward2021, Harker2022} increases the radiative equilibrium temperature of the Mg-olivine by a factor of 1.7 over that predicted using the optical constants in order to fit the Hale-Bopp spectra at 2.75~\au. 
\HB 's Mg-pyroxene is spectrally discernible by its sharp peaks more so at epochs near perihelion than at \rh \ =2.75~\au,  which is attributed to its transparency compared to the other dust compositions \citep{Wooden1999}. The same jets are active at the two epochs at \rh = 1.2~\au \ and 0.93~\au  \ \citep{Hayward2000}, and the thermal models fitted to their IR SEDs reveal minor differences in their mineralogy such that Mg-pyroxene has a greater relative abundance compared to Mg-olivine at 1.2~\au \ than at 0.93~au: if the relative strength of the Mg-pyroxene and Mg-olivine features only are attributed to a temperature increase at smaller heliocentric distances then Mg-pyroxene would be expected to be enhanced relative to the Mg-olivine peaks at 0.93~\au \ but instead by visual comparison the Mg-pyroxene peaks are enhanced at 1.2~\au \ compared to 0.93~\au. 
Also, shown in Fig.~\ref{fig:halebopp}  is the distinct contribution of the Mg-pyroxene 9.3~\micron \ peak to the short wavelength shoulders of the broad 10~\micron \ feature from Mg:Fe amorphous pyroxene and Mg:Fe amorphous olivine.

\begin{figure}[t!]
\centering
\includegraphics[width=0.40\textwidth]{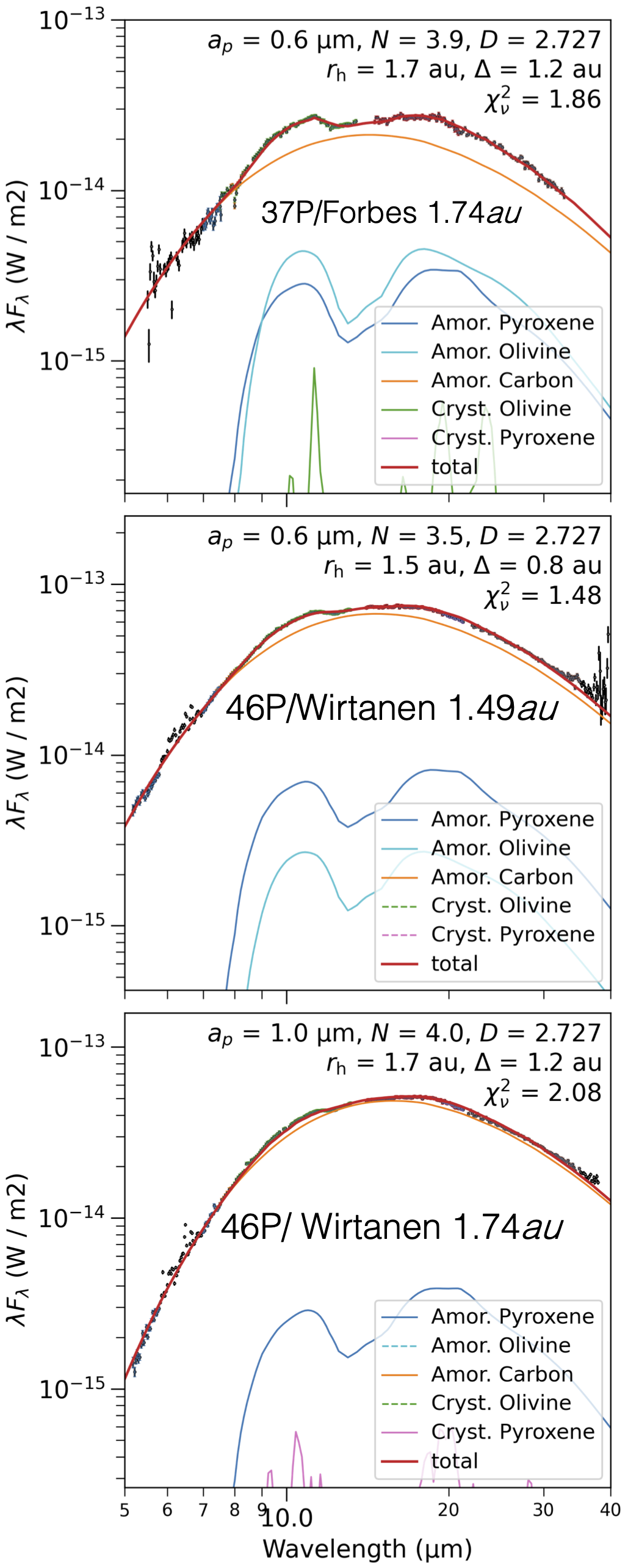}
\caption{{\it Spitzer}+IRS SEDs of JFCs including 37P/Forbes at \rh = 1.74~\au \ (2005-10-14.63~UT), and two epochs of 46P/Wirtanen showing IR SEDs with decreasing constrast silicate features and decreasing ratios of silicate-to-amorphous carbon in their fitted thermal models \citep{Harker2022}. Amorphous silicates and no crystalline silicates are constrained in the thermal model fit to 46P at \rh=1.49~\au \ (2008-04-24.64~UT). For comet 46P/Wirtanen at \rh = 1.74~\au \ (2008-05-24.06~UT), weak features from Mg-pyroxene (crystalline) are evident at 9.3, 10.5~\micron \ but more difficult to discern in the far-IR and where the Mg-pyroxene resonances overlap in spectral wavelength with the far-IR broad Mg:Fe amorphous pyroxene feature. 
}
\label{fig:Spitzer_3SEDs_JFCs}
\end{figure}

\begin{figure}[t!]
\centering
\includegraphics[width=0.45\textwidth]{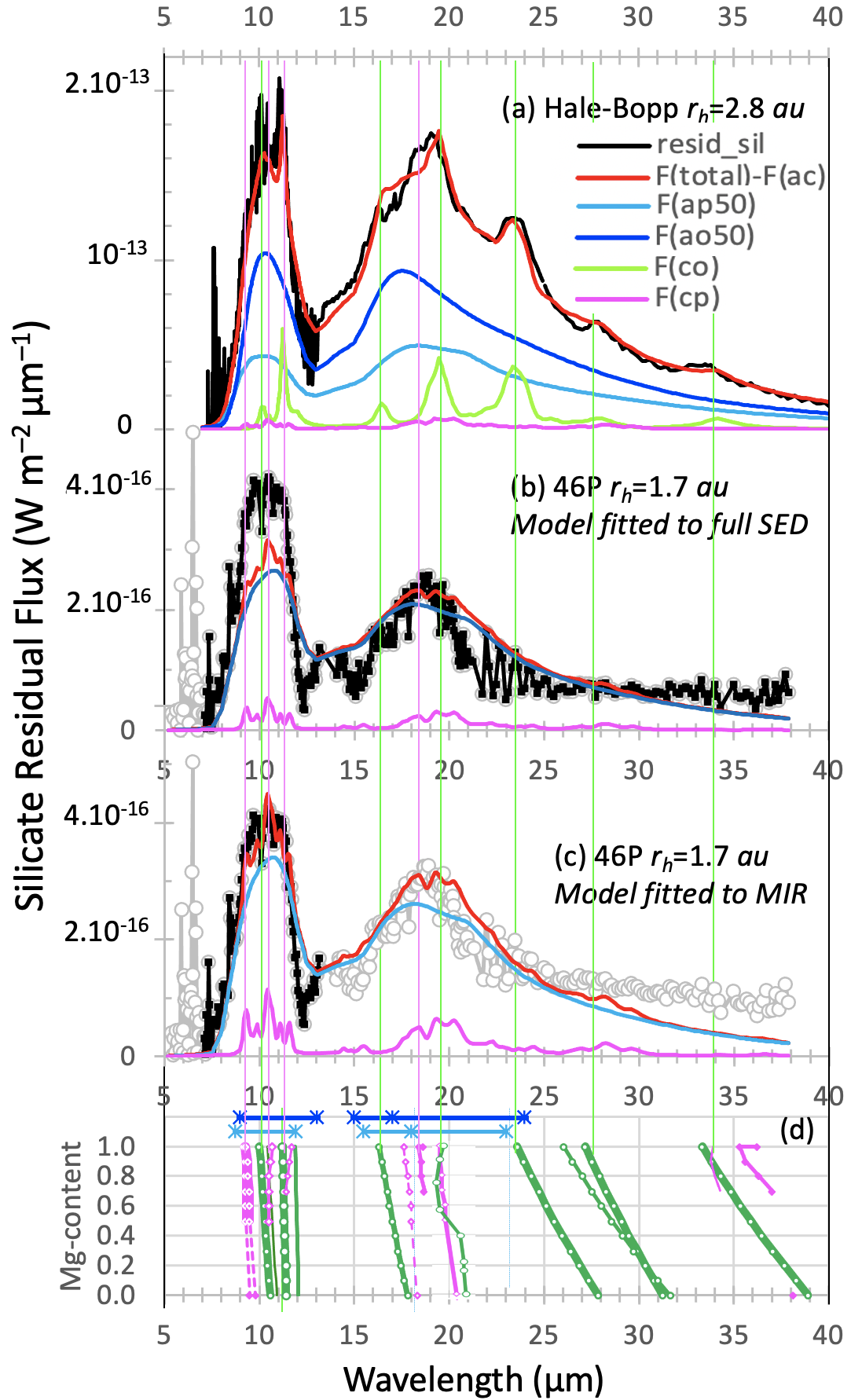}
\caption{Silicate residual flux spectra (a-c) \citep{Harker2022}, shown to highlight the wavelengths of observed features of (crystalline) Mg-olivine (co, in {\it green}) and (crystalline) Mg-pyroxene (cp, in {\it magenta}) and to compare with (d) the wavelength dependencies of the laboratory spectra versus Mg-content=(1 $-$ Fe-content). The silicate residual fluxes are the data and model with amorphous carbon (ac) subtracted, i.e.,  F$_{data}$ - F$_{model}$(ac) and F$_{model, total}$-F$_{model}$(ac), respectively, for: (a) Hale-Bopp at \rh =2.8~\au \ (1996-10-11~UT) (Fig.~\ref{fig:halebopp}), (b) 46P at \rh =1.7~\au \ (2008-05-24.06~UT) fitted over the full SED (5--35~\micron) {\it black points} (as fitted in Fig.~\ref{fig:Spitzer_3SEDs_JFCs}), and (c) 46P at \rh =1.7~\au \ ((2008-05-24.06~UT) and fitted only over the MIR 5--13~\micron{}, which shows an increase in Mg-pyroxene relative abundance compared to the model in (b). (d) Wavelength dependencies of Mg-olivine ({\it green}) and Mg-pyroxene ({\it magenta}) from laboratory work by \citet{Chihara2002, Koike2003}; thickness of lines relates to strengths of observed peaks. Also are shown are Mg:Fe amorphous olivine ({\it royal blue}) and Mg:Fe amorphous pyroxene ({\it cyan}). The observed and modeled crystalline feature wavelengths align with Mg=1.0--0.8.}
\label{fig:sil_resid_wave_vs_Mgcontent}
\end{figure}

\subsubsection{Degeneracies in compositions derived from thermal models}

Possible degeneracies in dust compositions that can be overcome by fitting multiple spectral peaks include: (a)~the 11.1-11.2~\micron \ feature of Mg-olivine and the 11.2~\micron \ PAH emission band; (b)~wavelength shifts due to crystal shape \citep{Koike2010,Lindsay2013} or increasing Fe-contents \citep{Koike2003} because the 23.5~\micron \ FIR feature of Mg-olivine is more sensitive to shape (olivine with greater than 40\% Fe causes the entire FIR spectrum to change morphology \citep{Koike2003}); (c)~at \rh \gtsimeq 3--3.5~\au, the MIR short wavelength shoulder can be fitted with Mg:Fe amorphous pyroxene or by the warm featureless emission from amorphous carbon but fitting the FIR pyroxene band removes this degeneracy; (d)~the Mg-pyroxene 9.3~\micron \ peak and the short wavelength shoulder of the Mg:Fe amorphous pyroxene feature occur at 9.3~\micron \ but sufficent SNR to detect the Mg-pyroxene 10.5~\micron \ peak as well as measuring the far-IR resonances can distinguish between these two compositions. If we contrast Mg-pyroxene with Mg-olivine, Mg-olivine may be assessed even when the 11.1-11.2~\micron \ peak is not clearly discernible against the broad Mg:Fe amorphous olivine or Mg:Fe amorphous pyroxene features because Mg-olivine's primary peak in the MIR occurs at the longer wavelength side of the broad `10~\micron \ silicate feature' so that if a broad silicate feature appears to not be declining at 11-12.5~\micron \ but instead appears `flat-topped' then Mg-olivine is the candidate. 

The FIR the peaks from Mg-olivine are at distinctly different wavelengths than the FIR Mg:Fe amorphous olivine feature, thereby allowing for FIR Mg-olivine peaks to not compete with FIR Mg:Fe amorphous olivine in the $\chi^2$-minimization between the model and the data, and the effect is that Mg-olivine FIR peaks are spectrally discernible even when the FIR peaks are weak.

In contrast, the FIR sharp peaks from Mg-pyroxene and the FIR broad feature of Mg:Fe amorphous pyroxene are at nearly the same wavelengths. Moreover, there are some variations in the predicted wavelengths for the Mg-pyroxene FIR peaks partly because of there are variations in the optical constants, the peaks depend on shapes of the crystals, and laboratory measurements of feature wavelengths depend on the embedding medium \citep{Tamanai2009}.  We have yet to observe a cometary IR SED that has strong Mg-pyroxene peaks in the FIR to confirm the choice of optical constants and crystal shapes in the models.
We show for comet 46P/Wirtanen at \rh = 1.74~\au , the coincidence of features of Mg-pyroxene and Mg:Fe amorphous pyroxene in the FIR (see  Fig.~\ref{fig:sil_resid_wave_vs_Mgcontent}(b,c); {\it Bottom Panel} Fig.~\ref{fig:Spitzer_3SEDs_JFCs}). In observed cometary IR SEDs, Mg-pyroxene peaks appear more discernible at MIR wavelengths than in the FIR, and this may contribute to Mg-pyroxene having lower relative abundance relative to the other dust components when the MIR and FIR are fitted compared to when only the MIR is fitted. As a demonstration, compare model fitted to full SED of 46P at 1.7~\au \  (Fig.\ref{fig:sil_resid_wave_vs_Mgcontent}~b) with  model fitted only to MIR  (Fig.\ref{fig:sil_resid_wave_vs_Mgcontent}~c). This aspect of the thermal models seems contrary to the aim of fitting multiple resonances over the fullest wavelength range and why spectral features of Mg-pyroxene in the FIR are of lower contrast than predicted by thermal models, given the resonances in the MIR, is a puzzle and may depend on their optical properties (see section \ref{sec:two.dust_five_comp}). When weak Mg-pyroxene features are modeled then there is a significant relative mass fraction (\gtsimeq 30\%) of this dust component because its lower absorptivity ($Q_{abs}$) \citep{Harker2022} yeilds cooler temperatures and hence lower fluxes relative to warmer dust components (i.e., a greater relative mass is required to fit the observed features when the particles are cooler) \citep{Wooden1999}. Mg-pyroxene FIR peaks may or may not be fitted by $\chi^2$-minimization of models to the data when the stronger broad feature from Mg:Fe amorphous pyroxene broad feature dominates the flux. These factors need to be considered when contrasting the relatively low number of cometary IR SEDs fitted with Mg-pyroxene compared to the frequent identification of Mg-pyroxene in cometary samples. 

Comet 46P at \rh =1.47~\au \ has a weak silicate feature (Fig.~\ref{fig:Spitzer_3SEDs_JFCs} (b)). The fitted thermal model has a DSD that peaks at $a_p$=0.6~\micron \ so submicron particles are present in the coma, which can produce distinct features when composed of silicates. The weak silicate feature thus is modeled by a high relative mass fraction of amorphous carbon compared to the amorphous silicates whose resonances also are fitted. This epoch of comet 46P is similar to comet C/2017 US$_{10}$ (Catalina) that has a high wt\% amorphous carbon. The elemental C/Si ratio for comet C/2017 US$_{10}$ (Catalina) is similar to comet \CG , which has organic IOM-like matter, and both of which have C/Si ratios close to the ISM value \citep{Woodward2021}.

\subsubsection{Dust compositions not yet firmly detected in IR SEDs}
\label{sec:two.dust_not_detected_yet}

As stated above, the five dominant compositions assessed from IR SEDs have analog dust species in cometary IDPs, in \Stardust \ samples, and in \emph{in situ} compsitional studies of \Halley \ and \CG \ particles but  the contrary is not true: there are dust compositions in cometary samples that are not detected in IR SEDs that include inorganics and organics. 

As the Fe-content of olivine increases from 20\% to 40\%, the MIR peak shifts from 11.2 -- 11.4~\micron \ \citep{Koike2010} but this wavelength also depends on crystal shape \citep{Lindsay2013}. Detections of peaks in the 16--27~\micron \ FIR region are required to discern Mg:Fe$~\approx$~60:40 from the peaks we observe and model in comets that span Mg:Fe proportions from 100:0 to 80:20 (see Fig.~\ref{fig:sil_resid_wave_vs_Mgcontent} and \citet{Koike2003}). Due to the enhanced \qabsvis \ of Fe-olivine, thermal models also will need to demonstrate increased raditative equilibrium temperatures of Fe-olivine in contrast to Mg-olivine.

Fe sulfides are present and abundant in many cometary samples and UCAMMs. FeS may be identified through a very broad 23~\micron \ feature, which is much broader compared to Mg-olivine 23.5~\micron \ feature, and only is expected for submicron FeS grains \citep{Keller2000}. However, different measurements produce different predictions for \qabsir \ that range from strong FIR resonances \citep{Begemann1994, Henning1997} to no FIR resonances \citep{Hofmeister2003}. A single reference suggests a 3.63~\micron \ feature in the NIR \citep{Tamanai2003}, 
and dirth of FeS data in the NIR is represented by a dashed line in the \citet{Pollack1994}'s protoplanetary disk opacities. If FeS has a spectral feature near 3~\micron \ then FeS could not provide the opacity needed to explain the ubiquitous featureless thermal emission (NIR `pseudo-continuum') from warm particles, which currently is ascribed to a highly absorbing cabonaceous component and well-fitted by amorphous carbon. 

The MIR spectral features of phyllosilicates overlap with the 10~\micron \ features of anhydrous amorphous silicates, e.g.\ Mg:Fe amorphous olivine and Mg:Fe amorphous pyroxene, but the 20~\micron \ features from phyllosilicates are significantly different \citep{Wooden1999}. Phyllosilicates including montmorillonite \citep{Wooden1999}, smectite and serpentine are abundant in hydrated chondritic IDPs \citep{Bradley1989}. At most 5\% montmorillonite may be present in the coma of \HaleBopp \ \citep{Wooden1999}. The spectral features of phyllosilicates (smectite nontronite) are claimed for \tempelone \ at +45~{\it min} post-\DI \ \citep{Lisse2007}, as well as for comet \HaleBopp  \ but other well-fitted thermal models are without phyllosilicates \citep{Harker2002,Harker2004,Harker2002a, Min2005} (Fig.~\ref{fig:halebopp}).

The low density amorphous silicates in cometary IDPs, the 
Glass with Embedded Metal and Sulfides (GEMS) \citep{Bradley2022}, have limited spectral data \citep{Bradley1999, Ishii2018} and are presumed to be spectrally analogous to Mg:Fe amorphous silicates in thermal models, which were derived from rapid-cooling of melts \citep{Dorschner1995}. Amorphous silicates are discussed in the context of IR spectra of cometary IDPs \citep[e.g.][]{Brunetto2011}. Experiments have created amorphous silicates, e.g.\ via particle bombardment of crystals or via the solgel method \citep{Jaeger2003, Brucato2004, Demyk2004, Wooden2005, Wooden2007, Jaeger2016} but how the structure of the experimental amorphous silicates or the “amorphization rate” affects the IR signature of the analogs used in thermal models has not been  assessed.

Mg-carbonates are relatively rare in cometary samples (section \S\ref{sec:two.mineral}) and are not definitively detected in cometary IR SEDs. Carbonates (siderite and magnesite) are discussed for comet \tempelone \ \citep{Lisse2007} but the simultaneous occurrence of water vapor emission lines in the overlapping wavelength region, when modeled, yields a marginal detection of the 7.00~\micron \ carbonate feature and an abundance that is 2 to 3 times lower \citep{Crovisier2008}. Water vapor lines must be modeled concurrently with the dust thermal emission, and the 5--8~\micron \ spectral region calls for higher SNR studies such as will be available from \jwst .

\begin{figure}[h!]
\centering
\includegraphics[width=0.45\textwidth]{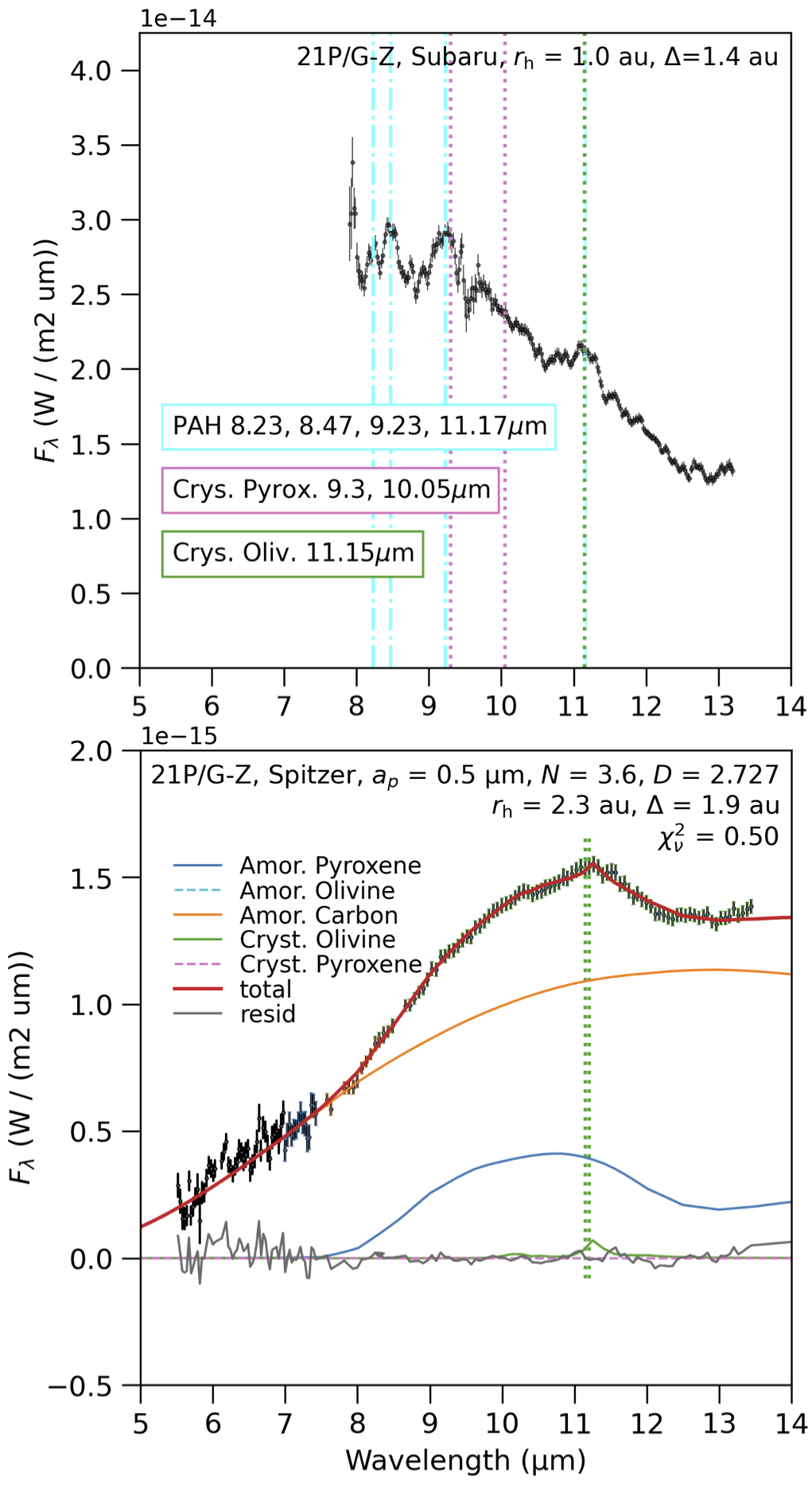}
\caption{Comet 21P/Giacobini-Zinner (21P/G-Z) at two different heliocentric distances (\rh{}) and with higher {\it versus} lower spatial resolution.\ {\it Upper Panel}.\ For 21P/G-Z at \rh{}=1.0~\au (2005-07-05~UT), {\it Subaru}+COMICS spectroscopy of the inner coma reveals emission bands attributed to PAHs (8.23, 9.23, 11.17~\micron{}) that contribute in excess to the thermal model of amorphous carbon, Mg:Fe amorphous olivine, Mg:Fe amorphous pyroxene, Mg-olivine (Crys.\ Oliv.) and Mg-pyroxene (Crys.\ Pyrox.) \citep{Ootsubo2020}. Data courtesy of T.~Ootsubo. {\it Lower (SED)}.\ \spitzer{}+IRS spectroscopy of 21P/G-Z at \rh=2.3~\au, fitted with the 5-mineral thermal model providing constraints for amorphous carbon, Mg:Fe amorphous pyroxene, and Mg-olivine (crystalline). 
21P is only fitted through 14~\micron\ because of possible calibration issues associated with the data taking sequence \citep{Harker2022, Kelley2021}. Compared to 2.3~\au, PAH bands are present in the coma of 21P at \rh{}=1~\au \ as well as hotter particles, which can be seen by the spectral slope change between 1.0~\au \ and 2.3~\au.}
\label{fig:21P}
\end{figure}

Aromatic macromolecules or PAHs are efficient emitters of IR photons and their spectra can be predicted \citep{Astrochem2022}.  
Unlike the free-flying PAH macromolecules, we are asking: what are the excitation mechanisms that could produce emission features from aromatic-bonded carbon within organic-rich particles? Can photoprocessing of PAHs, which can alter their size distribution, occur within organic-rich particles \citep{Clemett2010}? Currently, we rely on computing emission spectra of particles using optical constants that have been determined for a limited number of organic residues (tholin-like materials). Measurements of absorption spectra are more numerous than the suite of available optical constants ($n$, $k$) from which we can compute \qabsvis \ and \qabsir \ for thermal emission models. UCAMMs offer a template for the potential wavelength positions of IR spectral features because UCAMMs have so much organic material, including polyaromatic bonds, that their absorption spectra can be obtained without acid dissolution of the silicates \citep[e.g.][]{Dartois2018}, which is done for meteoritic IOM \citep{Alexander2017} and that is known to alter some of the organic fraction in IDPs \citep[e.g., see][]{Matrajt2005}. However, UCAMMs contain a N-rich organic phase that is too N-rich to be akin to meteoritic IOM and \CG{}, so UCAMMs can not serve to predict the presence of organic features in comets such as \CG{}.

The PAH spectral features sought are their skeletal vibration bands in the 5--8~\micron \ wavelength region and by their peripherial C-H bonds near 3.28~\micron \ and at longer wavelengths where the skeletal structure affects the feature locations. 
PAHs are discussed for comet \tempelone \ \citep{Lisse2007}. However, thermal models of the same IR SED that simultaneously model the water vapor emission lines and the PAHs in the 5.25--8.5~\micron \ wavelength region produce lower abundances of PAHs by at least 2--3 times \citep{Crovisier2008}. 

Distinct resonances from PAH-like organics contribute to features at $\sim$8.5~\micron \ and  $\sim$11.2~\micron \ and possibly aliphatic hydrocarbons at $\sim$9.2~\micron \ (Fig.~\ref{fig:21P}) in the inner coma of comet 21P/Giacobini-Zinner at \rh =1.04~\au \ \citep{Ootsubo2020} as well as Mg-rich amorphous silicates and Mg-silicates. However, the organic bands are not detected in the coma of 21P post-perihelion by lower spatial resolution \spitzer +IRS at the larger heliocentric distance of \rh = 2.292~\au \  \citep{Harker2022} (Fig.~\ref{fig:21P}).

Organics with aliphatic bonds may produce a 3.4~\micron \ emission feature such as seen in CA-IDPs \citep{Matrajt2005a} and in some \Stardust \ particles \citep{Matrajt2008} and are suggested for comet \Hartley \ \citep{Feaga2021, Wooden2011} as well as other comets by \citet{Bockelee-Morvan1995} who noted a strong correlation with CH$_3$OH production rates. Lines of sight through the diffuse ISM, e.g, towards the Galactic Center, reveal the 3.4~\micron \ feature. By comparison of the two primary components of this feature from $-$CH$_3$ ($\sim$2960~\wvnum , 3.38~\micron) and $-$CH$_2$ ($\sim$2930~\wvnum, 3.41~\micron ), the diffuse ISM has shorter chains and is more processed than in cometary matter \citep{Matrajt2013}. 

This solid state emission feature from organics is in the same spectral region as gaseous emission lines from ethane C$_2$H$_6$ and methanol CH$_3$OH that are observed at high spectral resolution and for which modeling is a challenge \citep[Fig.~5]{Bonev2021} so a method that combines observing or modeling the emission lines of the gaseous species \citep[as mentioned in ]{Feaga2021} with modeling the solid state material that typically has multiple resonances from $-$CH$_2$ and $-$CH$_3$ bonds is required to further assess the presence of this aliphatic carbon component. Absorption features from aliphatic bonds are common in CA-IDPs (\S\ref{sec:two.organics}). The difference between laboratory studies of IDP absorption spectra and cometary IR spectra is that the 3.4~\micron \ features arising from dust in comae are expected to be in emission.  Predicting the emission spectra is at the forefront of thermal modeling developments and we note that optical constants are lacking for carbonaceous materials dominated by aliphatic bonds as opposed to rich in aromatic bonds.

To date, most organics that have measured optical constants are dominated by aromatic bonds (the 3.28~\micron \ PAH feature and features in the 5--9~\micron \ region) and for modeling comets we need organic residues that are rich in aliphatic bonds, if we use CA-IDPs as our guide. The temperatures of the organic-bearing particles that produce the 3.4~\micron \ emission feature and its relative abundance will determine the strength or contrast of the feature relative to the strong NIR `continuum' emission from  highly absorbing carbonaceous matter that is modeled by optical constants of amorphous carbon.

\subsubsection{Discrete materials or mixed material aggregates?}

There are compelling reasons to treat cometary particles as aggregates of mixed materials. CA-IDPs show that particles are mixtures, i.e., unequilibrated aggregates of amorphous silicates (GEMS), crystalline silicates, organics of varying compositions (\S~\ref{sec:two.organics}), and iron sulfides. Some CA-IDPs are dominated by carbonaceous matter \citep{Thomas1993} while others like the Giant IDPs are intimate mixtures of mini-chondrules of Fe-olivine with minor Mg-olivine, as well as sulfides and GEMS \citep{Brownlee2017}. 
Alternatively, there are some observations that strongly motivate our thinking that the carbonaceous materials and the siliceous materials may be discrete components in cometary comae or at least they vary in their relative abundance ratios depending on what part of the nucleus is active.  Some examples are: (a) comet C/2001 Q4 (NEAT) revealed a significant drop in the silicate feature contrast in about a hour, which is the jet crossing time for the observing aperture \citep{Wooden2004}, (b) comet C/2017 US$_{10}$ (Catalina) had an increase in the amorphous carbon-to-silicate ratio between two epochs separated by about 6 weeks of time, where the observing epoch refers the UT date and time and ephemerides of the comet \citep{Woodward2021}, (c \& d) the inner coma studies of \nineP \ post-\DI \ by {\it Gemini}+Michelle spectroscopy show rapidly changing compositions \citep{Harker2007} and by {\it Subaru}+COMICS narrow band imaging show Mg-olivine, then amorphous carbon, and then Mg-olivine in the few hours following the \DI \ event, (e) the outburst of comet 17P/Holmes showed crystalline-rich material that then changed composition well after outburst \citep{Reach2010}, and (f) multi-epoch \Spitzer +IRS spectra of a handful of comets show variable compositions for multi-epochs \citep{Harker2022} (see also Fig.~\ref{fig:Spitzer_3SEDs_JFCs}). Complimenting remote sensing results, the \emph{in situ} measurements of cometary particles indicate from \Halley \ revealed that particles were siliceous-only, carbonaceous-only, and mixed particles \citep{Schulze1997, Fomenkova1992a, Lawler1992} and that carbonaceous-only particles dominated within 9000~km of the nucleus \citep{Fomenkova1994}. 

When computing mixed material aggregates for thermal models to be fitted to cometary IR SEDs, there are a lot of combinations of materials that have to be computed in order to provide the potential suite of relative abundances of discrete materials that will need to be used to fit the IR SEDs. Employing silicate crystals with edges is particularly pertinent to fitting cometary spectra \citep{Lindsay2013} because notably spheres do not fit the observed spectral features and ellipsoids are adequate \citep{Harker2002, Harker2004, Min2005, Moreno2003}. The work to compute a suite of potential materials in mixed material aggregates has begun \citep{Wooden2021} but is as yet insufficient in breath to constrain relative abundances of materials for a sample of cometary IR SEDs. A particle size distribution of an ensemble of discrete materials is currently the state of the art for thermal models.

\subsubsection{Thermal emission models, Dust equations}

Computing the scattered light component and the thermal emission component (typically starting at 3~\micron \ for \rh $<$2.5~\au ) of the flux observed from a cometary coma requires a set of equations (Table~\ref{tab:dusteq}) and a set of suppositions: suppose a single DSD, a chosen porosity $P$ and an ensemble of compositions represent the dust in the coma and then the models are assessed against the observations using standard minimization ($\chi^2$-minimization) techniques.

The emergent flux is a sum over the particle size distribution of the thermally emitted fluxes per particle of varying compositions and/or particle structures. Per particle, the thermal flux is set by the particles' dust temperature (T$_d$) that results from radiative equilibrium between absorbed sunlight and emitted thermal radiation.  Both the dust temperature and dust flux depend upon the product of the particle's absorptivity (\qabsir) and its cross sectional area \Ga . The particle's wavelength-dependent absorptivity (\qabs) depends on composition, radius ($a$), shape and porosity.

The optical properties are called the absorptivity \qabsir \ and scattering efficiency 
\qsca , which are computed from the real and imaginary optical constants, $n$ and $k$, using various methods. Porous aggregates of mixed compositions, such as amorphous silicates, amorphous carbon and FeS, can be well modeled by `mixing' optical constants and vacuum such as when Mie Theory is combined with Effective Medium Theory (EMT) or Brugemann Mixing Theory (BM); or by layering such as in Distribution of Hollow Spheres (DHS) \citep{Min2005}; or  Rayleigh-Gans-Debye \citep{Bockelee-Morvan2017a, Bockelee-Morvan2017b}. However, Mg-olivine cannot be well modeled by mixing optical constants with vacuum; the predicted spectral features do not come close to observed spectra or laboratory spectra. This presents the challenge of computing the optical properties of mixed material porous aggregates with crystal monomers with edges and faces, and some progress has been made using the Discrete Dipole Approximation with the DDSCAT code \citep{Moreno2003,Wooden2021}. Scattering efficiencies (\qsca) have been computed using DDSCAT or T-Matrix because aggregates scatter differently than spheres \citep{Kimura2016, Kolokolova2023}.

More absorbing particles are warmer and smaller particles are warmer, and the co-dependencies between these dust properties is mitigated by the assessment of the relative fluxes and wavelengths of spectral  resonances or spectral `features'. Spectral features only arise from particles composed of dielectric materials that are smaller than about 1--3~\micron -radii for solid 
(0\% porosity, fractal dimension $D=3$) particles and for \ltsimeq 10--20~\micron -diameter moderately porous particles ($\sim$65--85\% porosity, $D\sim2.86--2.7$, \citet{Harker2002,Harker2004}). 
For particles of the same composition, the dust temperature depends more weakly on effective radius than on heliocentric distance from the Sun (\rh). When of the same composition, smaller particles are hotter simply because quantum mechanically smaller particles are less efficient emitters of photons at wavelengths larger than their cross sections (\Ga); historically, this effect has been called {\it Superheat} \citep{Gehrz1992}. Cometary SEDs do not reveal a single color temperature (Planck function fitted to wavelengths outside resonant features); the color temperature is warmer at shorter NIR wavelengths and cooler at FIR wavelengths. For a color temperature $T_{color}(r_h)$ fitted to $\sim$7.5~\micron \ and $\sim$13~\micron , a longstanding relationship versus heliocentric distance (\rh) is given in the Table~\ref{tab:dusteq}. 

For particles of the same effective radius and \rh , there is strong dependence of dust temperature on the dust composition through significant variations in the absorptivity \qabsvis \ at wavelengths where sunlight is absorbed. To achieve the highest temperatures observed for comae particles (e.g., \S~\ref{sec:two.dust.67Poutbursts}), small and highly absorbing particles are required or large extremely porous aggregates that are composed of mainly highly absorbing materials and whose large particle temperatures are equivalent to the temperatures of their submicron monomers. 

Crystalline materials in the DSD are often are limited to radii of less than 1~\micron \ to a few micron because predicted resonances of larger crystals do not fit observed spectral features \citep[e.g., see][]{Lindsay2013}. Relative mass fractions are quoted for the up-to 1~\micron \ portion of the DSD. The emission spectra of silicate particles or of amorphous carbon particles of increasing larger radii than these 3~\micron \ and 20~\micron \ radii, respectively for solid and for moderately porous particles, have increasingly broader as well as significantly weaker contrast resonances (with respect to wavelengths outside of their resonances). 
Thermal model parameters are co-dependent but are not degenerate when dielectric materials like silicates are present in comae because, in practice, varying the composition, DSD, and particle porosities produces thermal models that are distinguishable when fitted against the observed SEDs with good signal-to-noise ratios using minimization metrics ($\chi^2$-minimization). 

There are three albedos of interest for assessing comae particle properties: 
the particle albedo ($A$), the geometric albedo, and the bolometric albedo. 

The geometric albedo is, by definition, assessed at zero degrees phase angle ($\alpha=0^\circ$ in the observer's frame) \citep{Hanner1981,Bockelee-Morvan2017a} ($\alpha=0^\circ$ translates to $\theta=180^\circ$ or opposition, where $\theta$ is the angle between incident sunlight and scattered ray in the particle's frame). The geometric albedo at non-zero phase angles may be extrapolated from comae observations of  $A_p(\alpha)=A_p(\alpha=0^\circ)\times j(\alpha)$  using the ``phase curve'' $j(\alpha)$ and $A_p(\alpha)$ can be compared to $A_p(\alpha)$ computed for spheres (Mie) and for porous aggregate particles of varying composition and porosity  \citep{Hanner1981,Kimura2016,Kimura2006,Kimura2003}.

Alternatively, to calculate at the observed phase angle ($A_p(\alpha)$), one can use the thermal model fitting parameters that include the product \Ga \qabs \ to derive the dust effective area by assuming $Q_{abs}=1$:
\[\rm{dust~effective~area}=K\int_{a_{min}}^{a_{max}} G(a)~n(a)~da\]

\noindent Then the geometric albedo is the scattered light flux density divided by the dust effective area \Ga \ \citep{Hanner1981, Tokunaga1986} at the observer's phase angle~$\alpha$. \Ga \ can be calculated from either the Wein side of the thermal emission in the NIR or by thermal models fitted to a broader wavelength IR SED. This technique of calculating the geometric albedo, which was popular when the dust emission was studied with narrow band filter photometry, avoids having to measure or assume knowledge of the dusty coma phase curve $j(\alpha)$.  

Finally, the bolometric albedo \citep{Gehrz1992} enables an empirical assessment of the particle properties in the comae of many comets at different phase angles \citep{Woodward2015, Woodward2021}. The bolometric albedo $A(\alpha)_{bolo}$ is an approximate measure of the scattered to total incident energy (sunlight) \citep{Woodward2015}, where the total incident energy is assessed by the sum of the thermal (re-emitted) and scattered energies ($\lambda~F_\lambda$), each measured at the wavelengths of their maximum energy output (Table~\ref{tab:dusteq}). The opportunity to tie the scattered light to the thermal emission potentially offers additional insights into dust properties and compositions for two reasons: the scattered light may be contributed to by higher albedo (and potentially cooler) dust components such as ice grains or organics, and particle structure affects scattering and thermal in distinct ways \citep{Tokunaga1986} (\S~\ref{sec:two.dust.67Poutbursts}, ~\ref{sec:two.dust.distributedsources}).

\begin{table*}[t!]
\begin{tabular}{ | l | p{3.5cm} | p{2.0cm} | }
\hline
Dust Equations & Designation & Notes \\
\hline

$P= 1 - f$, ~given fractional filled volume $f = (a/a_0)^{D-3}$ & Porosity, fractal dimension $D$ for 1.7\ltsimeq $D \le 3 $ & \citep{Harker2002, Woodward2021, Lasue2019c} \\

$Q_{abs}$, $Q_{sca}$ & Absorptivity, Scattering efficiency & \\

$C_{abs}=G(a)~Q_{abs}$, $C_{sca}=G(a)~Q_{sca}$ & Absorption, Scattering Cross Sections & \\

$Q_{ext} =(Q_{abs}+Q_{sca})$ & Extinction efficiency & \\

$F_{emiss}(\lambda) = \frac{1}{1}K\int_{a_{min}}^{a_{max}}G(a) Q_{\lambda, abs}(a)~\pi B_{\lambda}(T_{d}(a))~n(a)~da$ & thermal flux density &  $G(a)=\pi a^2$ for sphere \\

$\pi B_\lambda(T_d(a)={2 h c^2}{\lambda^{-5}}\left ( exp^{h c/\lambda k T_d} -1   \right )^{-1}$ & Planck Function &    \\

$n(a)=\left (1-\frac{a_0}{a}\right )^{M}\left(\frac{a}{a_0}\right )^{N}$ & differential grain size distribution (DSD, GSD) & Hanner GSD:  $a_p=\frac{(M+N)}{N}$ \\

$K=\frac{N_{dust}(a_0)}{4\pi (\Delta ~[cm])^2}$ ~or~ $K=\frac{N_{dust}(a_p)}{4\pi (\Delta ~[cm])^2}$ for HGSD 
& flux scaler, at $a_0$ (DSD) or at $a_p$ (HGSD, $M\ne 0$) &   \\

$F_{sca}(\lambda,\alpha) = \frac{F_\lambda(T_\odot)}{(r_h~[au])^2}K\int_{a_{min}}^{a_{max}}G(a) Q_{\lambda, sca}(a)~p_{\lambda,sca}(a,\alpha)~n(a)~da$ & scattered & $F_\lambda(T_\odot)$ solar flux at 1~\au \\

$\theta({\rm deg})$ & angle from incoming to outgoing ray  & \\
$\alpha=180^\circ - \theta$ & phase angle betw.\ incident sunlight and observer & \\

$\int_{4\pi}{ p_{\lambda,sca}}~{d\Omega}=1=\frac{1}{C_{\lambda,sca}(a)}\int_{4\pi} \frac{d (C_{\lambda,sca}(a;\theta,\phi))}{d\Omega} d\Omega$ & ``phase function'' $p_{\lambda,sca}$, normalized differential scattering cross section &  $C=G(a)Q$ \\

$j_\lambda (a,\alpha^\prime)=\frac{p_\lambda (a,\theta=180^\circ - \alpha^\prime)}{p_\lambda(\theta=180^\circ)}=\frac{p_\lambda (a,180^\circ - \alpha^\prime)}{p_\lambda(\alpha^\prime=0^\circ)}$ & ``phase curve'', phase fn normalized at backscattering angle $\theta$=180$^\circ$ & \citep{Hanner1981} $\alpha^\prime$=0$\equiv$$\theta$=180$^\circ$  \\ 


$A$=$Q_{sca}/Q_{ext}$ & Albedo of particle & ref \\

$A_p(\alpha^\prime)=A_p(\alpha=0^\circ)\frac{p_\lambda (a,180^\circ - \alpha^\prime)}{p_\lambda (a,180^\circ)}\equiv A_p(0)j(\alpha^\prime)$ & geometric albedo, ratio of energy {\it backscattered} to that of Lambertian surface of equal area & phase angle $\alpha^\prime \equiv 180^\circ - \theta$ \\

$A_\lambda (\theta)_{bolo}= \frac{f(\theta)}{(1+f(\theta))}$, ~$f(\theta)= \frac{[\lambda F_{sca}(\lambda,\alpha)]~|_{\lambda=\lambda _{max,sca}}}{[\lambda F_{abs}(\lambda)]~|_{\lambda=\lambda _{max,abs}}}$
& bolometric albedo, ratio of scattered to sum of scat. \& thermal energy at $\lambda_{max,sca}$ \& $\lambda_{max,emiss}$
& \citep{Woodward2015} \\

$\int_{0}^{\inf} F_{\lambda_{VIS},abs}(Q_{\lambda_{VIS},abs})~d\lambda  =\int_{0}^{\inf} F_{\lambda_{IR},emis}(Q_{\lambda_{IR},abs},T_d)~d\lambda $ & Radiative Equilibrium (absorption = emission)  & rad~eq~temp $T_d$ \\

$\int_{\lambda_{min}}^{\lambda_{max}}F_{\lambda_{VIS},abs}(Q_{\lambda_{VIS},abs})~d\lambda = \int_{\lambda_{min}}^{\lambda_{max}} \frac{F_\lambda(T_\odot)}{r_h^2}~G(a) Q_{\lambda,abs}(a) d\lambda$ & Sunlight absorbed (L) &   \\

$\int_{\lambda_{min}}^{\lambda_{max}}F_{\lambda_{IR},emis}(Q_{\lambda_{IR},abs},T_d)~d\lambda = \int_{\lambda_{min}}^{\lambda_{max}} \pi B_\lambda(T_d)~G(a) Q_{\lambda,abs}( a) d\lambda $ & Thermal emitted (R) &   \\ 

$g_{col}(\lambda) \equiv \left(\frac{\lambda}{\lambda_{ref}} \right)^{-p_{col}} \propto \int G(a)~ Q_{\lambda,sca}(a)~p_{\lambda,sca}(a,\alpha)~n(a)~da$ & scat. light color $g_{col}$ & power $p_{col}$ \\

$S_{color}^\prime = \frac{2}{(\lambda_{ref}-\lambda)[nm]}~ \frac{g_{col}(\lambda) - g_{col}(\lambda_{ref})}{g_{col}(\lambda)+g_{col}(\lambda_{ref})}$ & color gradient (slope) [\%/100~\nm] &   \\

$T_{d,color} \approx {1.1\times 278~[K]}{(r_h/[au])^{-0.5}} $ & dust color temperature, typical behavior & \citep{Hanner1997} \\

\hline
\end{tabular}
\caption{\it \small 
Table of equations including dust thermally emitted and scattered fluxes, and parameters that quantify observed dust properties.} 
\label{tab:dusteq}
\end{table*}

\subsubsection{Dust composition and DSD from combined scattered \& NIR thermal: 67P/C-G outbursts}
\label{sec:two.dust.67Poutbursts}

The \Rosetta +{VIRTIS-H} spectra of the quiescent coma and of two short duration outbursts (2015-09-13T13.645 and 2015-09-14T18.828~UT) from comet \CG \ as modeled and presented by \citet{Bockelee-Morvan2017a, Bockelee-Morvan2017b} provide an excellent demonstration of how models fitted to dust scattering together with dust thermal emission can predict $S_{color}^\prime$, $T_{d,color}$ and  $A(\theta)_{bolo}$, and thereby provide constraints on dust compositions and DSD parameters. 

The scattered light color ($S^\prime$) (whether the color is 'blue', `neutral' or `red' relative to reflected sunlight) provides information about the composition of the dust particles  \citep{Storrs1992, Zubko2015b, Hyland2019, Kulyk2021, Li2014}. The dust color temperature, derived in this case from fitting a scaled Planck function to the 2--5~\micron \ dust continuum measurements, represents either the DSD-weighted temperatures of the warmest particles, i.e., the smallest and/or most highly absorbing ones, or the DSD-weighted temperatures of aggregate mixed particles, or a combination thereof. 
The bolometric albedo is an approximate measure of the scattered to total incident energy (sunlight) \citep{Woodward2015} and is assessed at the observers phase angle $\alpha$ = 180$^\circ$--$ \theta$ = 108$^\circ$,~99$^\circ$ for the two dates. 
The geometric albedo $A_p(\alpha)$ requires the dust effective area be calculated from dust thermal models fitted to the IR \citep{Hanner1981,Hanner1985,Tokunaga1986}.  
When the spectra provide the three metrics  of dust scattered light color slope $S^\prime$, NIR color temperature $T_{d,color}$ from the thermal emission, and either the bolometric albedo  $A_\lambda(\theta)_{bolo}$ or the geometric albedo $A_p$, then these four dust model parameters can be constrained, via $\chi^2$-minimization against the metrics derived from the spectra: two DSD parameters (smallest  particle radius $a_0$ and DSD slope $N$), the particle porosity $P$ that is parameterized by a fractal dimension $D$, and the dust composition quantified by $q_{frac}$ where $q_{frac}^3$ is the volume fraction of inclusions of lesser opaque material within a matrix of more highly opaque material.  

First, we summarize the properties of the quiescent coma of \CG \ on the two dates prior to the coma outbursts (Fig.~1 in \cite{Bockelee-Morvan2017a}).
The quiescent coma presents a NIR `red' color slope of $S_{color}^\prime$=2.6$\pm0.3$\%/100~\nm , 2.3$\pm0.4$\%/100~\nm \ for $\lambda_{ref},\lambda=$2.0~\micron, 2.5~\micron \ on the two respective dates. Compared to the NIR color slope, the visible wavelength scattered light spectra have a steeper color slope 15--18$\pm 3$\%/100~\nm \ with $\lambda_{ref},\lambda$=0.45--0.8~\micron \ \citep{Rinaldi2017}, which is typical of cometary comae. 
The quiescent coma has an estimated bolometric albedo of $A_{\lambda}(\theta)_{bolo}=0.13\pm 0.2$, and a dust color temperature of $T_{d,color}\approx 300~K$. 

A set of models that fit the data include porous spheres of 2-compositions of Mg:Fe amorphous olivine in a matrix of amorphous  carbon ($q_{frac}$=0.7) and with porosity limited to be $P\leq 0.5$ \citep[MIDAS `compact particles']{Mannel2016} and DSD slope of $N\leq 3$. The composition ($q_{frac}$=0.7) translates to 66~vol\% amorphous carbon and 34~vol\% amorphous  olivine, with ranges of coupled model parameters that include ($a_{0}$, $N$,$q_{frac}$, $D$, $P$) = (0.3\micron , 2.5, 0.7, 2.5, $\leq$0.5) and = (0.9\micron , 3.0, 0.7, 2.5, $\leq$0.5). The model limits $P\leq0.5$, so this limits particles bigger than 0.4~\micron \ to have the porosity that is independent of radii $a$ and this porosity is lower that the porosity equation (Table~\ref{tab:dusteq}); if $D$ was applied over the full DSD then  for $D=2.5$ the porosity values would be, e.g., $P$(0.4~\micron)=0.5, $P$(1~\micron)=0.68, $P$(3~\micron)=0.82. 

Alternatively, for a similar porosity prescription and DSD parameters, the data also can be fitted using compositions of pure carbon, carbon and a lower  $q_{frac}$ of more transparent Mg-rich amorphous pyroxene or Mg-olivine (forsterite). However, for pure silicate grains or silicates mixed with 6~vol\% FeS (consistent with \citet{Fulle2016a}), DSD parameters sufficient to produce the measured $T_{d,color}$ yield NIR neutral colors (as opposed to the observed NIR red colors) and bolometric albedos higher than measured.  

Lastly, the quiescent coma also is modeled successfully with 25\% by number of extremely porous aggregate particles ($D$=1.7), which provide less than 1.5 per cent of the particle albedo at 2~\micron \ as well as provide thermal emission  \citep{Bockelee-Morvan2017a}. Extremely porous aggregate particles, often called Ballistic Cluster Cluster Aggregates (BCCA), have particle temperatures and spectral resonances similar to their (small and hotter) monomers, i.e., independent of their particle size  \citep{Bockelee-Morvan2017a, Tazaki2016, Kimura2016, Kolokolova2007}, and the relative contributions of BCCA to the scattered light albedo is extremely minimal \citep{Kimura2006}. 
Models for the quiescent comae, which have similar dust compositions as cited for lower porosity particles ($D$=2.5, $P\leq0.4$), with 25\% BCCA and with steeper size distributions $N\geq3$ \citep{Bockelee-Morvan2017b}, are more commensurate with studies of \CG's dust by GIADA \citep{Lasue2019c, Fulle2015b}, by GIADA and OSIRIS \citet{Fulle2016a}, by MIDAS \citep{Mannel2016, Bentley2016}, and by particle topologies \citep{Langevin2016}. 

In contrast to the quiescent coma, the outbursts are characterized by sudden increase in dust thermal emission that peaks in a few minutes and then decays towards nominal comae levels in about 30~minutes, with the outbursting comae having hotter particles $T_{d,color}=550~K, 640~K$, bluer NIR scattered light colors (extreme values of  $S_{color}^\prime=–$10\%/100~\nm \ with modeled color slope power of $p_{col}=2.30$), and higher bolometric albedos $A_\lambda (\theta)_{bolo}=0.6$, where all three properties change with time as the outbursts evolve. In the first outburst, all three metrics trend together and follow the light curve. In contrast, during the second outburst $A(\theta)_{bolo}$ remains high (and increasing) as the light curve and $T_{d,color}$ and $S_{color}^\prime$ decay towards quiescent coma values. During outbursts, the visible scattered light colors remain in ranges of 10–15\%/100~\nm \ and 6–12~\%/100~\nm , for the respective two dates \citep{Rinaldi2018}. The particles are moving relatively fast compared to dust nominal speeds for \CG, $v$($a<$10~\micron)$\geq$ 30~$m~s^{-1}$, so subsequent \emph{VIRTIS-H} spectra sampled different sets of particles. 

Particles ejected in the outburst, as modeled, are higher in carbon (smaller $q_{frac}$), and have smaller minimum particle radii as well as having a steep DSD slope ($a_0$=0.1~\micron , 4$\leq N \leq$ 5), which means submicron particles dominate. Two of three metrics  $S_{color}^\prime$ and  $T_{d,color}$ are fitted but the observed high $A(\theta)_{bolo}$ of $\sim$0.6 is not fitted by submicron carbonaceous (dark, i.e., low particle albedo, and hot) particles. Somewhat larger moderately porous particles ($a_{0}\sim 0.5$\micron ) are in the second outburst but the highest values of the bolometric albedo also are not explained. Pure and submicron Mg:Fe olivine grains can account for the bolometric albedo but the modeled compact ($P$=0) olivine spheres have a color temperature of $<$530~$K$, which is too low to match the observed color temperatures ($>$600~$K$). If 6~vol\% FeS is mixed with Mg:Fe amorphous olivine then the color temperatures increase to 575~$K$ but the $S_{color}^\prime$ cannot be fitted. BCCA may explain the high color temperatures but certainly not the high albedos during outburst. A water ice band cannot be more than $\sim$10\% in depth so ice particles are not likely driving the high bolometric albedo. Another possibility is that there could be two distinct compositions of submicron grains in the outbursts, one that is cold and bright complimented by dark and hot. \citet{Bockelee-Morvan2017a} also propound that some limited  lifetime organics may be mixed with the submicron Mg:Fe amorphous olivine such that the rapid degeneration of organics may contribute to the enhanced temperatures and enhanced albedos observed during outburst, especially at the onset of the first outburst and the sustained albedo during the second outburst. In summary, a dramatic increase in the numbers of dark (carbonaceous) and smaller particles characterize the outbursts because the ``high color temperatures and blue colors imply the presence of Rayleigh-type scatterers in the ejecta, i.e. either very small grains or BCCA type agglomerates'' \citep{Bockelee-Morvan2017a}. However, the higher $A(\theta)_{bolo}$ in the outbursts are not explained.

\subsubsection{Scattered light, limited lifetime distributed sources}
\label{sec:two.dust.distributedsources}
Scattered light observations at UV through NIR wavelengths provides constraints on the structure of dust particles.  Polarization \citep{Kolokolova2023},

the scattered light color, and the surface brightness spatial distribution can provide insights into dust composition for those dust components that contribute to scattered light. Here are some examples that do not involve polarization measurements. 

Limited-lifetime organic species were discovered following outburst of \WMone \ and the \DI \ impact of \nineP \ because they produce wavelength-dependent scattered light `colors' (slope in \%/nm) akin to a combination of organics created in the laboratory by irradiating ice mixtures \citep{Jenniskens1993}. (Limited lifetime species associated with the dust, or `extended sources' or `distributed sources' also are known to occur in some comae by the molecular production rates having more extended spatial distributions than the water vapor and these molecules may include CO, formaldehyde (H$_2$CO), NH$_3$, and recently discovered C$_2$H$_2$ \citep{Dello-Russo2022}.)

Another metric is the total scattering cross section (SA), related to the number and intrinsic color of the dust particles:
\[ \rm{SA}=\int\left(\Sigma~ A_\lambda(\alpha) ~f\right) d\rho\]
\citet{Tozzi2004} and \citet{Tozzi2015} used this approach to assess limited lifetime dust that contributed the changes in the scattered light for comets \WMone\ and \nineP. The significant difference between the two comets is that \WMone \ has a sublimating component that scatters in the visible and \nineP \ does not, so nature of the sublimating grains are different between the two comets. 

For \WMone \ at \rh=1.2~\au , scale-lengths for column densities of limited lifetime organics are assessed for two components with 12250$\pm$1625\km \ and 940$\pm$150\km,  and adopting a dust velocity $v_{dust}\approx0.2~km~s^{-1}$ implies 1.7~hr and 17~hr coma lifetimes, respectively \citep{Tozzi2004}. The scattered light colors were similar to minimally irradiated and long-exposure irradiated organic residues that may be similar to the UV-irradiated residues created on NASA's orbiting Skylab (EURECA \citep{Li1997}. An impulsive event offers a better opportunity to assess the changes in surface brightness distribution and color, that then can be used to determine lifetimes in the coma. 

Prior to the \DI \ event, \nineP \ has a non-sublimating component and sublimating component (with scale length of 6300~\km, assuming $v_{dust}\approx0.2~km~s^{-1}$ implies 11~hr at \rh=1.5~\au) and these two sublimating components differ in their NIR colors. The NIR colors are assessed by differencing their brightness measurements between pairs of the 2MASS photometric bands of H, J, and Ks (defined to exclude telluric absorptions and centered at 1.25~\micron{}, 1.65~\micron{}, and 2.17~\micron{} \citep{Bessell2005}).
Details about post-\DI \ observations include that the color was neutral from $J$ to $H$ band but increases by 25\% from $H$ to $K_s$. Three hours later, the scattering efficiency increases by 84\% between $J$ and $K_s$ so the reddening increased but the total brightness declined. Also, there was no correlation in the quiescent coma versus \DI \ ejecta cloud.

One may wish to consider the ``quiescent'' coma of comet \nineP \ with the post\DI \ coma. In the hours following the \DI \ encounter, polarization images uniquely reveal an expanding front of polarizing particles that are not detected in non-polarized images so these particles are submicron-size (from their ejected velocities) and 'dark' carbonaceous grains \citep{Kadono2007}. Spatial imaging via IR photometry as well as scattered light colors and structures in the coma reveal properties about the dust but with more degeneracies in the derived dust composition than with IR spectroscopy thermal modeling because of the IR spectral resonances directly probe the composition of the dust particles contributing to the thermal emission for those compositions that have resonances.

\section{\textbf{PHYSICAL PROPERTIES}}
\label{sec:phys}

\subsection{Sizes and size distribution}
\label{sec:1sizedist}

Cometary dust is ejected from the active cometary nuclei and expand in space following initially a quasi-spherical shell within about 10,000~km from the nucleus, named the coma, and further creating an expanding tail in the direction opposite to the Sun as illustrated in Fig.~\ref{fig:comet}
\citep[see e.g.][]{finson1968}. The striae visible in the solar direction of the impressive 
comet C/2006 P1 McNaught (Fig.~\ref{fig:comet}), named synchrones, represent dust  ejected 
at different times along the orbit of the comet, and that disperse in space due to the effect of 
the $\beta$ parameter representing the ratio of the forces of radiation and gravity acting on the dust
particles. 

\begin{equation}
\beta = \frac{3 L_{\circ}}{16 \pi c G M_{\circ}} \frac{Q_{pr}}{\rho s}
\end{equation}
where $L_{\circ}$ and $M_{\circ}$ are the luminosity and the mass of the Sun, $c$ is the speed of light, 
$G$ the gravitational constant and $Q_{pr}$ the radiation pressure efficiency of the dust grain 
having bulk density $\rho$ and an effective radius $s$. $\beta$ therefore depends on the dust grain's  
composition, shape, structure and size but is generally proportional to $\frac{1}{\rho s}$, the small 
grains being easily pushed away by the radiation pressure, while the largest grains, typically 1 micrometer
in size and above,  will tend to follow the orbit of the cometary nucleus around the Sun \citep{burns1979a}. 

The extent of cometary trails is evidence that cometary dust grains present a large range of sizes. 
Cometary dust size distributions are traditionally estimated by inverting the cometary tails \citep[see e.g.][]{finson1968}. 
While cometary dust size distributions are typically inverted bin size by bin size, they are canonically 
represented by a simpler power-law distribution as it represents well the  properties of the observed clouds of dust. 
This is represented as a {\it Differential Size Distribution} (DSD) with power index, $N$ (also called $\alpha$). $N$ is related to $\gamma$, the power index of the mass distribution of particles by $N = -3 \gamma -1$, assuming a constant density for the particles.
If $N > -3$, both the mass and brightness depend on the largest ejected grains. 
Brightness and mass become decoupled if $-4 < N < -3$ in which case the dust mass depends 
on the largest ejected grains, while the brightness depends on the micrometer-sized grains \citep{fulle2004a}.
Typical values for the mean power index, $N$, range from about -3 to about -4 
\cite[See table 1 in][for a review of ground-based derived DSD]{fulle2004a}.
But one has to recognize that the actual dust size distributions in comets are more complex 
and can be quite variable with time and space due to outburst and activity \citep{fulle1987}.

With the advent of space missions to active cometary nuclei, the ground-based observations and models 
of cometary dust size distributions of the coma are now complemented by direct measurements close to the nucleus, 
or laboratory measurements from returned samples (see Table~\ref{tab:DSD}). However, those measurements cannot directly be compared as their relationship is complicated by fragmentation of particles, sublimation of volatiles, differential speed of
ejection, and surface inhomogeneous activity \citep[see ][and references therein]{agarwal2007}.

In situ data on the dust mass distribution was obtained for 1P/Halley by the {\it Vega 1}, {\it Vega 2} and {\it Giotto} missionsin 1986 \citep{divine1988}. \citet{mcdonnell1987} used the dust impact detection system (DIDSY) on-board {\it Giotto} to derive  a double size distribution at the nucleus of 1P/Halley with 
$N = -4.06$ for small particles ($m<10^{-8}$~kg) and $N = -3.13$ for larger particles ($m>10^{-8}$~kg).
The average DSD prior to close approach was estimated at the nucleus to be $N = -3.49\pm0.15$ \citep{mcdonnell1986}.
\citet{fulle1995} developed a model of 1P/Halley dust emissions to fit the DIDSY fluences and 
obtained a constant DSD index of $N = -3.5 \pm 0.2$ for grains larger than $20~\mu m$.  
Later combined models of optical and impact measurements resulted in a value of $N=-2.6\pm 0.2$ \citep{fulle2000a}.
The interpretation of Halley data did not lead to a general agreement regarding the dust size distribution
at the nucleus of the comet, however it is generally admitted that the coma is dominated by 
millimetre-sized and larger particles \citep{agarwal2007}.
The {\it Giotto} space mission continued its exploration of comets with a close fly-by of comet 26P/Grigg–Skjellerup in 1992. 
The dust distribution detected in that case corresponds to $N = -1.81$ for particles with mass larger than  $m>10^{-9}$~kg
indicating for this comet a coma dominated by large particles \citep{mcdonnell1993a}.

In 2005, the Deep Impact mission collided with comet 9P/Tempel 1, excavating a crater about 150~m large and 
several 10s of meters deep, which helped decipher the composition and low strength of 
the near subsurface layers of cometary nuclei. The impact generated a strong ejection of fresh ice and 
dust particles that was akin to a cometary activity outburst \citep{ahearn2005}.
Similarly, the DSD index measured in the coma after the impact ($N = -4.5 \pm0.2$ \citep{Lisse2006}) demonstrates a strong increase in the DSD slope as compared to the pre-impact coma value ($N = -3.0 \pm0.45$ \citep{Lisse2005a}) corresponding to a sharp increase in the number of small particles present in the coma liberated by the impact. Evidence from the changes in activity, in gas composition, and in dust particles DSD were used to argue that pristine cometary material was present a few 10s of meters below the surface of cometary nuclei \citep{ahearn2005}. 
The Deep Impact mission was then diverted towards comet 103P/Hartley 2. This very active small comet nucleus ejects very large particles, some of which are made of pure water ice \citep{ahearn2011}. Observations from the ground indicated relatively steep DSD indices, from $-3.2 \pm 0.1$ \citep{epifani2001} to $-3.91 \pm 0.3$ \citep{bauer2011}. The in situ dust size distribution obtained by the space probe was even steeper with a value ranging from $-6.6$ to $-4.7$, but applicable to the largest dust particles sizes detected by photometry, from 1~cm to 20~cm in the case of icy particles \citep{kelley2013}.

In 2006, the \Stardust \ mission delivered samples from comet 81P/Wild 2 to Earth for laboratory analysis \citep{Brownlee2006}. During the comet fly-by, the Dust Flux Monitor Instrument (DFMI) monitored the dust impacts with a high temporal resolution and gave evidence of fragmentation of dust particles within the coma.  
The size distributions detected by the probe are quite variable with a best fit value of $-3.25$ for particle masses lower than $10^{-8}$~kg at around closest approach, but with values as high as $-1.99$ and as low as $-4.39$ depending on the coma region probed \citep{tuzzolino2004, green2004}. 
However, additional information could be retrieved from the laboratory analyses of the tracks in the aerogel and the craters on the aluminum foils of the return capsule, using calibrations made on Earth by impacting with analog compact particles. The resulting DSD index corresponds to $-2.71$  \citep{Horz2006}. Further recalibration on the ground taking into account impacts by aggregates of silica particles slightly revised this value to a lower one of $N  = -2.89$ for particles larger than 10 microns,  more compatible with the average value given by DFMI \citep{price2010}. 

Finally, the {\it Rosetta} space mission was the only cometary space probe that could survey a comet nucleus over a major part of its orbital trajectory \citep{glassmeier2007a} for a period spanning 2 years and a half.
The ground-based size distribution  of comet 67P/Churyumov-Gerasimenko was determined to be $N = -3.4 \pm 0.2$ from an average of previous observations since its discovery \citep[][and references therein]{fulle2004a}. 
During the {\it Rosetta} mission survey, the dust size distribution showed a strong time-evolution, 
with the optical cross-section dominated by the largest ejected dust far from perihelion with $N \approx -3$, while the smallest ejected dust dominate around perihelion with $N \approx -3.9$  \citep{moreno2017}. 
	
\begin{figure*}[t!]
\centering
\includegraphics[width=0.9\textwidth]{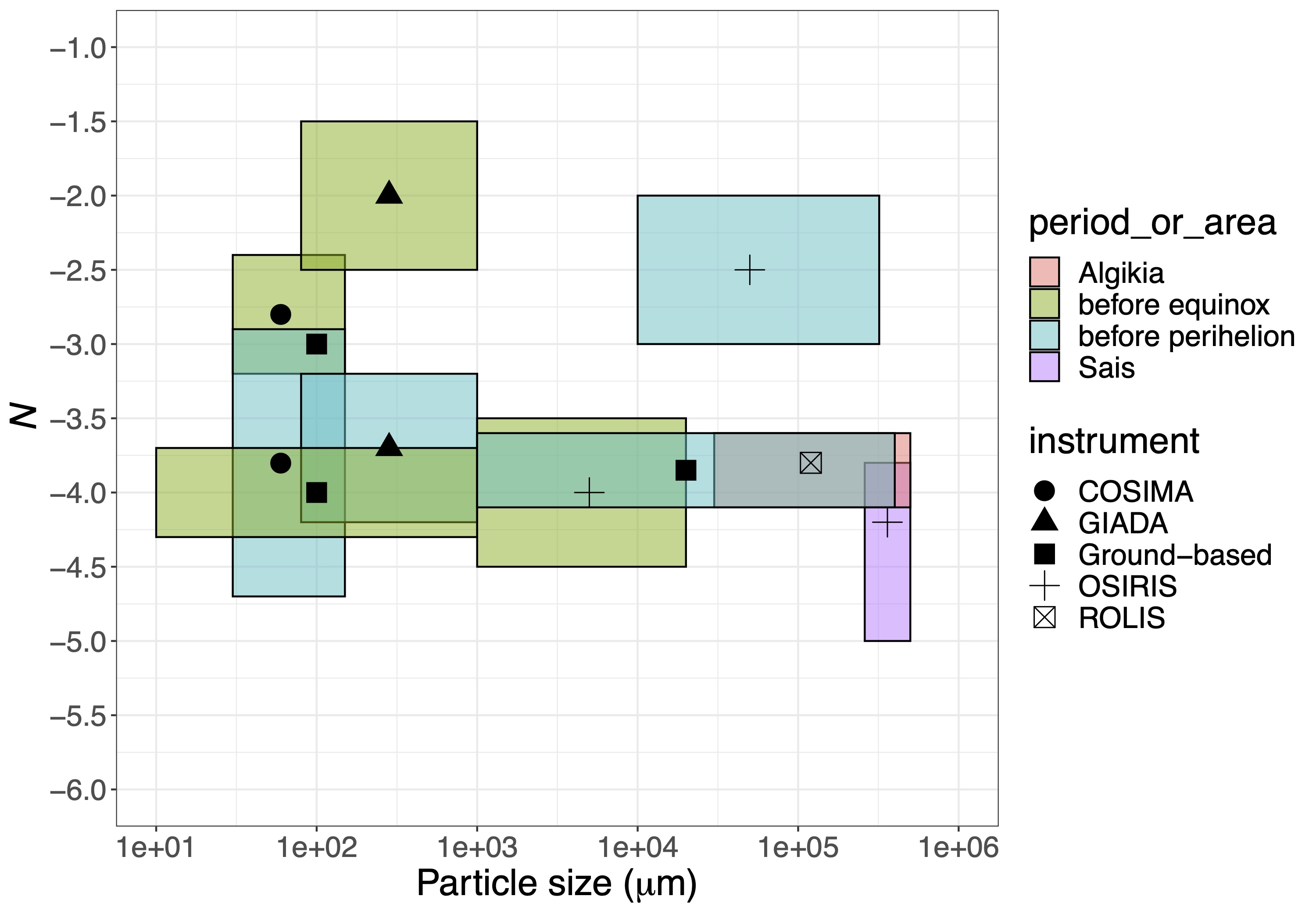}
\caption{\it \small 
Power index of the differential size distribution of 67P/C-G. Blue rectangles: power index before the 2015 equinox. Green rectangles: power index around the 2015 perihelion.
The ranges along the x axis show the instrument size range sensitivity, the ranges along the y axis are given by the uncertainty of the power index. (Data taken from  \citet{merouane2017,Rotundi2015,Fulle2016b,ott2017,pajola2017,moreno2017}. Figure adapted from \citet{Levasseur-Regourd2018b}). Agilkia was the initial landing site of Philae, and Sais was the final landing site  of the Rosetta spacecraft.}
\label{fig:67Psd}
\end{figure*}

The {\it Rosetta} space mission also used its many dust analysis instruments (GIADA, COSIMA, Osiris, ROLIS) to determine the dust size distribution in many complementary size ranges from $1~\mu$m to 1~m. 
Fig.~\ref{fig:67Psd} summarizes the results obtained for all these different measurements before the 2015 equinox, around the perihelion time in August 2015 and the size distribution of boulders on the surface of 2 landing areas studied for Philae. 
Overall the DSD index  measured by {\it Rosetta} is consistent with an average value around $-4$, with variations due to the timings of measurements with smaller particles typically ejected around perihelion time.

\begin{table*}[t]
\begin{tabular}{| p{4.5cm} | p{3cm} | l | p{5cm} |}
  \hline
 \bf{Comet} & \bf{Instrument} & \bf{DSD index $N$} & \bf{Reference} \\
  \hline
  1P/Halley & DIDSY & $-3.49 \pm 0.15$ (avg) & \citet{mcdonnell1986}\\
  1P/Halley & DIDSY & $-4.06 (<10^{-8}~kg)$ & \citet{mcdonnell1987}\\
  1P/Halley & DIDSY & $-3.13 (>10^{-8}~kg)$ & \citet{mcdonnell1987}\\
  1P/Halley & DIDSY & $-3.5 \pm 0.2 (>20~\mu m)$ & \citet{fulle1995}\\
  1P/Halley & OPE + DID & $-2.6 \pm 0.2 (>10^{-12}~kg)$ & \citet{fulle2000a}\\
  9P/Tempel 1  & IRAS & $-3.0 \pm 0.45$ (pre-impact) & \citet{Lisse2005a}\\
  9P/Tempel 1  & Spitzer & $-4.5 \pm 0.2$ (post-impact) & \citet{Lisse2006}\\
  26P/Grigg-Skjellerup & Ground-based & $-4.0 < N  < -3.0$ & \citet[][and references therein]{fulle2004a}\\
  26P/Grigg-Skjellerup & DIDSY & $-1.81_{-0.6}^{+0.39} $ & \citet{mcdonnell1993a}\\
  67P/Churyumov-Gerasimenko & Ground-based & $-3.4 \pm 0.2$ (avg) & \citet[][and references therein]{fulle2004a}\\
  67P/Churyumov-Gerasimenko (far from perihelion)& Ground-based (tail)& $-3$ & \citet{moreno2017}\\
  67P/Churyumov-Gerasimenko (around perihelion)& Ground-based (tail)& $-3.7 < N < -4.3$ & \citet{moreno2017}\\
  67P/Churyumov-Gerasimenko & Ground-based (trail) & $-3.6 < N < -4.1$ & \cite{moreno2017}\\
  81P/Wild 2 & DFMI & $ -3.25 $ (closest approach) & \citet{tuzzolino2004, green2004}\\
  81P/Wild 2 & DFMI & $ -1.99 < N < -4.39$ & \citet{tuzzolino2004, green2004}\\
  81P/Wild 2 & laboratory analysis & $-2.72; -2.89$ & \citet{Horz2006, price2010}\\
  103P/Hartley 2 & ISOCAM & $-3.2\pm0.1$ & \citet{epifani2001}\\  
  103P/Hartley 2 & WISE/NEOWISE & $-3.91\pm0.3$ & \citet{bauer2011}\\  
  103P/Hartley 2 & Deep Impact photometry & $ -6.6 < N  < -4.7$ & \citet{kelley2013}\\  
  2I/Borisov & Ground-based & $-3.7\pm1.8$ & \citet{guzik2020}\\  
  2I/Borisov & Ground-based & $-4.0\pm 0.3$ & \citet{cremonese2020}\\  
  \hline
\end{tabular}
\caption{\it \small 
Summary table of cometary dust size distribution indices $N$ retrieved by space missions and ground-based observations. 
A full table of $N$ values determined from ground-based observations (range and averages) and models is available in \citet{fulle2004a}.}
\label{tab:DSD}
\end{table*}

In addition, over the last five years, the Solar System was visited by 2 interstellar objects (1I/Oumuamua and 2I/Borisov). Fortunately, interstellar comet 2I/Borisov was active enough that its coma and tail could be studied by telescopic observations. In that case, the DSD is also ranging from $-3.7$ to $-4$ (see Table~\ref{tab:DSD})
which is consistent with most active comets and also 'fresh' comets such as C/1995 O1 (Hale-Bopp) or C/1996 B2 (Hyakutake)
\citep[][and references therein]{fulle2004a}

In summary, cometary dust size distributions usually correspond well to power laws, with typical indices ranging from $-3$ to $-4$. However large variations are detected depending on the  activity of the nucleus, the timing of measurements and the techniques used. The consistency of all measurements made so far are certainly consistent with probing the DSD of primordial building blocks of comets as the DSD index for active comets is consistent over all types of comets, including 'fresh' ones and even interstellar comet 2I/Borisov. 

Further analysis of cometary dust ejected at different times will certainly improve our knowledge 
of the dust size properties and how they may relate to primordial solar nebula materials.

\subsection{Optical and thermal properties}
\label{sec:2optical}

Light scattering has historically been the main provider of information on the physical properties of cometary dust 
through telescopic studies \citep[for more details see][]{kolokolova2015, Kolokolova2023}
The observations made in several domains have been useful in constraining the size, size distribution, 
structure and optical indices  of the dust particles. 
The observations in the visible domain have been used to model the extension of the coma and 
tail of comets \citep{haser2020,finson1968,finson1968a} and deduce from it the surface activity 
of the nucleus. 
In the case of the dust particles in the coma of comet 67P/C-G, they present a specific scattering 
phase function with a u-shape and minimum at intermediate phase angles. This is different from the
phase function that was usually considered for cometary dust \citep{kolokolova2004}.
The color is consistent with the average of the nucleus surface below $30^{\circ}$ of phase angles.
There is negligible phase reddening at phase angles $<90^{\circ}$ indicating a coma dominated 
by single scattering \citep{bertini2017c}. Such a phase curve shape may be consistent with crushed
primitive meteorites, but even more with analogues developed to simulated the scattering properties 
of interplanetary dust particles \citep{levasseur-regourd2019b}.
Photometric studies of single grains with the OSIRIS 
camera filters from 535 to 882~nm indicate slopes covering the ranges of slopes detected 
over the reddest to bluest regions of the nucleus \citep{frattin2017a}.
Assuming that the majority of dust particles in the 
zodiacal cloud come from comets, their average geometric albedo towards the Gegenschein has been determined  
to be $0.06\pm0.01$, similar to the low albedo detected for cometary nuclei \citep{ishiguro2013a}.

The coma and tails of comets are astronomical objects that present some of the largest polarization 
detected in the solar system \citep[for more details see][]{kolokolova2015, Kolokolova2023}. 
Polarimetric observations of comets give complementary information 
to the scattering properties of the dust particles, in particular on their optical indices and morphologies. 
Initial polarimetric observations of comet Halley combined with Mie light scattering simulations and laboratory work 
comparisons already indicated that the scattering particles were likely large, rough with a low albedo, 
and it was shown that material from the Orgueil meteorite was a good scattering analogue
\citep{kikuchi1988,mukai1988,dollfus1989}. It was also recognized that polarimetric properties 
of dust particles were significantly changed during outbursts \citep{dollfus1988a} 
and varied related to jet structures in the coma when the {\it Giotto} mission crossed them 
\citep{levasseur-regourd1999}. Since then, improved models of cometary dust particles 
have been developed that may include a diversity of material mixtures (silicates, organics, ices, etc.)
and morphologies, such as hollow spheres, irregular particles, spheroids and aggregates thereof
\citep[see e.g.][and references
therein]{hanner2003a,Min2005,lasue2009b,kolokolova2010a,Zubko2015a}.

Polarimetric observations of 67P/C-G have been performed during both 2008 and 2015 perihelion 
passages \citep{hadamcik2010,hadamcik2016,rosenbush2017}. 
Agglomerates of sub-micrometer-sized grains best fit the higher polarization observed in cometary jets and after fragmentation
or disruption events, while a mixture of porous agglomerates of submicrometer-sized Mg-silicates, Fe-silicates, and carbon black 
grains mixed with compact Mg-silicates grains is generally needed to fit whole comae observations 
\citep{hadamcik2006,hadamcik2007a}.
Simple geometric shapes of dust particles are generally poor fits to the observational cometary data \citep{kolokolova2004}.
Numerical simulations  strongly suggest that cometary dust is a mixture of (possibly fractal)
agglomerates and of compact particles of both non-absorbing silicate-type materials and more absorbing organic-type materials
\citep[see e.g.][]{lasue2009b,kiselev2015a}.
The variety, structure and size distribution of agglomerates and grains 
is consistent with the general description of dust particles detected at 67P-C-G by {\it Rosetta} \citep{guttler2019a, mannel2019}.

Observations in the infrared and thermal wavelength ranges give specific information on the dust 
particle size distribution and their spatial distribution and dynamics in the cometary coma 
and tail \citep{agarwal2007}. In the particular case of 67P/C-G, VIRTIS observations of the dust
in the coma from 2 to 5~$\mu$m, 
generally show a temperature a few per cents above the equilibrium, 
but it increases 3 to 4 fold during outbursts. This may be related to the ejection of much 
smaller dust particles during outbursts (size~$<100$nm) \citep{Bockelee-Morvan2017a}.
Such small particles are not detected by other {\it Rosetta} instruments, indicating a collecting 
bias or a dearth of such particles in the cometary environment. 

A general model of light scattering and emission by dust particles consistent with   
all the observed constraints remains to be elaborated. 

\subsection{Morphology} \label{Morpho}
\label{sec:3morpho}

The morphology of dust particles corresponds to the spatial arrangement of their constituting 
components and is often described by parameters such as the porosity, which indicates
the ratio of voids to occupied volume, or the fractal dimension, a measure of the self-similarity
at different scales of the assemblages. It is critical to study the morphology of cometary 
dust particles as these properties are essential to better understand the physical properties 
of the dust, such as strength, thermal and light scattering properties of the dust, but may also 
be witness to the primitive aggregation processes in the primordial nebula. 
Initially, polarimetric observations of comets have suggested that the basic morphology of 
cometary dust particles was best explained by including a combination of porous aggregates 
and more compact dust particles \citep[see e.g.][]{lasue2009b, kolokolova2015, Kolokolova2023}
As the \Stardust \ samples were analyzed in the laboratory, the impacts on the aluminium foils 
and the aerogel demonstrated the presence of both compact and porous, easily fragmented 
dust particles \citep{Brownlee2014}. However, the samples were altered by the speed of collection
reaching about 6~m.s$^{-1}$. IDPs collected from the stratosphere also indicate a diversity of 
morphologies including very porous fragile aggregates as well as more compact particles, 
however, the effect of atmospheric traverse, collection by plane and unknown specific origin
of the particles makes it difficult to relate their precise morphologies to a particular Solar System 
body or physics phenomenon \citep[see e.g.][]{brownlee2016}. 

To date, the microscopes on-board the {\it Rosetta} space mission have provided the best in situ 
morphological analysis of cometary dust particles collected at low speeds of 1 to 15~m.s$^{-1}$
\citep{Fulle2015b} and distances from the nucleus lower than 500~km. 
The scale ranges accessible to both the atomic force microscope MIDAS and 
the microscope and mass spectrometer COSIMA were $<$~1~$\mu$m to 1~mm with 
topographic information also available \citep{bentley2016c,mannel2019, hilchenbach2016a}. 
These scales also correspond to the smallest (1~$\mu$m, \citet{bentley2016c, mannel2019})
and largest (350~$\mu$m, \citet{Langevin2016}) particles detected. All particles present 
textural substructures identified as aggregated monomers which classifies them generally 
as ``compact aggregates'' \citep{guttler2019a}. Additionally, it has been shown that 
collected particles  have a tendency to breakup as they impact the plates 
and produce clusters of fragments with a variety of morphologies ranging from shattered, 
flattened particles to rubble piles \citep{Langevin2016}.
Such a diverse set of morphologies can originate from a single type of aggregates 
and be related to different incoming velocities and  tensile strength of the particles 
\citep{hornung2016,ellerbroek2017,ellerbroek2019,Lasue2019c}.
In fact, many compact aggregates analyzed by COSIMA fragmented into smaller constituents
during analysis due to electrostatic forces \citep{hilchenbach2017}, demonstrating the relationship
between the fragments and their parent compact aggregate particles.
Similar fragmentation upon collection have been shown to occur for MIDAS as well
\citep{bentley2016c}. The surface features detected by the two microscopes on-board
at their respective scales are reminiscent of those  found in chondritic porous 
interplanetary dust particles as illustrated in Fig.~\ref{fig:comet}. 

The extension of the morphological analysis to the lowest scales accessible by MIDAS have
shown that the smallest aggregate dust particles units are down to 8~nm in size with most 
of the dust particles being fragile agglomerated dust particles and small micrometer-sized dust 
particles also formed of those subunits \citep{mannel2019}. A fragile agglomerate was also determined
to have a  fractal dimension of about 1.7 \citep{Mannel2016}
consistent with very fluffy dust particles measured by the impact GIADA instrument \citep{Fulle2015b}. 
Even though they do not present a significant fraction of the mass of the cometary nucleus, 
such fragile particles would not survive most impacts of the early solar system accretion phase
and their detection favors an accretion of planetesimals under the gravitational collapse 
of pebble model \citep{fulle2017c,blum2017b}.

In summary, we find that the cometary dust particles present an apparent scale invariance of properties
similar to those that would result from a fractal aggregation process, consistent with 
the one that would be expected to be at work during the early stages of the planetary formation 
in the early Solar System \citep[see e.g.][]{blum2018}.
The results the microscopes obtained on dust collected from 67P/C-G 
together with many other in situ measurements by the {\it Rosetta} space mission provide the first view
of the hierarchical structure of dust in comets as reviewed in \citet{guttler2019a}.

\subsection{Tensile strength}
\label{sec:4tensile}

Laboratory simulations of macroscopic agglomerates of small silica dust particles 
(diameters ranging from 0.1 to 10 $\mu$m)
by ballistic deposition were realized to simulate early aggregation of dust particles similar to the ones forming comets. 
The tensile strength of such resulting aggregates, the size of which may reach several centimeters, 
range from 1 to 6~kPa \citep{Blum2004a, guttler2009, meisner2012}. A more realistic set of simulations using 
silica dust particles and water ice particles under low temperatures ($\approx 150$~K) showed that
the tensile strength decreases linearly with the particles diameter, ranging from 4kPa to 18kPa
in agreement with previous estimates \citep{gundlach2018}. Additionally, the experiments demonstrated 
that under low temperatures, the tensile strength of water ice aggregates was comparable to 
the  data for the silica spheres. This means that at low temperatures water ice presents a specific 
surface energy similar to the one of silica, which was not expected. Perhaps at temperatures above 
150~K, the surface energy of water ice increases steeply, or sintering effects take place. 
Few direct measurements of the ejected solid material of comets are available, however, 
the tensile strength of dust particles ejected from 67P/C-G was estimated to be of the order 
of $\approx 1$~kPa from the study of the fragments distribution observed with the COSIMA experiment 
on-board {\it Rosetta} \citep{hornung2016}. 
Similarly, the meteor showers breakup observed in the Earth atmosphere can give estimates of cometary 
dust tensile strengths since they are associated with parent comets \citep{jenniskens2006a}. 
The derived tensile strengths are again extremely low, and of the same order of magnitude of 
the other estimates depending on the parent comet: from 40 to 1000~Pa in \citet{trigo-rodriguez2006} 
and from 0.4 to 150~kPa as presented in Table~2 of \citet{blum2014}.
A more detailed synthesis of tensile strength values for cometary materials at different scales 
of the cometary nucleus can be found in Fig.~10 of \citet{groussin2019}.

\subsection{Density}
\label{sec:5density}

The density of cometary dust particles is closely related to their morphology and their 
porosity,  defined as 1 minus the volume filling factor of the particle. 
As described in \S\ref{sec:3morpho}, two main morphological types of dust particles 
have been detected from cometary ejection: 
compact dust particles and very fluffy aggregated dust particles following the classification 
made by \citet{guttler2019a}.
This was seen first by the \Stardust \ samples brought back to Earth, where
high-speed impacts of 81P/Wild 2 particles generated carrot-like aerogel tracks 
for compact particles (65\% of tracks) and bulbous tracks consistent with the 
disruption of fluffy aggregates (35\% of tracks) \citep{Brownlee2014, burchell2008a, trigo-rodriguez2008}

These observations are consistent with the particle detections of GIADA 
for which compact and fluffy dust particles are detected \citep{dellacorte2015}. 
The strength of the compact particles is consistent with a microporosity ranging 
from 34\% to 85\% \citep{Levasseur-Regourd2018b}. 
Dust showers observed by GIADA can only be explained by fractal aggregates with 
dimensions lower than 2 getting fragmented a few meters from the spacecraft \citep{Fulle2015b}, 
this represents about 30\% of the dust detected \citep{fulle2017c}. 
MIDAS also detected an extremely porous particle with fractal dimension $D_f = 1.7 \pm 0.1$
which would translate to a porosity around 99\% \citep{Mannel2016, mannel2019, fulle2017c}.
Following \citet{guttler2019a}, fluffy aggregated particles are expected to have a microporosity 
$>90\%$. Studies of the mean free path of light through particles fragments detected by COSIMA
have also indicated a microporosity $>50\%$ \citep{langevin2017a}.

The GIADA measurements combine both the geometric cross section of the particles 
and the momentum of impact, which allows to retrieve the average density of the particles. 
The value obtained over 271 compact particles detections gives 
$\rho = 785^{+520}_{-115}$~kg~m$^{-3}$, at $1\sigma$ confidence level \citep{fulle2017d}.
With a dust microporosity estimated to be $59\pm8 \%$, this corresponds to a  bulk density of compacted 
and dried dust of $1925^{+2030}_{-560}$~kg~m$^{-3}$, at $1\sigma$ \citep{fulle2017d}.  
Additionally, a significant fraction of dust particles detected by GIADA has a bulk density 
larger than $4000$~kg~m$^{-3}$. Those are interpreted as single grain minerals similar to 
single mineral tracks in \Stardust \ \citep{burchell2008a}.

\subsection{Electrical properties}
\label{sec:5electrical}

As dust particles are ejected from a comet, they get exposed to space plasma and UV radiations
and become electrically charged, which influences their motion \citep{horanyi1996}.  
While there were evidence of particle charging in the coma of 1P/Halley from the calculations of particles trajectories \citep{ellis1991}, the {\it Rosetta} mission provided the first direct evidence of electrically charged nanodust particles in a cometary coma \citep{burch2015,llera2020}. 
It is typically estimated that a $10^{-19}$~kg dust particle will be disrupted by its charging 
if its tensile strength is less than about 0.5 MPa \citep{mendis2013}. 

The dust showers detected by GIADA have been modelled in terms of mm-sized fluffy dust aggregates
charged by the flux of secondary electrons from the spacecraft decreasing their electric potential by 7 to 15V \citep{Fulle2015b}. The particles are disrupted by their interaction with the electric field of the spacecraft, their deceleration provides the appropriate kinetic energy to explain the RPC/IES charged nanodust detections of 0.2 to 20keV \citep{Fulle2016c}. The predicted fractal dimension of such fluffy aggregates is about 2, consistent with the fractal dimensions measured by MIDAS on some particles 
\citep[][$D_f=1.7 \pm 0.1$]{Mannel2016, mannel2019}. 

During COSIMA TOF-SIMS measurements, partial charging of the collected dust particles allowed
the direct determination of their bulk electrical properties like their specific resistivity
($> 1.2 \times 10^{10} \Omega~m $) and the real part of their relative electrical permittivity
($< 1.2$) \citep{Hornung2020}. These values are consistent with a dust porosity larger than 80\%.

\section{FUTURE DIRECTIONS}
Analysis of the 81P/Wild 2 coma grains in particular has revealed how similar inorganic comet solids are to some asteroidal materials, at least for comet Wild 2. The observation of crystalline silicates in Hale-Bopp, and the presence in the coma of comet Wild 2 of a significant fraction of coarse-grained inorganic mineral grains, including very refractory materials, was largely unexpected, and has profoundly influenced models of early solar system dynamics. The apparent lack of a significant fraction of amorphous and presolar materials in Wild 2 was another surprise, although the collection process in aerogel may have significantly destroyed these materials.  The nature of cometary organics was not a major goal of the \Stardust \ mission, and thus great uncertainty remains for this topic. The {\it Rosetta} mission contributed to the analysis of cometary organics, confirming that cometary particles can contain a large proportion of organics (as first seen in the CHON grains in comet Halley), and that this organic matter bears some similarities with the refractory organics present in carbonaceous chondrites. The organic matter present in CA-IDPs and in CP-MMs also bear similarities with that of carbonaceous chondrites, as for two of the three organic phases identified in UCAMMs. The formation of the N-rich organic phase in UCAMMs could have been made by irradiation of N-rich ices in the outer regions of the protoplanetary disk, by Galactic cosmic rays. The formation and incorporation mechanisms of mineral and organic components in comets are not yet fully understood. Cometary dust (or the ice that initially contained it) also carry soluble organics like amino acids. Cometary dust thus could have contributed to the input of prebiotic matter on the early Earth. The presence of a ``continuum" between asteroidal and cometary matter is now considered seriously, asteroidal components being found in cometary material, and cometary activity being observed in asteroids (the so-called "main-belt comets"). 

Thus, despite great recent progress in our understanding of comets, critical gaps remain concerning the formation and processing of organics, condensation of inorganic volatiles, nature and role of presolar dust in the evolution of early nebular solids, timing and location of condensation and processing of cometary materials, possible role of radiogenic nuclides including $^{26}$Al... The comparison between properties of cometary dust deduced by astronomical observations or analyzed by different techniques also sometimes shows discrepancies. In order to address these issues, a cryogenic sample return from a comet nucleus would be a dream, with well preserved cometary samples available for analysis in terrestrial laboratories \citep{Bockelee-Morvan2019}. For the time being, it however remains just that - a dream.

\vskip .2in
\noindent \textbf{Acknowledgments} \\

The micrometeorite collections are performed thanks to the logistic support of the French and Italian Polar institutes (IPEV and PNRA). C.E. acknowledges the support in France of ANR (COMETOR ANR-18-CE31-0011), CNRS, IN2P3, LabEx P2IO, DIM-ACAV+ and CNES (Rosetta and MIAMI-H2). J.L. acknowledges the support in France of the Programme National de Planétologie (PNP) of CNRS/INSU, and of CNES for the Rosetta mission. D.H.W. thanks D.E.~Harker for sharing thermal models prior to publication in \citet{Harker2022} as well as that article's co-authors C.E.~Woodward and M.S.P.~Kelley regarding many subtleties and details discussed here, T.~Ootsubo for sharing data on 21P/G-Z, and NASA Ames Space Science and Astrobiology Division for support for time on this chapter. 
M.E.Z. thanks NASA for support of astromaterials research and curation, and the Cosmic Dust Program and \Stardust \ Mission. The authors thank the two reviewers, D. Baklouti and B. T. De Gregorio for their helpful and constructive comments.

\end{document}